\documentstyle[12pt,fleqn,epsf]{report}
 
\begin{document}
\pagestyle{plain}
%\pagenumbering{roman}
%
% title
%
\begin{titlepage}
\centerline{\large\bf Isospin Physics in Heavy-ion Collisions at Intermediate
Energies}
\vspace{1.5cm}
\centerline{\bf Bao-An Li\footnote{email: Bali@comp.tamu.edu}$^{\rm a}$, 
Che Ming Ko\footnote{email: Ko@comp.tamu.edu}$^{\rm a}$ and 
Wolfgang Bauer\footnote{email: Bauer@nscl.msu.edu}$^{\rm b}$}
\vspace{0.5cm}
\centerline{{\rm\bf $^a$}Department of Physics and Cyclotron Institute}
\centerline{Texas A\&M University, College Station, TX 77843-3366, USA}
\vspace{0.3 cm}
\centerline{{\rm\bf $^b$}National Superconducting Cyclotron Laboratory }
\centerline{and Department of Physics and Astronomy}
\centerline{Michigan State University, East Lansing, MI 48824-1321, USA}
\vspace{1cm}
\centerline{\bf Abstract}
\vspace{0.5cm}                                                

In nuclear collisions induced by stable or radioactive neutron-rich nuclei 
a transient state of nuclear matter with an appreciable isospin asymmetry 
as well as thermal and compressional excitation 
can be created.  This offers the possibility to 
study the properties of nuclear matter in the region between 
symmetric nuclear matter and pure neutron matter.
In this review, we discuss recent theoretical studies of the 
equation of state of isospin-asymmetric nuclear matter 
and its relations to the properties of neutron stars and radioactive nuclei. 
Chemical and mechanical instabilities as well as the liquid-gas 
phase transition in asymmetric nuclear matter are investigated.  The 
in-medium nucleon-nucleon cross sections at different isospin states are 
reviewed as they affect significantly the dynamics of 
heavy ion collisions induced by radioactive beams. We then discuss
an isospin-dependent transport model, which includes different mean-field
potentials and cross sections for the proton and neutron,
and its application to these reactions. Furthermore, we review the 
comparisons between theoretical predictions 
and available experimental data.  In particular, we discuss the study 
of nuclear stopping in terms of isospin equilibration, the dependence of 
nuclear collective flow and balance energy on the isospin-dependent 
nuclear equation of state and cross sections, the isospin dependence of 
total nuclear reaction cross sections, and the role of isospin in 
preequilibrium nucleon emissions and subthreshold pion production.

\end{titlepage}
 
\tableofcontents
\clearpage
%\listoffigures
%\listoftables
%\clearpage
%\doublespacing
\chapter{Introduction}

Recent advances in experiments using radioactive beams with
large neutron or proton excess have made it possible to create
not only nuclei at the limits of stability, but also a transient state 
of nuclear matter with appreciable isospin asymmetry, thermal excitation, 
and compression. 
Extensive reviews have been written on the properties of 
these nuclei, and they can be found in Refs. 
\cite{iso91,hus91,boyd92,ms93,tanihata95,gei95,hansen95,spiral}. In this 
article we will review instead the role of the isospin degree of freedom
in heavy-ion collisions at intermediate energies. The main goal of such
studies is to probe the properties of nuclear matter in the 
region between symmetric nuclear matter and pure neutron matter. 
This information is important in understanding the explosion mechanism of
supernova and the cooling rate of neutron stars. For example, the prompt 
shock invoked to understand the explosion mechanism of a type {\sc ii}
supernova
requires a relatively soft equation of state ( {\sc eos}) \cite{bck}, which 
can be 
understood in terms of the dependence of the nuclear compressibility on 
isospin. In the model for prompt explosion \cite{kahana}, 
the electron-capture reaction drives the star in the latest stage of collapse
to an equilibrium state where the proton concentration is about $1/3$,
which, according to microscopic many-body calculations, reduces the 
compressibility by about 30\% compared to that for symmetric nuclear 
matter. Moreover, the magnitude of proton concentration at $\beta$ equilibrium 
in a neutron star is almost entirely determined by the isospin-dependent 
part of the nuclear  {\sc eos}, i.e., the symmetry energy. 
The proton fraction affects not only the stiffness of the  {\sc eos} but also 
the cooling mechanisms of neutron stars \cite{lat91,sum94} and the 
possibility of kaon condensation ($e^{-}\rightarrow K^{-}\nu_e$)
in dense stellar matter\cite{chlee}.
If the proton concentration is larger than a critical value of about 15\%, 
the direct {\sc urca} process 
$(n\rightarrow p+e^-+\bar{\nu}_e,~p+e^-\rightarrow 
n+\nu_e)$ becomes possible, and would then enhance the emission of neutrinos, 
making it a more important process in the cooling of a neutron 
star \cite{lat91}. 

Usually, the properties of asymmetric nuclear matter are inferred from 
that of symmetric and pure neutron matter by using a parabolic 
approximation \cite{bru}, and this seems to be supported by 
studies based on detailed many-body calculations. Although different 
assumptions have been introduced in these theoretical approaches, they all
predict similar results for the asymmetric nuclear matter.
In particular, the density and compressibility 
at saturation are predicted to decrease as the nuclear matter becomes 
more neutron-rich. Also, the instability that leads to the liquid-gas phase 
separation in asymmetric nuclear matter is found to arise from the  
chemical instability due to fluctuations in isospin-asymmetry 
instead from the mechanical instability as a result of fluctuations in 
baryon density as in symmetric 
nuclear matter \cite{serot}. Furthermore, the nature of liquid-gas 
phase transition in asymmetric nuclear matter is predicted to be 
different from that in symmetric nuclear matter. 
For example, M\"uller and Serot have recently shown that 
the liquid-gas phase transition is second order 
rather than first order as in symmetric nuclear matter \cite{muller}. 

To relate the theoretical predictions to the experimental observations 
in heavy ion collisions with radioactive beams, the isospin-dependent 
nuclear transport model, which uses different mean-field
potentials and cross sections for the proton and neutron, has 
been found to be very successful.  Based on this model, it has been shown that
many observed phenomena are mainly determined by the 
isospin-dependent nuclear equation of state and in-medium nucleon-nucleon 
cross sections. Other models such as the percolation and lattice gas models 
have also been found useful in understanding some of the phenomena. 
We shall review the applications of these models in studying several 
topics of isospin physics.  These include the equilibration of the 
isospin degree of freedom as a probe of nuclear stopping, the isospin 
dependence of the nuclear collective flow and balance energy,
preequilibrium nucleon emissions, and subthreshold pion production as well
as the determination of the radii of neutron-rich nuclei from the total
nuclear reaction cross sections,  An overview and outlook 
will be given at the end.

\chapter{Isospin dependence of the nuclear equation of state}
Since the early work of Brueckner {\it et al.} \cite{bru} and 
Siemens \cite{siemens70} on isospin asymmetric nuclear matter in 
the late 60's, there have been many studies on this subject
based on different two-body 
and three-body forces or interaction Lagrangians.  These include the 
non-relativistic Brueckner approach \cite{sjo74,cug87,bom91}, relativistic 
Brueckner approach \cite{mut87,har87,sumi92,hub93}, variational
many-body approach \cite{fri81,laga81,wiringa88}, relativistic mean-field 
theory \cite{serot,sumi92,chin77,horo87,glen82,hir91,sug94}, 
relativistic and non-relativistic Hartree-Fock \cite{lopez88,wer94,kho96} or
Thomas-Fermi approximation \cite{kar85,band90}, and the chiral sigma model 
\cite{pra87}.  Instead of discussing details of these models and their 
predictions, we review the most important, common features of the
 {\sc eos} of asymmetric nuclear matter predicted by these models. Also,
we shall point out the most obvious, qualitative difference among 
the model predictions. For example, at densities greater than the normal
nuclear saturation density, it may become important to include the 
relativistic effects \cite{bro90}, which seem already to be needed in
obtaining the correct nuclear matter binding energy and saturation density.
In addition, causality is violated
at high densities in many non-relativistic models. 
We shall therefore also distinguish the predictions between  
relativistic and non-relativistic approaches.  

\begin{figure}[htp]
\vspace{-5.0truecm}
\setlength{\epsfxsize=10truecm}
\centerline{\epsffile{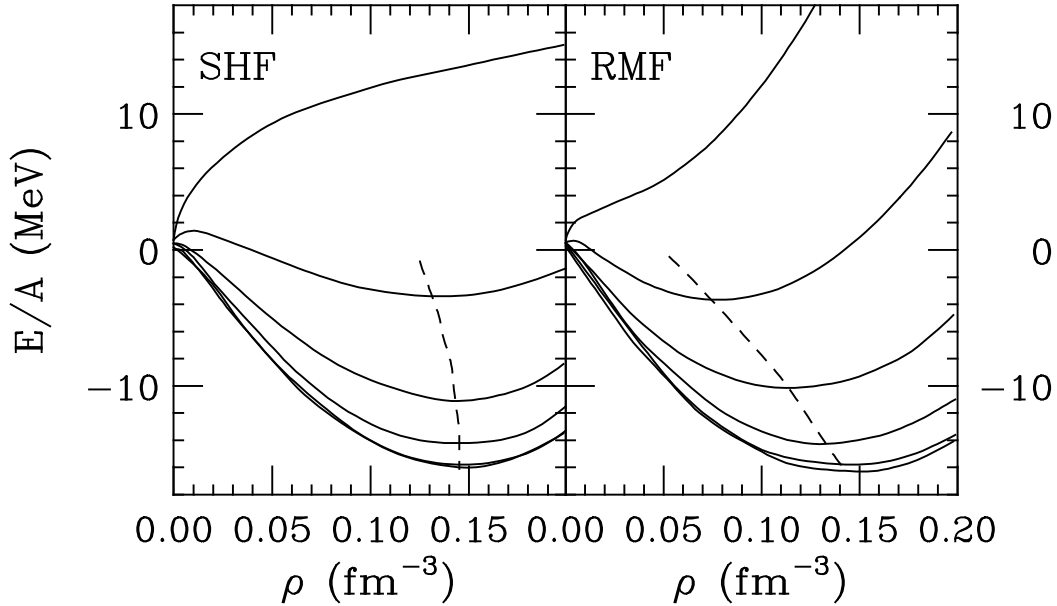} 
\caption{The equation of state of asymmetric nuclear matter from 
the Skyrme-Hartree-Fock (left) and relativistic mean-field (right) model
calculations. The solid curves correspond to neutron-to-proton ratios, 
$\rho_n/\rho_p$, of (from top to bottom) 0, 0.2, 0.4, 0.6, 0.8, and 1.
Results taken from Ref.\ \protect\cite{tanihata96}.}\label{1eos}} 
\end{figure}
\begin{figure}[htp]
\vspace{-5.0truecm}
\setlength{\epsfxsize=10truecm}
\centerline{\epsffile{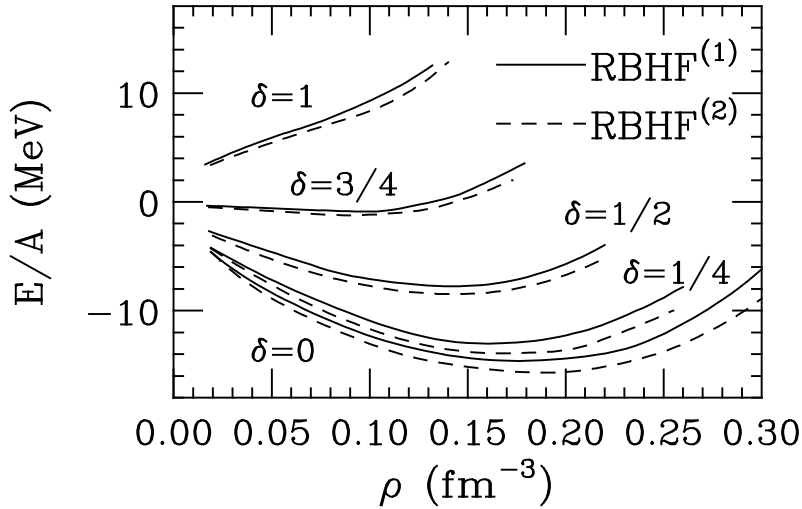}
\caption{Same as Fig. 2.1 from
the relativistic Brueckner-Hartree-Fock calculations. The solid and dashed 
curves correspond, respectively, to the case with and without momentum 
dependent self-energies. Taken from Ref.\ \protect\cite{hub93}.}\label{2eos}} 
\end{figure}

  
Fig.\ \ref{1eos} and Fig.\ \ref{2eos} show three typical predictions for the
 {\sc eos} of asymmetric nuclear matter from the non-relativistic 
Skyrme-Hartree-Fock ({\sc shf}) model using the parameter set 
{\sc siii} \cite{tanihata96}, the relativistic mean-field ({\sc rmf}) 
model using the parameter set {\sc tm1} \cite{sug94} and the relativistic 
Brueckner-Hartree-Fock approach with or without the momentum-dependent  
self-energies \cite{hub93}. The isospin asymmetry is indicated for each 
curve by the neutron to proton ratio $\rho_n/\rho_p$ in Fig.\ \ref{1eos} and 
the relative neutron excess 
\begin{equation}
\delta \equiv \frac{\rho_n-\rho_p}{\rho_n+\rho_p}
       \equiv \frac{\rho_n-\rho_p}{\rho}
\end{equation}
in Fig.\ \ref{2eos}. A common prediction from these studies is 
that the asymmetric nuclear matter is less stiff and bound at saturation. 
The minimum in the equation of state, i.e., the energy per nucleon versus 
density, disappears before the pure neutron matter limit is reached,
and the compressibility at saturation thus decreases 
as nuclear matter becomes more neutron-rich.
Also, the saturation density is generally reduced with increasing 
neutron/proton ratio. Although all these models give the correct 
saturation properties for symmetric nuclear matter, their predictions 
for the  {\sc eos} of asymmetric nuclear matter, such as
the saturation density, are quantitatively different, 
In the {\sc shf} model the saturation density depends weakly on the isospin
asymmetry, while in the {\sc rmf} model the dependence is much stronger.
This difference, as we shall see in 
section \ref{radio}, results in significant differences in the density 
distribution and the thickness of neutron skins of radioactive nuclei.  
More detailed 
comparisons among results from the relativistic mean-field theory, 
relativistic Brueckner Hartree-Fock approach, and Skyrme-Hartree-Fock approach 
can be found in Refs.\ \cite{sumi92,sumi93}.

\section{Isospin dependence of the nuclear matter compressibility, 
binding energy, and density at saturation}\label{saturation}

From the  {\sc eos} of asymmetric nuclear matter, i.e, the energy per nucleon 
$e(\rho,\delta)\equiv E/A$, the compressibility 
can be calculated from 
\begin{equation}
K(\rho,\delta)=9\frac{\partial P(\rho,\delta)}{\partial \rho},
\end{equation}
where the pressure $P$ is given by
\begin{equation}\label{pressure}
P(\rho,\delta)=\rho^2\frac{\partial}{\partial \rho}e(\rho,\delta).
\end{equation}
Here, $\rho$ and $\delta$ are the baryon density and the relative neutron 
excess, respectively.
In most models the compressibility $K$ decreases as 
the matter becomes more neutron-rich. But the rate of decrease varies widely 
among models, in particular at high densities. 

\begin{figure}[htp]
%\vspace{10.5cm}
\vspace{-5.0truecm}
\setlength{\epsfxsize=10truecm}
\centerline{\epsffile{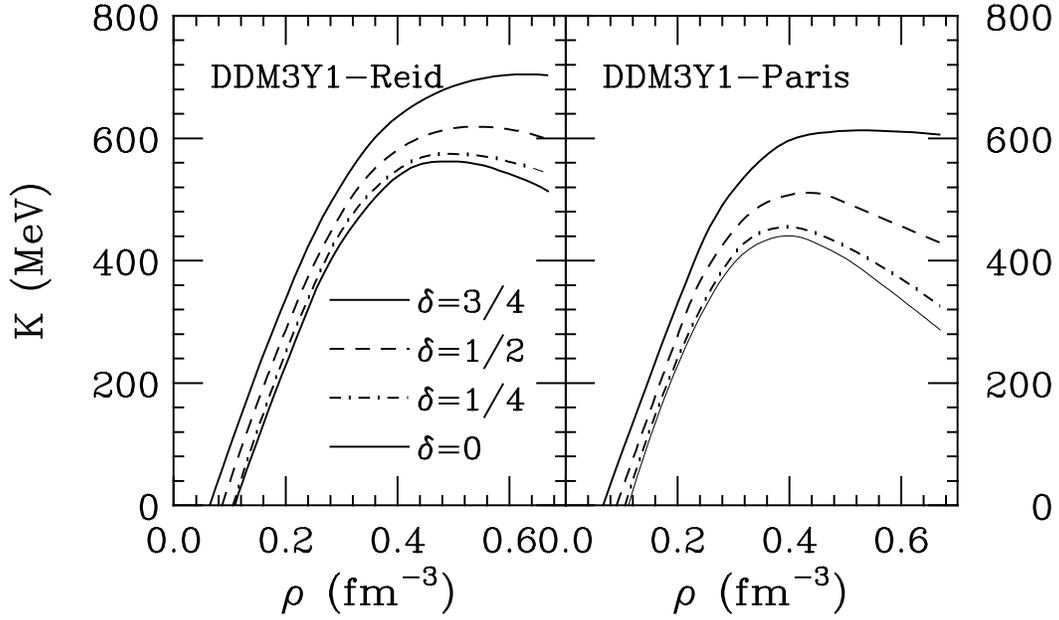}}
\caption{The nuclear compressibility $K$ obtained from 
the non-relativistic Hartree-Fock calculations using the density-dependent
Reid (left window) and Paris (right window) nucleon-nucleon interactions, 
respectively. Results taken from Ref.\ \protect\cite{kho96}.}\label{kho} 
\end{figure}  
Fig.\ \ref{kho} shows the results from a non-relativistic 
Hartree-Fock calculation \cite{kho96}. The upper and lower windows are
obtained using the density-dependent Reid and Paris nucleon-nucleon 
interactions, respectively.  One finds from these studies that neutrons 
tend to make the nuclear  {\sc eos} stiffer. Also, at densities above about
$1.5\rho_0$ the predicted density dependence of nuclear compressibility
depends sensitively on the 
isospin independent part of the nucleon-nucleon interaction \cite{kho96}.
This is because the strength of isospin-dependent interaction is generally
much weaker than the isospin-independent one. 
The saturation density at each relative neutron excess
$\delta$ is indicated by the solid dot in Fig.\ \ref{kho} and is seen to 
shift towards a lower density as the nuclear matter becomes 
more neutron-rich. A similar decrease is seen for the nuclear compressibility 
at saturation density as a result of its strong density dependence.
\begin{figure}[htp]
%\vspace{10cm}
\vspace{-5.0truecm}
\setlength{\epsfxsize=10truecm}
\centerline{\epsffile{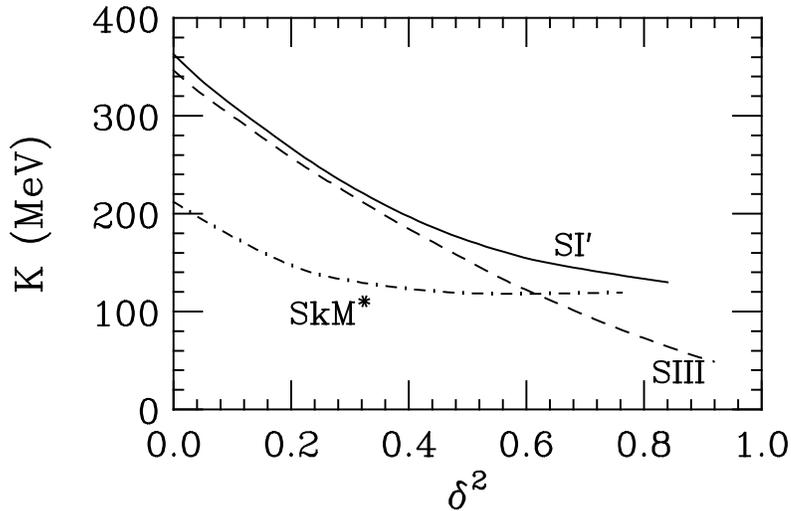}}
\caption{Nuclear compressibility as a function of relative neutron
excess, obtained from 
the extended Thomas-Fermi calculations using three different 
Skyrme interactions. Results taken from Ref.\ \protect\cite{kar85}.}\label{kar} 
\end{figure}  

A non-relativistic study of the isospin dependence of the nuclear 
compressibility and density at saturation has been carried out by 
Kolehmainen {\it et al.}\ \cite{kar85} using the Skyrme-type interaction 
in an extended Thomas-Fermi approach. Their prediction on the
isospin dependence of the nuclear compressibility at saturation $K_0(\delta)$ 
is shown in Fig.\ \ref{kar}, and their results can be parameterized as
a quadratic function of $\delta$
\begin{equation}\label{k0d}
K_0(\delta)=K_0(0)(1-a\delta^2),
\end{equation}
\begin{table}
\caption{Parameters $K_{0}$ and $a$ for the isospin dependence 
of the compressibility at saturation.}
\label{tableofak0}
\medskip
\centerline{
\begin{tabular}{cccccccc}
\hline\\
%\hline\hline\\
\multicolumn{1}{c}{Force} &\multicolumn{1}{c}{Paris}
&\multicolumn{1}{c}{{\sc skm}$^{*}$}&\multicolumn{1}{c}{SI$^{'}$}
&\multicolumn{1}{c}{{\sc siii}}&\multicolumn{1}{c}{{\sc av14+uvii}}
&\multicolumn{1}{c}{{\sc uv14+uvii}}&\multicolumn{1}{c}{{\sc fp}}\\\\
\hline\\
\multicolumn{1}{c}{$K_0(0)$} &\multicolumn{1}{c}{185}
&\multicolumn{1}{c}{216.6}&\multicolumn{1}{c}{370.3}
&\multicolumn{1}{c}{355.3}&\multicolumn{1}{c}{209}
&\multicolumn{1}{c}{202}&\multicolumn{1}{c}{238}\\\\
\multicolumn{1}{c}{$a$} &\multicolumn{1}{c}{2.027}
&\multicolumn{1}{c}{1.988}&\multicolumn{1}{c}{1.272}
&\multicolumn{1}{c}{1.275}&\multicolumn{1}{c}{2.196}
&\multicolumn{1}{c}{2.049}&\multicolumn{1}{c}{1.01}\\\\
\multicolumn{1}{c}{$K_{0}(\frac{1}{3})$} &\multicolumn{1}{c}{143.3}
&\multicolumn{1}{c}{168.7}&\multicolumn{1}{c}{318.0}
&\multicolumn{1}{c}{305.0}&\multicolumn{1}{c}{158}
&\multicolumn{1}{c}{156}&\multicolumn{1}{c}{212}\\\\
\hline\\
%\hline\hline
\end{tabular}
}
\end{table}
For small $\delta$ this 
parameterization works well and has been 
confirmed by other studies. For example, Bombaci and Lombardo \cite{bom91}
obtained a similar result using  a non-relativistic 
Brueckner-Bethe-Goldstone approach with the Paris interaction. We summarize
in Table \ref{tableofak0} the parameters $K_0(0)$ in units of MeV
and $a$ extracted from these studies \cite{bom91,fri81,kar85}
and those from fitting the  {\sc eos} calculated by Wiringa {\it et al.} 
\cite{wiringa88} using the variational many-body ({\sc vmb}) approach with 
the Argonne and Urbana two-body potentials {\sc av14} and {\sc uv14}
augmented by 
the Urbana three-body potential {\sc uvii}, as well as 
that from Friedman and Pandharipande ({\sc fp}) \cite{fri81} are also listed. 
As one expects, both $K_0(0)$ and $a$ extracted from different models or 
the same model but different nuclear interactions vary appreciably.

\begin{figure}[htp]
%\vspace{8.cm}
% fig 2.5
\vspace{-5.0truecm}
\setlength{\epsfxsize=10truecm}
\centerline{\epsffile{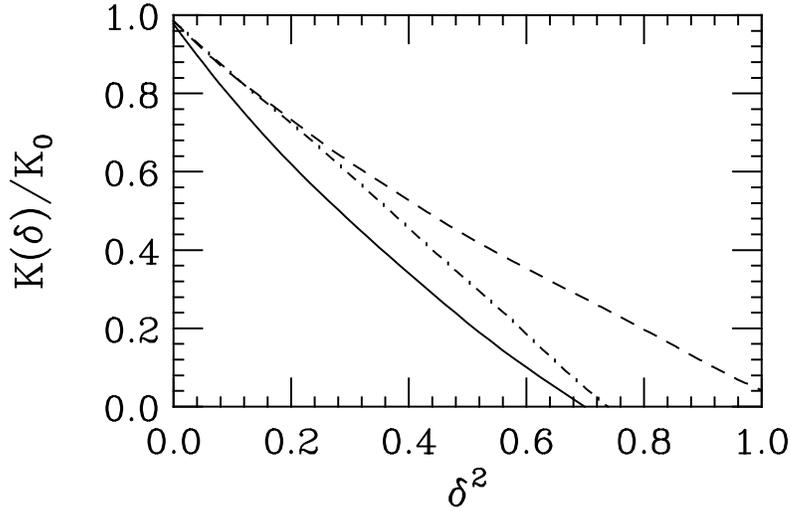}}
\caption{Relative compressibility at saturation obtained from 
the Dirac-Hartree-Fock calculations using isoscalar mesons only (dashed line),
both isoscalar and isovector (solid line), and the non-relativistic Hartree-Fock
calculations using the Skyrme {\sc siii} interaction (dot-dashed line). 
Results taken from 
Ref.\ \protect\cite{lopez88}.}\label{lopez1} 
\end{figure}  
We also show in Table 2.1 the compressibility $K_0(\frac{1}{3})$ 
at $\delta=1/3$, which is about 
20-30\% below $K_0(0)$ and is of special interest in 
nuclear astrophysics. According to the prompt explosion model of 
supernova \cite{bck}, a neutron-rich system is formed during the 
collapse of massive stars as a result of neutrino trapping, and the isospin 
asymmetry $\delta$ of a star stays almost at a constant value of 1/3. 
In this study, Eq.\ (\ref{k0d}) with the parameter set of {\sc skm}$^*$ was 
used, and it was found that the softening of nuclear  {\sc eos} due to 
the isospin asymmetry is responsible for generating a successful, prompt 
explosion. To our best knowledge, 
this is probably the only astrophysical evidence that supports the 
predicted isospin dependence of nuclear compressibility at saturation. 
In laboratory experiments, a 30\% reduction in compressibility 
seems too small to have a measurable 
effect even on nuclear collective flow which 
is believed to be the most sensitive probe to the compressibility. 
However, the corresponding changes in the 
mean-field potentials for neutrons and protons are opposite in sign, 
and this makes certain observables, such as the neutron to proton ratio 
of preequilibrium nucleons in heavy-ion collisions, sensitive to the 
isospin dependence of the  {\sc eos}. We shall discuss this 
in more detail in section \ref{npratio}.      

There are also many calculations of the nuclear compressibility using 
a relativistic approach. Although it is not easy to compare different 
model calculations as they usually involve  many parameters 
and assumptions, the comparison made in Ref.\ \cite{lopez88} 
is very instructive. Shown in
Fig.\ \ref{lopez1} is a comparison of the isospin dependence of the 
ratio $K_0(\delta)/K_0(0)$ calculated in the 
framework of a Dirac-Hartree-Fock approach using only isoscalar 
$(\sigma, \omega)$ mesons (curve b) and both isoscalar and  
isovector $(\pi, \rho)$ mesons (curve e) as well as from the 
non-relativistic Hartree-Fock approach using the Skyrme {\sc siii}
interaction (curve {\sc siii}). The general trend of decreasing
compressibility with
increasing isospin asymmetry $\delta$ is common to all three models. The 
rate of decrease is, however, sensitive to the details in the model. 
\begin{figure}[htp]
%\vspace{8.cm}
\vspace{-5.0truecm}
\setlength{\epsfxsize=10truecm}
\centerline{\epsffile{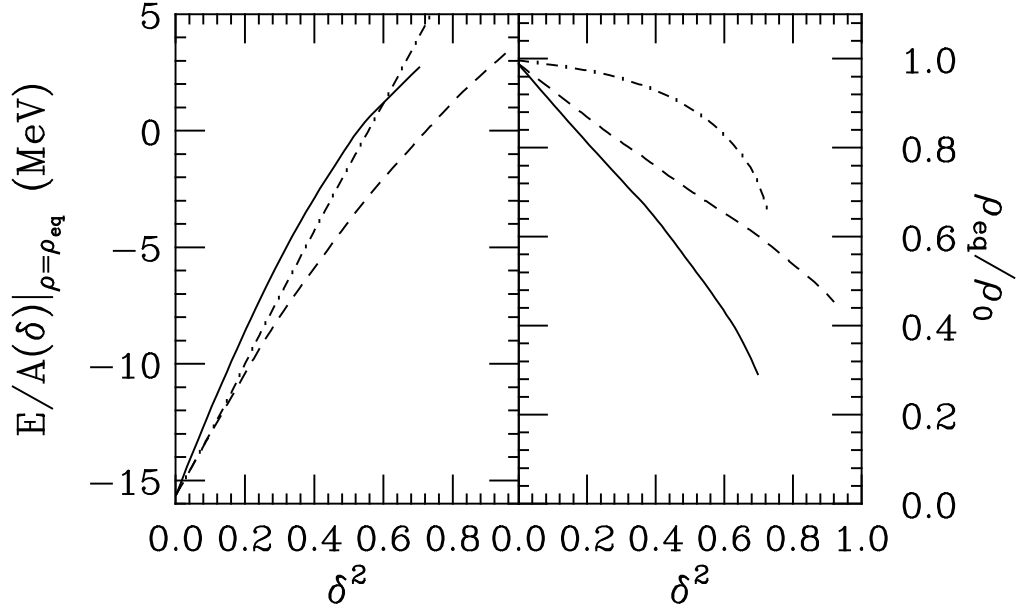}}
\caption{Same as Fig. 2.5 for the energy (left) and density (right) at 
saturation.  Results taken from Ref.\ \protect\cite{lopez88}.}\label{lopez2} 
\end{figure}  

Other properties at saturation, such as the binding energy, 
pressure and density, also show dependence on the isospin 
asymmetry $\delta$ of nuclear matter. Shown in Fig.\ \ref{lopez2} are the 
binding energy and density at saturation as functions of $\delta^2$ 
\cite{lopez88}. As for the compressibility at saturation, 
the saturation density can be parameterized as
\begin{equation}\label{sden}
\rho_{eq}(\delta)=\rho_0(0)(1-b\delta^2),
\end{equation}
In Table \ref{tableofb} we summarize the values of the parameters 
$b$ and $\rho_0(0)$ extracted from several 
models. Results from the Dirac-Hartree-Fock approach are 
denoted by {\sc dhf} (b) and {\sc dhf} (e) for using isoscalar mesons only and 
both isoscalar and isovector mesons, respectively. From 
tables \ref{tableofb} and \ref{tableofak0} one can clearly see a 
large variation of the parameters describing the isospin dependence 
of the compressibility and saturation density. Obviously, 
experimental data are needed to test these models and thus to 
narrow down the isospin-dependence of the nuclear  {\sc eos}. 
In this respect, collisions of 
neutron-rich or radioactive nuclei are promising tools. We shall come 
back to this point later.
  
\begin{table}
\caption{Parameters $\rho_{0}$ and $b$ in the isospin dependence 
of saturation density.}
\label{tableofb}
\medskip
\centerline{
\begin{tabular}{cccccccc}
\hline\\
%\hline\hline\\
\multicolumn{1}{c}{Force} &\multicolumn{1}{c}{Paris}
&\multicolumn{1}{c}{{\sc skm}$^{*}$}&\multicolumn{1}{c}{SI$^{'}$}
&\multicolumn{1}{c}{{\sc siii}}&\multicolumn{1}{c}{{\sc dhf} (b)}
&\multicolumn{1}{c}{{\sc dhf} (e)}\\\\
\hline\\
\multicolumn{1}{c}{$\rho_0(0)$} &\multicolumn{1}{c}{0.289}
&\multicolumn{1}{c}{0.1603}&\multicolumn{1}{c}{0.1553}
&\multicolumn{1}{c}{0.1453}&\multicolumn{1}{c}{0.1484}
&\multicolumn{1}{c}{0.1484}\\\\
\multicolumn{1}{c}{$b$} &\multicolumn{1}{c}{1.115}
&\multicolumn{1}{c}{0.634}&\multicolumn{1}{c}{0.286}
&\multicolumn{1}{c}{0.084}&\multicolumn{1}{c}{0.65}
&\multicolumn{1}{c}{0.9}\\\\
\hline\\
%\hline\hline
\end{tabular}
}
\end{table}
 
\section{Empirical parabolic law and the nuclear symmetry energy}\label{eoslaw}

\begin{figure}[htp]
%\vspace{9.5cm}
\vspace{-5.0truecm}
\setlength{\epsfxsize=10truecm}
\centerline{\epsffile{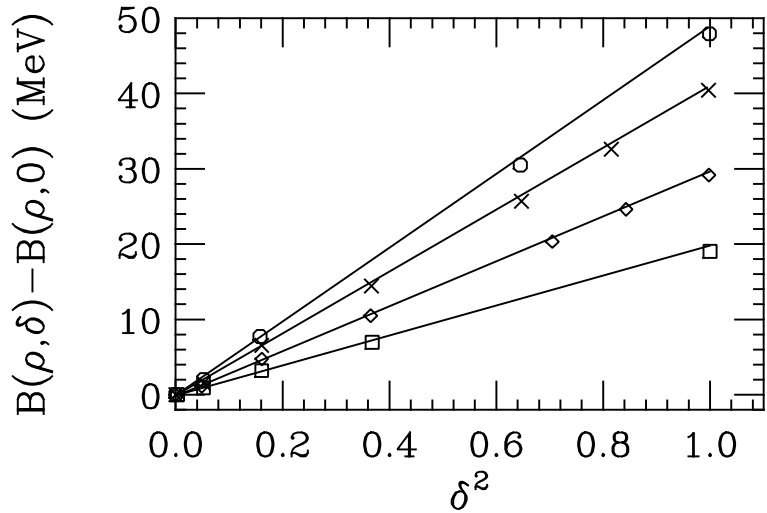}}
\caption{Total binding energy per nucleon as a function of $\delta^2$ for
densities (from top to bottom) of 0.4, 0.3, 0.17, and 0.076 fm$^{-3}$,
obtained from the non-relativistic Brueckner-Bethe-Goldstone approach.
Solid lines are the fits using the parabolic law to that from 
numerical calculations. Taken from Ref.\ \protect\cite{bom91}.}
\label{elaw1} 
\end{figure}

\begin{figure}[htp]
%\vspace{8cm}
\vspace{-5.0truecm}
\setlength{\epsfxsize=10truecm}
\centerline{\epsffile{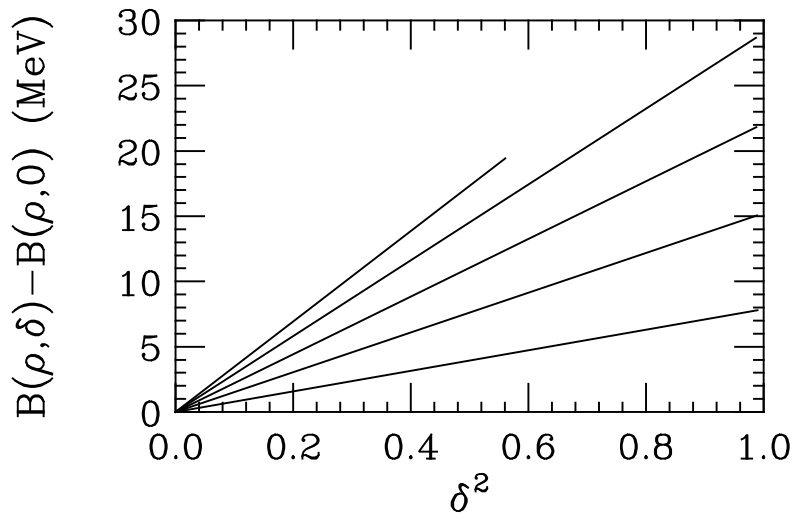}}
\caption{Same as Fig. 2.7
obtained from the relativistic Brueckner-Hartree-Fock approach.
The lines correspond to values of the density of (from top to bottom)
0.18, 0.14, 0.1, 0.06, and 0.02 fm$^{-3}$.
Results taken from Ref.\ \protect\cite{hub93}.}\label{elaw} 
\end{figure}

For asymmetric nuclear matter at densities away the saturation density, 
various theoretical studies, 
e.g., Refs. \cite{lat91,siemens70,laga81,wiringa88,baym71,thor94,prak88}, 
have shown that the energy per nucleon can be well approximated by
\begin{equation}\label{aeos0}
e(\rho,\delta)=T_{F}(\rho,\delta)+V_0(\rho)+\delta^2 V_2(\rho),
\end{equation}
where 
\begin{equation}
T_F(\rho,\delta)=\frac{3\hbar^2}{20m}\left(\frac{3\pi^2\rho}{2}\right)^{2/3}
\left[\delta^{5/3}+(1-\delta)^{5/3}\right]
\end{equation}
is the Fermi-gas kinetic energy, $V_0(\rho)$ and $V_2(\rho)$ are the 
isospin-independent and -dependent potential energies.
Higher-order terms in $\delta$ are negligible. 
For example, the magnitude of the $\delta^4$ term at $\rho_0$ has been 
estimated to be less than 1 MeV \cite{siemens70,sjo74,laga81}. 
Eq.\ (\ref{aeos0}) can be further approximated by
\begin{equation}\label{aeos}
e(\rho,\delta)= e(\rho,0)+e_{\rm sym}(\rho)\delta^2,
\end{equation}
where $e(\rho,0)$ is the  {\sc eos} of symmetric nuclear matter 
and 
\begin{equation}\label{sym}
e_{\rm sym}(\rho)=\frac{1}{2}\frac{\partial^2 e(\rho,\delta)}
{\partial \delta^2}|_{\delta=0}=\frac{5}{9}T_F(\rho,0)+V_2(\rho)
\end{equation}
is the bulk symmetry energy. At normal nuclear matter density, the
two terms in the above equation have about similar magnitude.
Eq.\ (\ref{aeos}) is known as the empirical parabolic 
law for the  {\sc eos} of asymmetric nuclear matter and is considered  to be
valid only at small isospin asymmetries. However, many non-relativistic 
and relativistic calculations have shown that it is actually valid
up to $\delta=1$.  In Fig.\ \ref{elaw1} and Fig.\ \ref{elaw}, two examples 
from the non-relativistic Brueckner-Bethe-Goldstone approach \cite{bom91} 
and the relativistic-Brueckner-Hartree-Fock approach \cite{hub93}, 
respectively, are shown for the total binding energy as a function of 
$\delta^2$ at several densities $\rho$. In both cases, the fit using 
the parabolic law shown by solid lines is indeed found to be valid
in the whole range of $\delta$.       

\begin{figure}[htp]
% Fig. 2.9
%\vspace{9.5cm}
\vspace{-5.0truecm}
\setlength{\epsfxsize=10truecm}
\centerline{\epsffile{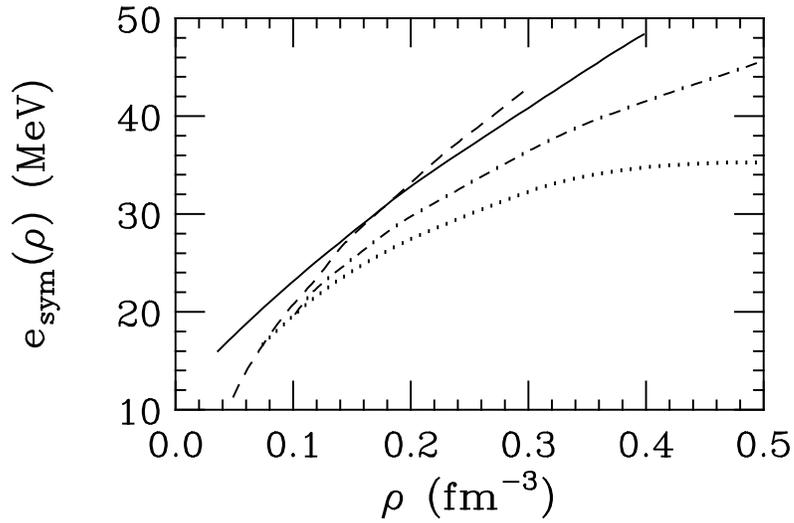}}
\caption{A comparison of the density dependence of the
symmetry energy predicted by several 
non-relativistic many-body calculations:  Paris (solid line), {\sc av14} 
(dotted line), {\sc uf14} (dot-dashed line), 
and Paris+3BF (dashed line). Results taken from ref.\
\protect\cite{cug87,bom91,wiringa88}.}
\label{esym} 
\end{figure}

Using the empirical parabolic law, one can easily extract the symmetry energy 
$e_{\rm sym}(\rho)$ from microscopic calculations. According to Eq.\ 
(\ref{aeos}) the bulk symmetry energy 
$e_{\rm sym}(\rho)$ can be evaluated from the two extreme cases of 
both pure neutron matter and symmetric nuclear matter via
\begin{equation}
e_{\rm sym}(\rho)=e(\rho,1)-e(\rho,0).
\end{equation}
Furthermore, the symmetry energy can be expanded around the
normal nuclear matter density $\rho_0$, i.e.,
\begin{equation}
e_{\rm sym}(\rho)=e_{\rm sym}(\rho_0)+\frac{L}{3}\left(\frac{\rho-\rho_0}
{\rho_0}\right)+\frac{K_{\rm sym}}{18}\left(\frac{\rho-\rho_0}
{\rho_0}\right)^2+\cdots.
\end{equation}  
In the above, $L$ and $K_{\rm sym}$ are the slope and curvature of the 
symmetry energy at normal density, i.e.,
\begin{eqnarray}
L&\equiv& 3\rho_0\frac{\partial e_{\rm sym}}{\partial \rho}|_{\rho
=\rho_0},\\
K_{\rm sym}&\equiv& 9\rho_0^2\frac{\partial^2 e_{\rm sym}}{\partial^2 
\rho}|_{\rho=\rho_0}.
\end{eqnarray}
\begin{table}
\caption{The slope and curvature of symmetry energy in units of MeV
at normal nuclear density.}
\label{tableofsym}
\medskip
\centerline{
\begin{tabular}{cccccccc}
\hline\\
%\hline\hline\\
\multicolumn{1}{c}{Force} &\multicolumn{1}{c}{Paris}
&\multicolumn{1}{c}{{\sc skm}$^{*}$}&\multicolumn{1}{c}{SI$^{'}$}
&\multicolumn{1}{c}{{\sc siii}}&\multicolumn{1}{c}{{\sc dhf} (b)}
&\multicolumn{1}{c}{{\sc dhf} (e)}\\\\
\hline\\
\multicolumn{1}{c}{$L$} 
&\multicolumn{1}{c}{68.8}&\multicolumn{1}{c}{45.78}
&\multicolumn{1}{c}{35.34}&\multicolumn{1}{c}{9.91}
&\multicolumn{1}{c}{132}&\multicolumn{1}{c}{138}\\\\
\multicolumn{1}{c}{$K_{sym}$} &\multicolumn{1}{c}{37.56}
&\multicolumn{1}{c}{-155.9}&\multicolumn{1}{c}{-259.1}
&\multicolumn{1}{c}{-393.7}&\multicolumn{1}{c}{466}
&\multicolumn{1}{c}{276}\\\\
\hline\\
%\hline\hline
\end{tabular}
}
\end{table}
With this expansion, the predicted symmetry energy at normal nuclear matter 
density can be compared with that determined from the mass formula,
which gives a value of $e_{\rm sym}(\rho_0)$ in the range of 27-36 MeV 
\cite{mass}.  Most models are tuned to give an $e_{\rm sym}(\rho_0)$ within 
this range.  For example, in the non-relativistic Hartree-Fock theory  
\cite{farine,pear}, the predicted value of $e_{\rm sym}(\rho_0)$ is
between 27 and 38 MeV depending on the nuclear interactions used in the 
calculation.  However, the relativistic mean-field ({\sc rmf} ) theory 
\cite{chin77,sug94,rein88,rufa88,shar93,scha96} usually       
gives higher values of $e_{\rm sym}(\rho_0)$ in the range of 35-42 MeV. 
What distinguishes these models around normal nuclear matter density
are thus the slope $L$ and curvature $K_{\rm sym}$. This is clearly 
illustrated in 
Fig.\ \ref{esym} where results from several non-relativistic approaches  
\cite{bom91,cug87,wiringa88} are compared.
Although all predict a $e_{\rm sym}(\rho_0)$ in the range 
of 27-30 MeV (in agreement with that from the empirical mass formula), 
it is seen that the slope and
curvature at $\rho_0$ are very different. 
As shown in Table \ref{tableofsym}, $K_{\rm sym}$ ranges from positive 
466 MeV to negative 400 MeV. The most recent
data from giant monopole resonances indicate that $K_{\rm sym}$
is between $-566\pm 1350$ MeV to $34\pm 159$ MeV \cite{shl93} depending on 
the mass region of nuclei and the number of parameters used in
parametrizing the compressibility of finite nuclei. The $K_{\rm sym}$ data 
thus still cannot distinguish these model calculations. 
\begin{figure}[htp]
%\vspace{9.5cm}
\vspace{-5.0truecm}
\setlength{\epsfxsize=10truecm}
\centerline{\epsffile{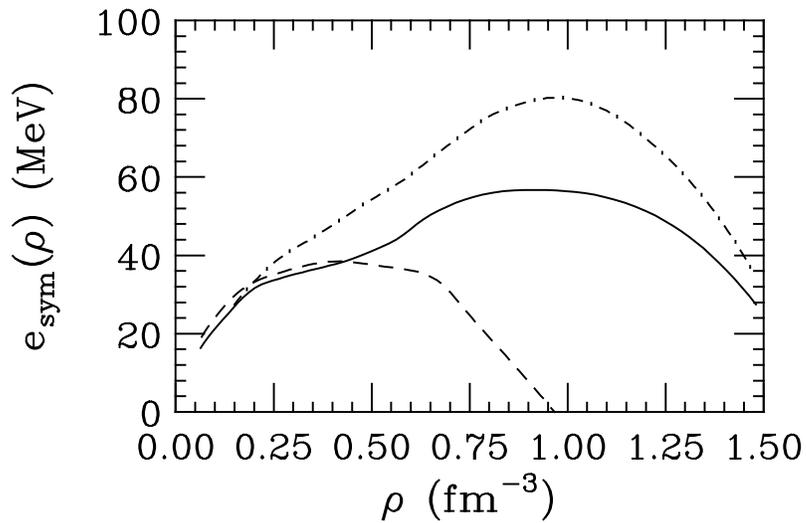}}
\caption{Symmetry energy predicted by variational 
many-body calculations: {\sc av14+uvii} (solid line), {\sc uv14+uvii}
(dot-dashed), and {\sc uv14+tni} (dashed). 
Results taken from Ref.\ \protect\cite{wiringa88}.}
\label{wiringa} 
\end{figure}  

In radioactive nuclei as well as neutron stars, the 
symmetry energy at densities away from the normal nuclear matter saturation 
density is relevant. Although the Bethe-Goldstone theory 
predicts a $\rho^{1/3}$ dependence for the symmetric energy, the density 
dependence is linear in the {\sc rmf} theory \cite{chin77,horo87,gle87}. 
Results from most other calculations cannot be simply described. 
In the most sophisticated calculations by
Wiringa {\it et al.} using the variational many-body ({\sc vmb}) 
theory \cite{wiringa88}, the density dependence of $e_{\rm sym}(\rho)$ 
has been calculated using either the Argonne two-body 
potential {\sc av14} or Urbana {\sc uv14} 
together with either the three-body potential 
{\sc uvii} or {\sc tni}, and the results differ appreciably as shown in 
Fig.\ \ref{wiringa}.
With three-body interactions the symmetry energy is seen to vanish
at high densities. Therefore, pure neutron matter was predicted by the
{\sc vmb} theory as the ground state of dense nuclear matter.  This prediction 
has important consequences on the structure and magnetic properties of 
neutron stars as stressed recently by Kutschera et al. \cite{kut94}. 
However, other models predict a nonzero symmetry energy at high densities.
\begin{figure}[htp]
%\vspace{8cm}
\vspace{-5.0truecm}
\setlength{\epsfxsize=10truecm}
\centerline{\epsffile{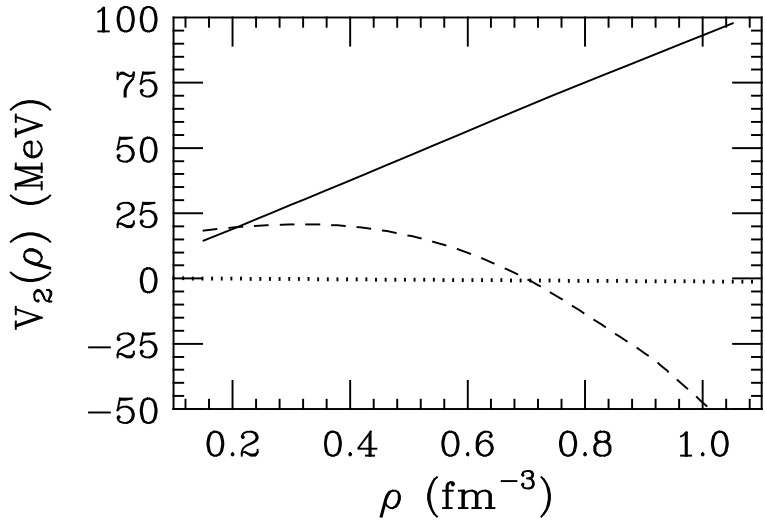}}
\caption{Contribution from nuclear interactions to the symmetry energy in 
the relativistic mean field ({\sc rmf}) theory (solid line) and the variational 
many-body ({\sc vmb}) theory (dashed line). 
Results taken from Ref.\ \protect\cite{kut94}.}
\label{v2} 
\end{figure}  
Among all models, the predictions from the {\sc rmf} theory and
the {\sc vmb} theory using the {\sc uv14+tni} interaction differ the most.
Since the kinetic contribution to the symmetry energy is about the same in
all model calculations, the differences among model predictions are mainly 
from the $V_2(\rho)$ term in Eq.\ (\ref{sym}). As shown in 
Fig.\ \ref{v2} the $V_2(\rho)$ term is always repulsive and increases
linearly with density in the {\sc rmf} 
theory, while in the {\sc vmb} theory it changes 
from repulsion to attraction as the density increases. The variation 
of $V_2(\rho)$ with density in the {\sc vmb} theory can be understood in terms
of the behavior of nuclear interactions in dense nuclear matter \cite{pan72}. 
At high densities the short-range repulsion dominates and is greater for a 
nucleon pair in an isospin singlet ($T=0$) than in an isospin triplet 
($T=1$) state.  Pure neutron matter is therefore more stable. At moderate 
densities the strong attractive isospin singlet tensor potential and 
correlations keep the isospin singlet pairs more bound, and symmetric 
nuclear matter is thus more stable  
than pure neutron matter. These features do not exist
in the {\sc rmf} theory where the asymmetry energy is due to the rho meson 
exchange, which leads to a repulsive $V_2(\rho)$ at all densities.
As a result, these two approaches predict very
different structures and chemical compositions for neutron stars \cite{kut94}. 
 
It is clear from the widely diverging model predictions
that despite great theoretical efforts 
our knowledge of the  {\sc eos} of asymmetric nuclear matter is rather limited.
In particular, the behaviour of the symmetry energy at high densities is
most uncertain among all properties of dense nuclear matter \cite{kut94}.

\section{Neutron stars and the  {\sc eos} of asymmetric 
nuclear matter}\label{star}
The mechanism for supernova explosion and the properties of neutron stars 
have been subjects of much interest; for recent reviews, 
see e.g., Refs. \cite{pet95,bethe90,pra96,pra96b,web96}. 
Here we only concentrate on the effects of the isospin-dependence of the
{\sc eos}, in particular
the density dependence of symmetry energy, on the properties
of neutron stars. Various studies have indicated that the symmetry energy
affects mainly the chemical composition of neutron stars
\cite{lat91,mut87,pra87,thor94,toki95,de97}. Other properties, such as the
cooling rate, lepton profiles and neutrino flux, which depend on the
chemical composition, will therefore also be affected.
\begin{figure}[htp]
%\vspace{8cm}
\vspace{-5.0truecm}
\setlength{\epsfxsize=10truecm}
\centerline{\epsffile{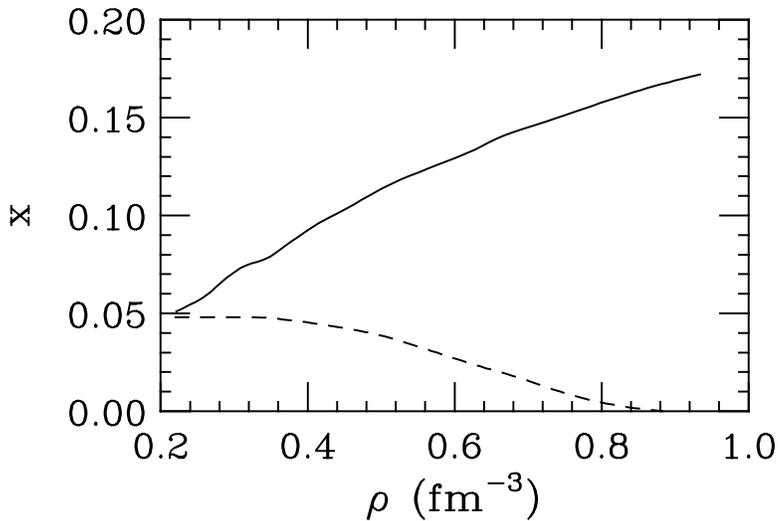}}
\caption{Equilibrium fraction of protons in neutron stars predicted by the
relativistic mean-field ({\sc rmf}, solid line) and variational 
many-body ({\sc vmb}, dashed line) theories. 
Taken from Ref.\ \protect\cite{kut94}.}
\label{xcom} 
\end{figure}  

\begin{figure}[htp]
%\vspace{9.5cm}
\vspace{-5.0truecm}
\setlength{\epsfxsize=10truecm}
\centerline{\epsffile{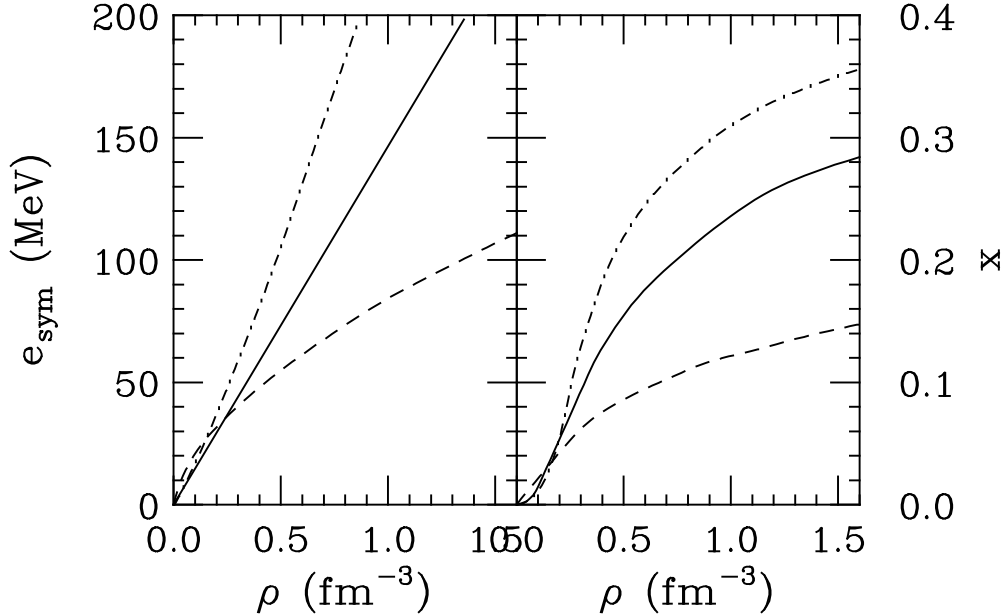}}
\caption{Left window: symmetry energy as a function of density.
Right window: corresponding equilibrium fraction of protons.
Solid lines: $F_2(u)$, dashed lines: $F_3(u)$, dash-dotted lines: $F_1(u)$. 
Results taken from Ref. \protect\cite{lat91}.}\label{lattimer} 
\end{figure}

In the most simple picture, a neutron star is composed of neutrons, protons
and electrons with a proton fraction of 
\begin{equation}
  x = {\textstyle\frac{1}{2}}(1-\delta).
\end{equation}
The condition for 
$\beta$-equilibrium in terms of the chemical potentials of electrons
$(\mu_e)$, neutrons $(\mu_n)$ and protons $(\mu_p)$ is 
\begin{equation}\label{xfraction}
\mu_e=\mu_n-\mu_p=-\frac{\partial e(\rho,\delta)}{\partial x}
=4e_{\rm sym}(\rho)(1-2x).
\end{equation}  
For relativistic degenerate electrons of density $\rho_e=\rho_p=x\rho$,
charge neutrality requires
\begin{equation}
\mu_e=(m_e^2+p_{F_e}^2)^{1/2}\approx \hbar c(3\pi^2\rho x)^{1/3},
\end{equation}
which together with Eqs.\ (\ref{xfraction}) and (\ref{aeos}) determine
an equilibrium proton fraction $x$ given by
\begin{equation}\label{fraction}
\hbar c(3\pi^2\rho x)^{1/3}=4e_{\rm sym}(\rho)(1-2x).
\end{equation}   
The equilibrium proton fraction $x$ is therefore determined solely by the
nuclear symmetry energy, $e_{\rm sym}(\rho)$. At high densities such that
$\mu_e\geq m_{\mu}$ both electrons and muons are present at 
$\beta$-equilibrium. The inclusion of muons mainly alters the value
of the equilibrium proton fraction, $x$, but not its density dependence.
Therefore, the difference in $x$ predicted by using different symmetry
energies is about the same with or without including muons
\cite{lat91,wiringa88}. As discussed in the previous section, the {\sc rmf} and
the {\sc vmb} theory with the {\sc av14+tni} interaction 
differ most in their
predictions on the symmetry energy at high densities. Their predicted
equilibrium proton fractions at high densities are therefore also very
different as shown in Fig.\ \ref{xcom}. We see that the {\sc rmf} theory
predicts a linear increase, while in the {\sc vmb} theory the proton fraction
in neutron stars gradually decreases as the density increases. The 
disappearance of protons in neutron stars is a common feature of the {\sc vmb}
theory, although the critical
density at which this occurs depends on the interaction used in the
calculation. As a result, there exists an instability in neutron stars
with respect to the separation of protons and neutrons at low densities.
This has thus led to the suggestion that polarons, which are localized
protons surrounded by neutron bubbles, can be formed in neutron stars
\cite{kut94}.  In Fig.\ \ref{lattimer} we compare the equilibrium proton
fractions (lower window) corresponding to different symmetry
energies (upper window)
\begin{equation}\label{esymfu}
e_{\rm sym}(\rho)=(2^{2/3}-1)\frac{3}{5}E_{F}^{0}[u^{2/3}-F(u)]
+e_{\rm sym}(\rho_{NM})F(u),
\end{equation} 
with $F(u)$ having one of the following three forms
\begin{eqnarray}\label{fu}
F_1(u)&=&\frac{2u^2}{1+u},\\\nonumber
F_2(u)&=&u,\\\nonumber
F_3(u)&=&u^{1/2},
\end{eqnarray}
where $u\equiv \rho/\rho_0$ is the reduced baryon density and 
$E_F^0$ is the Fermi energy. These forms of the symmetry energy 
resemble closely the three typical results from microscopic many-body
calculations discussed in the previous section. Although in all three
cases $x$ increases with density, the differences among them are still
appreciable.

The variation in symmetry energy, especially at high densities, results
in a corresponding variation in the proton fraction $x$. This has
significant effects on the cooling of neutron stars and the kaon 
condensation in dense stellar matter. In the so-called
standard model for neutron stars, their cooling is mainly due to the 
modified  {\sc urca} process
\begin{equation}
(n,p)+p+e^-\rightarrow (n,p)+n+\nu_e,~~
(n,p)+n\rightarrow (n,p)+p+e^-+\bar{\nu_e}.
\end{equation}
The direct  {\sc urca} process
\begin{equation}
n\rightarrow p+e^-+\bar{\nu_e},~~
p+e^-\rightarrow n+\nu_e
\end{equation}
is usually forbidden by energy-momentum conservation. However, it has been
shown by Lattimer {\it et al.} \cite{lat91} that if the proton fraction 
is higher than a critical value of about 1/9, the direct  {\sc urca} 
process can 
occur. This would then enhance the emission of neutrinos; so
the neutron star cooling rate is increased significantly.
From Fig.\ \ref{lattimer} it is seen that the symmetry energy determines 
entirely whether the proton fraction can exceeds the critical value and at what
density this happens. 

Another important effect of the symmetry potential is on the kaon condensation
in dense stellar matter. Since electrons with chemical potentials higher than
the kaon mass will trigger the process $e^{-}\rightarrow K^{-}\nu_e$, 
the condition for this to happen is determined by the equation 
$\mu_e\ge m_{k^-}$. It was shown in ref.\ \cite{chlee} that the critical 
density for kaon condensation depends sensitively on the form of the symmetry
potential.
\begin{figure}[htp]  
%\vspace{9.5cm}
\vspace{-5.0truecm}
\setlength{\epsfxsize=10truecm}
\centerline{\epsffile{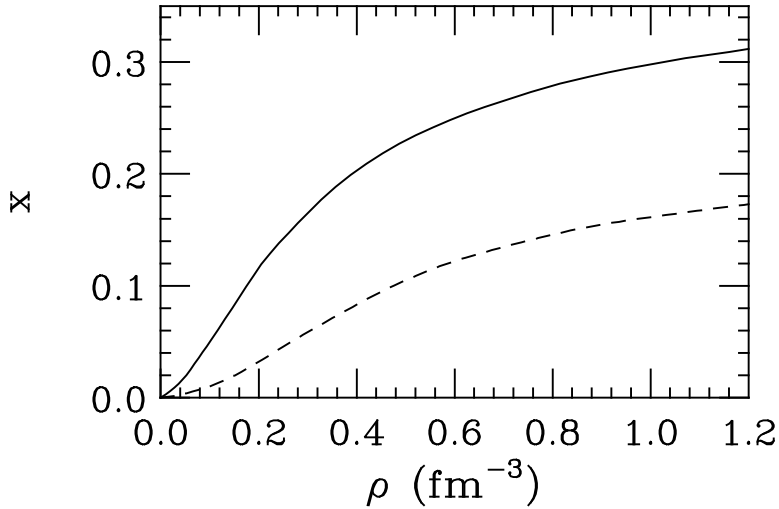}}
\caption{Equilibrium fraction of protons as a function of density obtained 
from the relativistic mean field theory. Solid curve: $g_p=5.507$,
dashed curve: $g_p=2.78$.  Results taken from 
Ref.\ \protect\cite{sumi92}.}
\label{sumi92gr} 
\end{figure}  

The dependence of equilibrium proton fraction on the underlying nuclear
interaction can be illustrated by the recent work of Sumiyoshi {\it et al.}
using the relativistic mean-field ({\sc rmf}) theory \cite{sumi92,toki95}.
The symmetry energy is found to vary almost linearly with density, and
its strength is related to the $\rho$ meson-nucleon
coupling constant $g_{\rho}$ via
\begin{equation}\label{e0sym}
e_{\rm sym}(\rho_0)=\frac{k_F^2}{6\sqrt{M^{*2}+k_F^2}}
+g_{\rho}^2\rho_0/2m_{\rho}^2,  
\end{equation}
where $M^*$ is the nucleon effective mass, $m_{\rho}$ is the $\rho$ 
meson mass, and $k_F$ is the Fermi momentum.
The first and second terms are the kinetic and potential 
contributions to the symmetry energy, respectively. From the above
expression and Eq.\ (\ref{fraction}), one
obtains the $g_{\rho}$ dependence of the proton 
fraction as shown in Fig.\ \ref{sumi92gr}.  As $g_{\rho}$ increases from 
2.78 to 5.507 the proton fraction is seen to increase by 
about a factor of two. 

Eq.\ (\ref{e0sym}) also shows that the effective mass of a nucleon, $M^*$,
affects the kinetic contribution to the symmetry energy. As nuclear
matter becomes heated, which is expected in neutron stars,
the nucleon effective mass may change. Donati {\it et al.} \cite{don94} have 
recently studied the coupling of the mean-field single-particle motion to 
the surface vibration within the quasiparticle random-phase approximation 
and found that $M^*$ decreases with increasing temperature for $T\leq 2$ MeV, 
which are the temperatures relevant for the pre-supernova collapse of 
massive stars. 
The corresponding increase of the symmetry energy was estimated to be 
about 2.5 MeV/nucleon, which would then increase 
the shock energy by about half of the observed explosion energy of SN1987a. 
The outcome of the explosion from
detailed numerical simulations thus depends sensitively on the symmetry
energy. However, more recent studies of Dean {\it et al.} \cite{dea95},
using the Monte Carlo shell model and taking into account pairing 
correlations, indicate that the symmetry energy changes by less than
0.6 MeV at the neutron star temperature.  The temperature dependence of 
the nuclear symmetry energy thus remains controversial. 
\begin{figure}[htp]
%\vspace{10cm}
\vspace{-5.0truecm}
\setlength{\epsfxsize=10truecm}
\centerline{\epsffile{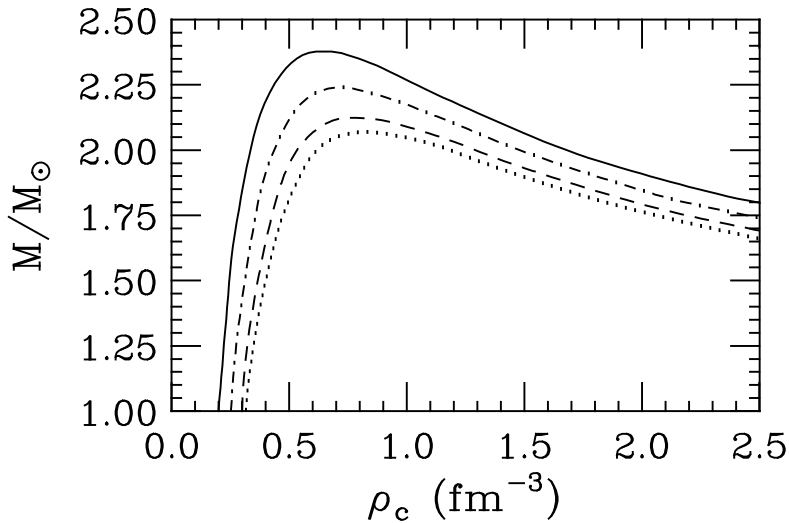}}
\caption{Mass of a neutron star $M/M_{\odot}$ (in solar mass units) 
as a function of its central density for various proton fractions, $x$.
Solid line: $x=0$, dot-dashed: $x=0.15$, dashed: $x=0.3$, dotted: $x=0.45$.
Results taken from Ref.\ \protect\cite{eng94}.}
\label{mass} 
\end{figure}
\begin{figure}[htp]
% Fig. 2.16
%\vspace{10cm}
\vspace{-5.0truecm}
\setlength{\epsfxsize=10truecm}
\centerline{\epsffile{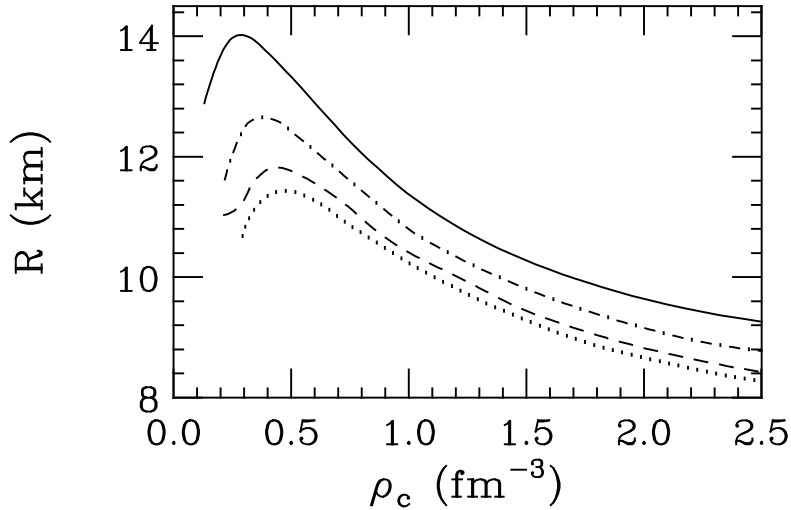}}
\caption{Radius of a neutron star as a function of its
central density for various proton fractions.
Solid line: $x=0$, dot-dashed: $x=0.15$, dashed: $x=0.3$, dotted: $x=0.45$.
Results taken from Ref.\ \protect\cite{eng94}.}
\label{radius} 
\end{figure}  

The structure or density profile of a neutron star with an isotropic 
mass distribution can be determined by solving the well-known 
Tolman-Oppenheimer-Volkov equation, i.e.,
\begin{equation}
\frac{dP}{dr}=-\frac{[\rho(r)+P(r)][M(r)+4\pi r^3P(r)]}{r^2-2rM(r)},
\end{equation}
where $P(r)$ is the pressure obtained from the  {\sc eos} via 
Eq.\ (\ref{pressure})
and $M(r)$ is the gravitational mass inside a radius $r$.
In Fig.\ \ref{mass} and Fig.\ \ref{radius} the mass and radius of a neutron 
star from a relativistic Brueckner Hartree-Fock approach are shown as a 
function of central density for various proton fractions \cite{eng94}.
A maximum mass of about $2.4~M_{\odot}$ and a radius of about $12~{\rm km}$ 
is obtained, and this is rather large compared to the 
observed value of about $1.9~M_{\odot}$, indicating 
that the underlying  {\sc eos} 
is too stiff. Pion and kaon condensates as well 
as effects of hyperons have
been suggested as possible mechanisms to soften the  {\sc eos} 
\cite{eng94}. Large 
variations of both neutron star mass and density profile are observed by
varying the proton fraction. However, the structure of a neutron star is 
indirectly affected by the symmetry energy through the equilibrium proton 
fraction. As shown in Refs.\ \cite{thor94,toki95,pra88},
the neutron star structure changes only slightly by varying $F(u)$ or 
$g_{\rho}$ in the ranges discussed above, and the stiffer the {\sc eos} is
the less 
important is the symmetry energy. This is mainly because the symmetry 
energy is much smaller compared to the 
energy per nucleon in symmetric nuclear matter. Nevertheless, as we 
have mentioned earlier, the stiffness of the nuclear  {\sc eos} is reduced 
by the neutrons. This reduction is important for generating a successful, 
prompt supernova explosion \cite{bck}.

\section{Radioactive nuclei and the
 {\sc eos} of asymmetric nuclear matter}\label{radio}

The radius, thickness of neutron skins, deformation, 
binding energy, density distributions, and other properties 
of radioactive nuclei near the
drip lines depend sensitively on the isospin-dependence of the
nuclear  {\sc eos}.  Tanihata has recently stressed the
possibility of extracting the  {\sc eos} of asymmetric nuclear matter through
the investigation of these properties \cite{tanihata96}. In this review
we discuss a complementary approach in extracting the isospin-dependent
{\sc eos} 
via reactions of neutron-rich nuclei at intermediate energies.
We shall first discuss the dependence of the properties
of radioactive nuclei on the isospin-dependent  {\sc eos} and then
demonstrate how these properties are sensitive to
the symmetry energy used in transport models for studying the dynamics 
of heavy-ion collisions.
\begin{figure}[htp]
% Fig. 2.17
%\vspace{9.5cm}
\vspace{-5.0truecm}
\setlength{\epsfxsize=9.5truecm}
\centerline{\epsffile{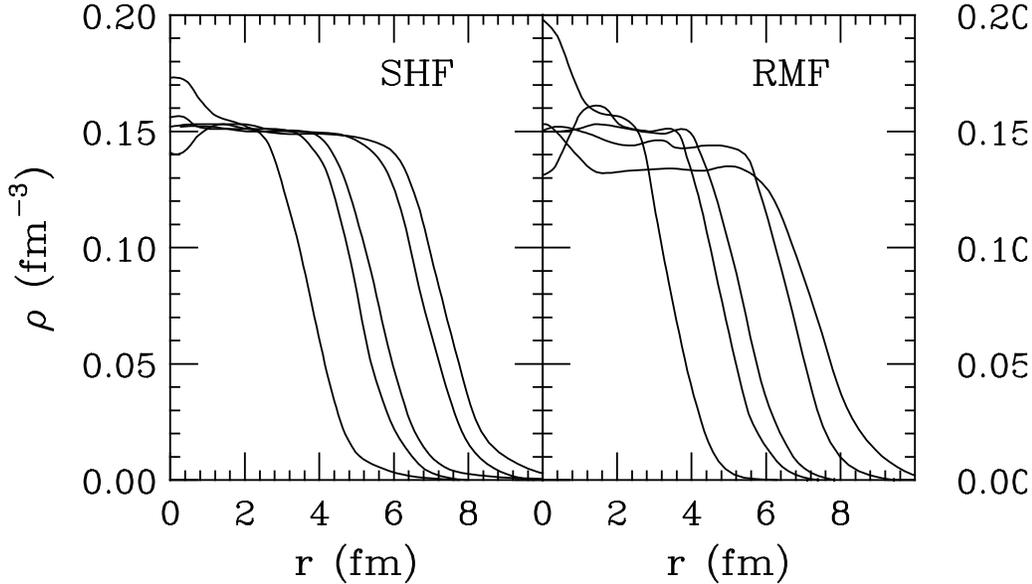}}
\caption{Density distributions for the nuclei $^{266}$Pb, $^{208}$Pb,
$^{120}$Sn, $^{90}$Zr, and $^{40}$Ca as 
predicted by the Skyrme-Hartree-Fock (left) 
and the Relativistic Mean-Field (right) model. 
Results taken from Ref.\ \protect\cite{tanihata96}.}
\label{density} 
\end{figure}  

In Fig.\ \ref{1eos} the predictions on the  {\sc eos} from the 
Skyrme-Hartree-Fock and the relativistic mean-field theories 
are compared. 
The properties of neutron rich nuclei have also been studied using these 
models. In Fig.\ \ref{density}, density distributions predicted by the two
models for several nuclei are shown \cite{tanihata96}. A comparison of
the results for $^{208}$Pb and $^{266}$Pb is particularly interesting.
One sees that, although the central density from {\sc shf} does not change 
from $^{208}$Pb to $^{266}$Pb, it decreases in {\sc rmf}. This is related to
the fact that in {\sc shf} the saturation density essentially does not change, 
but in {\sc rmf} it decreases rapidly as nuclear matter becomes more 
neutron-rich as shown by the thick dashed lines in Fig.\ \ref{1eos}. 
As a result, the {\sc rmf} gives a more prominent neutron skin for 
$^{266}{\rm Pb}$ than for $^{206}{\rm Pb}$. 
\begin{figure}[htp]
%\vspace{9.5cm}
\vspace{-5.0truecm}
\setlength{\epsfxsize=10truecm}
\centerline{\epsffile{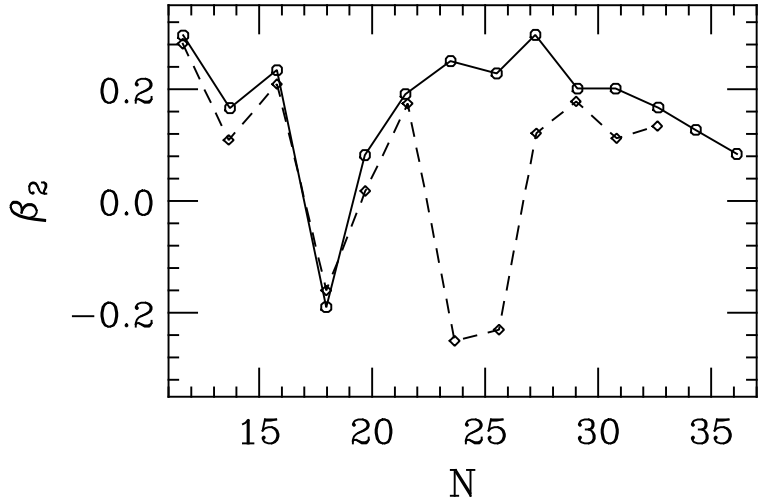}}
\caption{Quadrupole deformations of sulfur isotopes predicted by 
the {\sc shf} (dashed line connecting the squares) and 
{\sc rmf}  {\sc urca} (solid line connecting the circles) models. Results
taken from Ref.\ \protect\cite{wer94b}.}
\label{deform} 
\end{figure}  
\begin{figure}[htp]
%\vspace{8.cm}
\vspace{-5.0truecm}
\setlength{\epsfxsize=10truecm}
\centerline{\epsffile{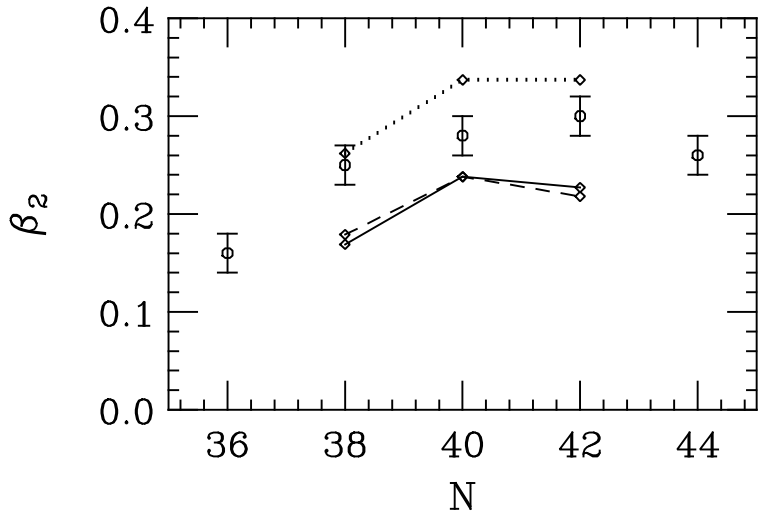}}
\caption{Comparison of the quadrupole deformations of sulfur 
isotopes. Points with error bars are the experimental data, while 
the dashed, solid, and dotted lines are the predictions 
of the {\sc shf}, {\sc rmf} and shell model, respectively. 
Results taken from Ref.\ \protect\cite{sch96,gla97}.}
\label{beta2} 
\end{figure}  

In Fig.\ \ref{deform}, we compare the quadrupole deformation, $\beta_2$,
of the ground states for 
sulfur isotopes predicted by the two models \cite{wer94b}.
The deformation for even-even isotopes $^{28-38}$S is fairly similar in both
{\sc shf} and {\sc rmf} models, i.e., prolate ($\beta_2>0$)
ground states in $^{28-32,38}$S, oblate minimum in $^{34}$S, and spherical
shape in the magic nucleus $^{36}$S with N=20. However, there are significant
differences in heavier isotopes, in particular, for $^{40,42}$S. They
are predicted by the {\sc rmf} to have prolate ground states 
with $\beta_2\approx 0.25$; the oblate minima with $\beta_2\approx -0.16$ lie
about 4 MeV above the ground state. In the {\sc shf} 
calculations, however, prolate
($\beta_2\approx 0.25$) and oblate ($\beta_2\approx -0.24$) configurations 
are essentially degenerate. 

A recent experiment at Michigan State University by Scheit {\it et al.} 
\cite{sch96}
has confirmed the predicted large deformation of $^{40,42}$S, but could 
not distinguish the sign of deformation. 
In this experiment the energies and $B(E2;0^+_{g.s.}\rightarrow 2^+_1)$ 
values of the $2^+_1$ states of $^{38,40,42}$S were measured by
Coulomb excitations.  In Fig.\ \ref{beta2} 
the measured deformation $|\beta_2|$ is compared to 
the predictions of the {\sc shf}, {\sc rmf} and shell model.  Both,
{\sc shf} and {\sc rmf} calculations 
slightly underpredict the deformation, while the 
shell model of Brown slightly overpredicts the deformation. Clearly,
additional information about the sign of deformation is needed to distinguish 
the prediction of these models about the nuclear  {\sc eos} .   
\begin{figure}[htp]
%\vspace{9.cm}
\vspace{-5.0truecm}
\setlength{\epsfxsize=10truecm}
\centerline{\epsffile{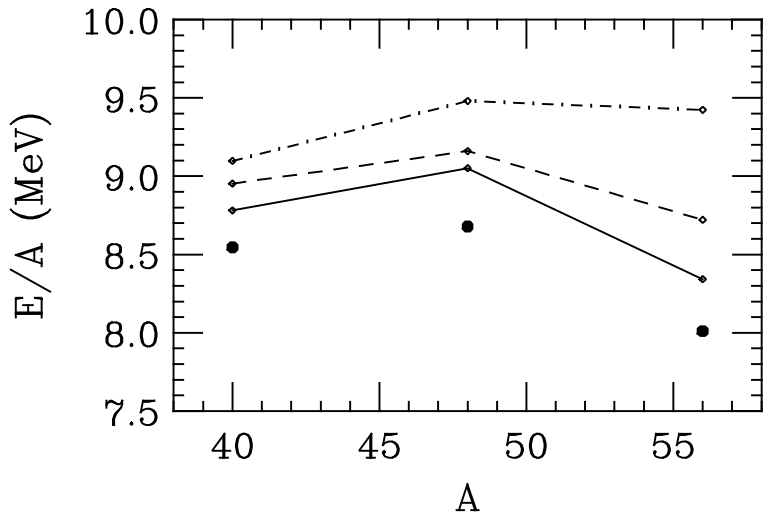}}
\caption{Binding energies of Ca isotopes. Solid points are the 
experimental data, while results from the Landau-Vlasov calculations with
various values of $c$ ($c=28$: solid line, $c=20$: dashed line,
and $c=0$: dot-dashed line) are shown as diamonds connected by lines.
Results taken from Ref.\ \protect\cite{far91}.}
\label{binding} 
\end{figure}  
\begin{figure}[htp]
%\vspace{9.cm}
\vspace{-5.0truecm}
\setlength{\epsfxsize=10truecm}
\centerline{\epsffile{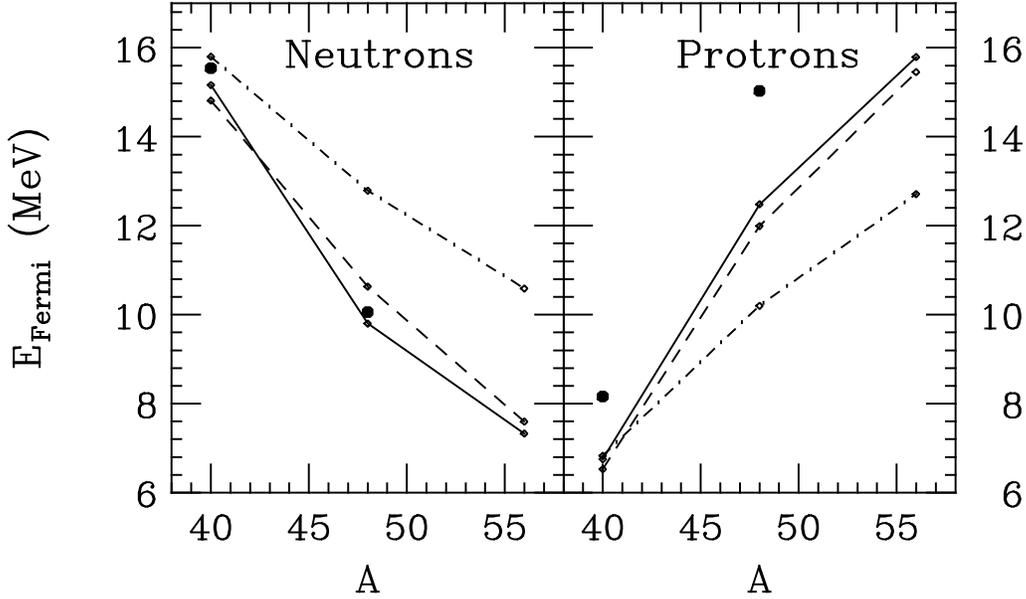}}
\caption{Same as Fig. 2.20 for the Fermi energies of protons and neutrons. 
Taken from Ref.\ \protect\cite{far91}.}
\label{fermi} 
\end{figure}  

Properties of neutron-rich nuclei have also been studied using the
Boltzmann-Uehling-Uhlenbeck ({\sc buu}) or Landau-Vlasov ({\sc lv}) models
\cite{far91,sob94}. These models,
as we shall discuss in more detail in section \ref{buu}, were originally
developed for studying the reaction dynamics of heavy-ion collisions. 
They are based on the semi-classical
approximation, and are thus suitable for studying the dependence of 
the gross properties of neutron rich nuclei on the isospin-dependent {\sc eos}.

\begin{figure}[htp]
%\vspace{9.5cm}
\vspace{-5.0truecm}
\setlength{\epsfxsize=10truecm}
\centerline{\epsffile{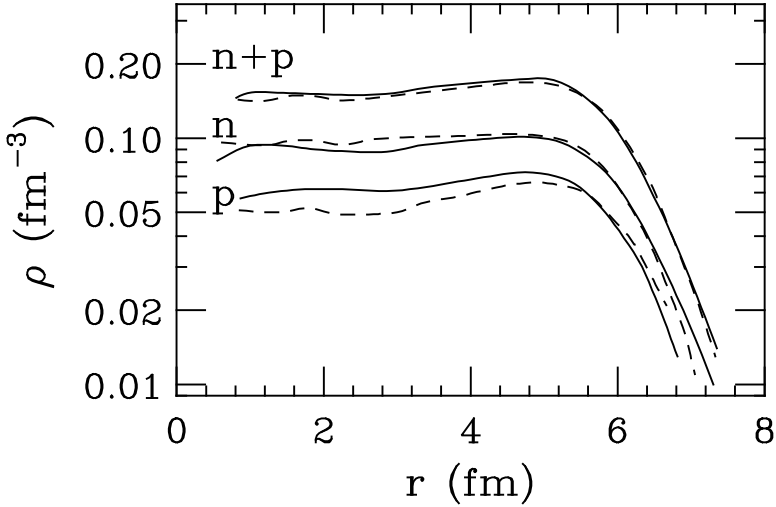}}
\caption{{\sc buu} model calculations of the neutron and proton densities for 
$^{197}Au$ with (solid lines) and without (dashed lines) symmetry energy. 
Shown are (from top to bottom) the total nucleon density, the neutron density,
and the proton density.
The calculations are from Ref.\ \protect\cite{sob94}.}
\label{lee} 
\end{figure}

To include the effects of isospin on nuclear dynamics, Farine
{\it et al.} have added to the Zamick potential energy density, which has
been widely used in standard {\sc buu/lv} 
models, an asymmetric term of the form
\cite{far91}
\begin{equation}\label{simple}
w_{\rm sym}=c\rho[(\rho_n-\rho_p)/\rho_0]^2, 
\end{equation}
where the coefficient $c=e_{\rm sym}(\rho_0)-\frac{1}{3}E_F^0$ is the 
the symmetry energy at normal nuclear matter density due to nuclear
interactions. They have studied the dependence of
ground state properties of neutron-rich Ca isotopes on the 
isospin-dependent  {\sc eos} by varying the value of $c$. 
Fig.\ \ref{binding} shows a comparison of the calculated binding energies of 
three neutron-rich Ca isotopes with the experimental data shown by
plus signs. The data for $^{56}$Ca do not exist, its binding energy 
is obtained using the extended Thomas-Fermi with 
Strutinsky-integral model (ETFSI) of Pearson {\it et al.} \cite{abo92}. 
As expected, nuclei become less bound with increasing $c$. 
Moreover, this effect is the strongest for the most neutron-rich 
nucleus $^{56}$Ca. However, the model tends to overbind 
nuclei for $c$ up to 28 MeV.  The comparison of the Fermi energies of
neutrons and protons for the three Ca isotopes are shown in
Fig.\ \ref{fermi}. General features of the Fermi
energies are seen to be correctly described by the {\sc lv} 
model. The results from
$c=0$ are clearly far from 
the data since the Coulomb energy of protons is then not properly balanced 
by the symmetry energy. As a result the density distribution of neutrons and
protons obtained with $c=0$ is incorrect. Since the symmetry 
potential is repulsive for neutrons and attractive for protons, 
a stronger symmetry potential would deplete more neutrons from the 
center to the surface of a nucleus as noticed 
in Refs.\ \cite{sob94,pawel,lir93,jou95}. The demonstration by Sobotka 
using the {\sc buu} model and an isospin-dependent  {\sc eos} 
based on the Skyrme {\sc ii}
effective interaction is instructive \cite{sob94}. The density
distributions for neutrons and protons in $^{197}$Au obtained from
this study are shown in Fig.\ \ref{lee} for the case with (solid line)
and without (dashed line) symmetry potential. 
Shown are (from top to bottom) the total nucleon density, the neutron density,
and the proton density.  In both cases, the total nucleon density 
distribution is the same.  However, there are differences in the
individual density distributions for protons and neutrons.
Without the attractive symmetry potential for protons, they are
pushed towards the surface by the Coulomb force, and there is 
almost no neutron skin.  Including the symmetry potential, a neutron skin
of about 1/3 fm is seen, and this is consistent
with the droplet model prediction \cite{mye74}.   

To summarize, properties of neutron-rich nuclei are sensitive to the 
the  {\sc eos} of asymmetric nuclear matter 
at saturation density. However, to extract properties of 
isospin asymmetric nuclear matter
at densities away from the saturation density collisions of neutron-rich
nuclei will be more useful. Since semiclassical transport models can
reproduce the gross properties of neutron-rich nuclei and their dependence
on the isospin-dependent  {\sc eos}, they thus provide a useful framework for
studying collisions of neutron-rich nuclei and for extracting information
about the isospin-dependence of the  nuclear {\sc eos}.

\section{Chemical and mechanical instabilities in hot asymmetric 
nuclear matter}\label{chemical}
Nuclear matter is not thermodynamically stable at all densities, $\rho$,
temperatures, $T$, and neutron excesses, $\delta$. The necessary and sufficient
conditions for the stability of an isospin-asymmetric nuclear matter can be
expressed by the following inequalities \cite{muller,lat78,bar80}:
\begin{eqnarray}
\left(\frac{\partial E}{\partial T}\right)_{\rho,\delta}&>& 0,\\\
\left(\frac{\partial P}{\partial \rho}\right)_{T,\delta}&\geq& 0,\\\
\left(\frac{\partial \mu_n}{\partial \delta}\right)_{P,T}&\geq& 0,\label{chem}
\end{eqnarray}
where $E, P$ and $\mu_n$ are the energy per nucleon, pressure and 
neutron chemical potential, respectively. The first condition ensures the 
thermodynamic stability and is satisfied by any reasonable 
equation of state; the second condition prevents the mechanical instability 
against the growth of density fluctuations; and the last one protects the 
diffusive stability against the development of neutronization. 
The last two conditions might be violated in certain regions in 
the $(\rho, T, \delta)$ configuration space. 
Nuclear multifragmentations initiated by various mechanical instabilities,
such as the volume instability, the surface instability of the Rayleigh kind,
and the Coulomb instability, have been extensively studied during the last 
decade \cite{shape1,shape2,shape3,shape4}; 
for a recent review, see, e.g., \cite{moretto}.
With the recent advance in radioactive ion beams, a nuclear matter with
a large isospin asymmetry can be created transiently in these reactions.
It is thus of interest to study the chemical
instability of isospin-asymmetric nuclear matter. 
\begin{figure}[htp]
%\vspace{10cm}
\vspace{-5.0truecm}
\setlength{\epsfxsize=10truecm}
\centerline{\epsffile{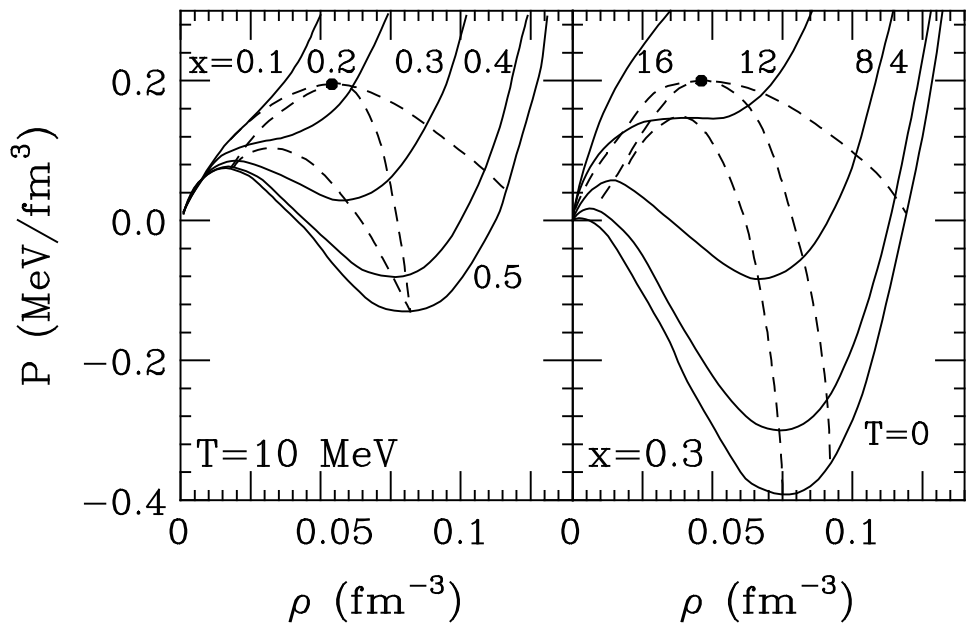}}
\caption{{\sc rmf} predictions of pressure as a function of density at 
fixed temperature $T$ for various proton fraction $x$ (left window), 
and at fixed $x$ for various temperature $T$ (right window).
Taken from Ref.\ \protect\cite{muller}.}
\label{mserot} 
\end{figure}  

To understand the respective roles of chemical and mechanical 
instabilities on the dynamics of nuclear matter, it is important to study
the relative locations of these instabilities in the 
$(\rho, T, \delta)$ configuration
space. Since the adiabatic spinodal is always enclosed by the isothermal
spinodal ({\sc its}), it is sufficient to compare the diffusion spinodal 
({\sc ds})
and {\sc its}. Obviously, ;the boundaries of these instabilities
depend on the equation of state. Using RMT theory M\"uller and Serot
have found that the diffusive spinodal encloses
more of the configuration space than the isothermal spinodal \cite{muller}.
Fig. \ref{mserot} shows typical results form their 
calculations at a constant temperature $T$ (left window) or a constant 
proton fraction $x$ (right window).  At currently reachable 
region of $x\geq 0.3$ in neutron-rich nuclei induced reactions, the 
separation between the isothermal and diffusive spinodals 
in $P-\rho$ configuration space is appreciable.
  
As discussed in previous sections, the equation of state of asymmetric 
nuclear matter varies significantly among models. In
particular, the {\sc rmf} theory predicts a characteristic linear density
dependence of the symmetry energy. It is therefore of interest
to examine how the conclusions based on the {\sc rmf} might change when
other equations of state for asymmetric matter are considered.  In the
following we discuss the results for the chemical and
mechanical instabilities in a thermal model using phenomenological 
equations of state. The advantage of a thermal model is that at modestly 
high temperatures, most quantities relevant to the discussion of chemical 
and mechanical instabilities can be evaluated analytically.
\begin{figure}[htp]
%\vspace{13cm}
\vspace{-5.0truecm}
\setlength{\epsfxsize=10truecm}
\centerline{\epsffile{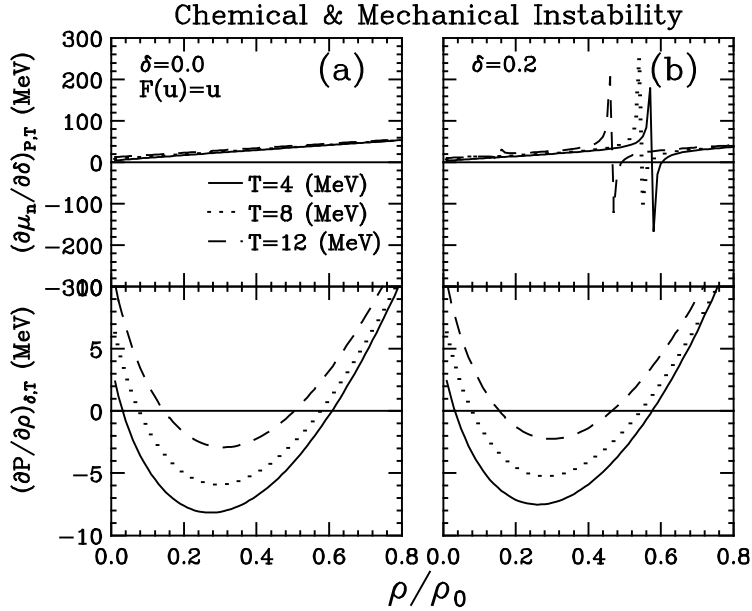}}
\caption{Chemical (upper window) and mechanical (lower window) stability 
conditions as functions of density at 
fixed temperatures $T=$4, 8 and 12 MeV for $\delta=$0.0 (left panel)
and $\delta=0.2$ (right panel).
Taken from Ref.\ \protect\cite{liko97}.}
\label{chem1} 
\end{figure}  
\begin{figure}[htp]
%\vspace{13cm}
\vspace{-5.0truecm}
\setlength{\epsfxsize=10truecm}
\centerline{\epsffile{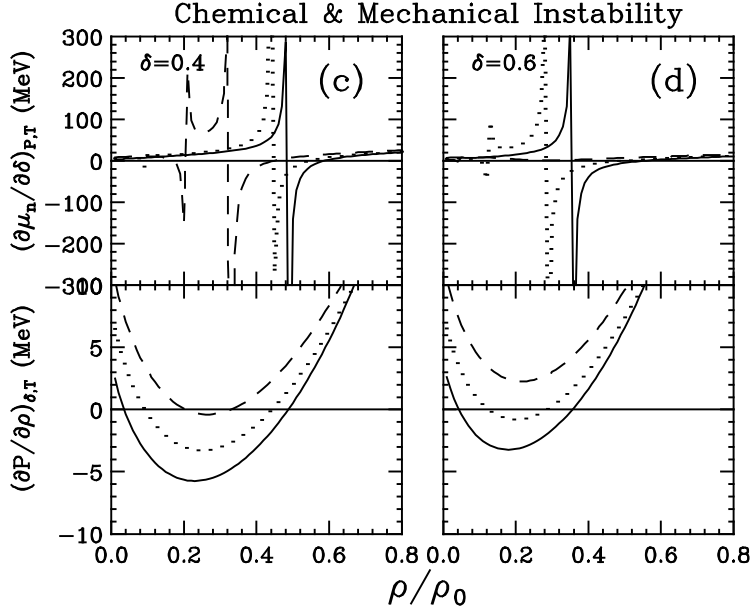}}
\caption{Same as Fig. 2.24 for $\delta=$0.4 (left panel)
and $\delta=0.6$ (right panel).
Taken from Ref.\ \protect\cite{liko97}.}
\label{chem2} 
\end{figure}  

The density $\rho_q$ of neutrons ($q=n$) or protons ($q=p$) is
\begin{equation}
\rho_q=\frac{1}{\pi^2}\int_{0}^{\infty}k^2f_q(k)dk,
\end{equation}
where
\begin{equation}
f_q(k)=[exp(e_q-\mu_q)/T+1]^{-1}
\end{equation}
is the nucleon Fermi distribution function, and $e_q$ is the single particle
kinetic energy. At modestly high temperatures ($T\geq 4 MeV$), this
equation can be inverted analytically to obtain the chemical 
potential \cite{brack,jaqaman1,jaqaman2}
\begin{equation}\label{muq}
\mu_q=V_q+T\left[{\rm ln}(\frac{\lambda_T^3\rho_q}{2})
+\sum_{n=1}^{\infty}\frac{n+1}{n}b_n(\frac{\lambda_T^{3}\rho_q}{2})^n\right],
\end{equation}
where 
\begin{equation}
\lambda_T=\left(\frac{2\pi \hbar^2}{mT}\right)^{1/2}
\end{equation}
is the thermal wavelength of the nucleon, and $b_n's$ are inversion 
coefficients.
In Eq. (\ref{muq}) $V_q$ is the single particle potential energy and can be
parameterized as
\begin{equation}
      V_{q}(\rho,\delta) = a (\rho/\rho_0) + b (\rho/\rho_0)^{\sigma}\ 
	+V_{\rm asy}^{q}(\rho,\delta).
\end{equation}
The parameters $a,~b$ and $\sigma$ are determined by the 
saturation properties and the compressibility $K$ of symmetric nuclear 
matter, i.e., 
\begin{eqnarray}
a&=&-29.81-46.90\frac{K+44.73}{K-166.32}~({\rm MeV}),\\
b&=&23.45\frac{K+255.78}{K-166.32}~({\rm MeV}),\\
\sigma&=&\frac{K+44.73}{211.05}.
\end{eqnarray}

The isospin-independent term should also contain a momentum-dependent part
which is important for certain dynamical observables in heavy ion collisions, 
such as the collective flow (e.g., \cite{gale87,zhang,gale90}), but is not
essential for observables such as the neutron to proton ratio of
preequilibrium nucleons \cite{lik96}. Effects of the momentum-dependent
interactions on thermal properties have been recently studied in Refs.
\cite{csernai92,fai}, and rather small effects on the global properties
of nuclei are found. Here we neglect the momentum-dependent interaction
and shall concentrate on investigating the effects of the isospin-asymmetric
potential $V_{\rm asy}^{q}$.  This potential is given by
\begin{equation}
V^{q}_{\rm asy}(\rho,\delta)=\partial w_a(\rho,\delta)/\partial \rho_{q},
\end{equation} 
where $w_a(\rho,\delta)$ is the contribution of nuclear interactions to the 
symmetry energy density, i.e.,  
\begin{equation}
w_a(\rho,\delta)=e_a\cdot \rho F(u)\delta^2,
\end{equation}
and
\begin{equation}
e_a\equiv e_{\rm sym}(\rho_{NM})-(2^{2/3}-1)\,{\textstyle\frac{3}{5}}E_F^0.
\end{equation}

Using the three forms of $F(u)$ discussed in section \ref{star}, i.e., 
$F(u)=u^2,~u$ and $u^{1/2}$,
the corresponding symmetry potentials are, respectively, 
\begin{eqnarray}\label{vasy}
V_{\rm asy}^{n(p)}&=&\pm 2e_a u^2\delta+e_a u^2\delta^2,\\\
V_{\rm asy}^{n(p)}&=&\pm 2 e_a u\delta,
\end{eqnarray}
and 
\begin{equation}
V_{\rm asy}^{n(p)}=\pm 2e_a u^{1/2}\delta-\frac{1}{2}e_au^{1/2}\delta^2.
\end{equation}

With the chemical potentials determined by Eq. (\ref{muq}), it is then
possible to obtain the total pressure of the nuclear matter by using the 
Gibbs-Duhem relation
\begin{equation}\label{gibbs}
\frac{\partial P}{\partial \rho}=\frac{\rho}{2}\left[(1+\delta)
\frac{\partial \mu_n}{\partial \rho}+(1-\delta)\frac{\partial 
\mu_p}{\partial \rho}\right].
\end{equation}
The pressure can be separated into three parts, i.e., 
\begin{equation}
P=P_{\rm kin}+P_0+P_{\rm asy},
\end{equation}
where $P_{\rm kin}$ is the kinetic contribution
\begin{equation}
P_{\rm kin}=T\rho\left[1+\frac{1}{2}\sum_{n=1}^{\infty}
b_n(\frac{\lambda_T^{3}\rho}{4})^n\left((1+\delta)^{n+1}
+(1-\delta)^{1+n}\right)\right],
\end{equation}
and $P_0$ is the contribution from the isospin-independent nuclear 
interaction
\begin{equation}
P_0=\frac{1}{2}a\rho_0(\frac{\rho}{\rho_0})^2+\frac{b\sigma}
{1+\sigma}\rho_0(\frac{\rho}{\rho_0})^{\sigma+1}.
\end{equation}
$P_{\rm asy}$ is the contribution from the isospin-dependent part of 
the nuclear interaction, and can be written as 
\begin{eqnarray}
P_{\rm asy}&=&2e_a\rho_0(\frac{\rho}{\rho_0})^3\delta^2,\\\
P_{\rm asy}&=&e_a\rho_0(\frac{\rho}{\rho_0})^2\delta^2,
\end{eqnarray}
and
\begin{equation}
P_{\rm asy}=\frac{1}{2}e_a\rho_0(\frac{\rho}{\rho_0})^{3/2}\delta^2
\end{equation}
for $F(u)=u^2,~u$ and $u^{1/2}$, respectively.

From the chemical potential and pressure, the stability conditions
can be determined. The determination of the mechanical stability condition
is straightforward. However, to calculate ${\partial \mu_n}/{\partial
\delta}$ at constant pressure $P$ and neutron excess $\delta$, the following
transformation is useful: 
\begin{eqnarray}\label{trans}
\left(\frac{\partial \mu_n}{\partial \delta}\right)_{T,P}&=& 
\left(\frac{\partial \mu_n}{\partial \delta}\right)_{T,\rho} 
-\left(\frac{\partial \mu_n}{\partial P}\right)_{T,\delta} 
\cdot\left(\frac{\partial P}{\partial \delta}\right)_{T,\rho},\\\
&=&
\left(\frac{\partial \mu_n}{\partial \delta}\right)_{T,\rho} 
-\left(\frac{\partial \mu_n}{\partial \rho}\right)_{T,\delta} 
\cdot\left(\frac{\partial P}{\partial \rho}\right)^{-1}_{T,\delta}
\cdot\left(\frac{\partial P}{\partial \delta}\right)_{T,\rho}.
\end{eqnarray}

Knowing $\left({\partial \mu_n}/{\partial \delta}\right)
_{T,P}$ and $\left({\partial P}/{\partial \rho}\right)_{T,\delta}$ 
as functions of $\rho$, $T$, and $\delta$, regions of chemical and/or
mechanical instability can then be identified in the configuration space.
First, we show in Fig.\ \ref{chem1} and Fig.\ \ref{chem2}
these two quantities as functions of $\rho$ at various $T$ and $\delta$.
In this calculation $K=200$ MeV and $F(u)=u$ are used. For comparison 
results for symmetric nuclear matter are shown in the left window of 
Fig.\ \ref{chem1}. In this case there is no chemical instability and the    
quantity $\left({\partial \mu_n}/{\partial \delta}\right)_{T,P}$ 
increases with both density and temperature. To understand this, we note that
\begin{eqnarray} 
\left(\frac{\partial P}{\partial \delta}\right)_{T,\rho}
&=&\frac{2}{\delta}P_{\rm asy}+\frac{T\rho}{2}\sum_{n=1}^{\infty}(n+1)b_n
(\frac{\lambda_T^3\rho}{4})^n\left[(1+\delta)^n-(1-\delta)^n\right],\\\
\left(\frac{\partial \mu_n}{\partial \delta}\right)_{T,\rho}
&=&\left(\frac{\partial V_{asy}^n}{\partial \delta}\right)_{T,\rho}
\nonumber\\
&+&T\left[\frac{1}{1+\delta}+\sum_{n=1}^{\infty}(n+1)b_n
\left(\frac{\lambda_T^3\rho}{4}\right)^n(1+\delta)^{n+1}\right],
\end{eqnarray}
and 
\begin{equation}
{\rm lim}_{\delta\rightarrow 0}
\left(\frac{\partial P}{\partial \delta}\right)_{T,\rho}=0.
\end{equation}
Then, it is easy to show that
\begin{eqnarray}
&&{\rm lim}_{\delta\rightarrow 0} \left(\frac{\partial 
\mu_n}{\partial \delta}\right)_{T,P}= 
{\rm lim}_{\delta\rightarrow 0} \left(\frac{\partial \mu_n}{\partial \delta}
\right)_{T,\rho},\\\
&&=2e_aF(u)+T\left[1+\sum_{n=1}^{\infty}(n+1)b_n
\left(\frac{\lambda_T^3\rho}{4}\right)^n\right],
\end{eqnarray}
which is positive as one expects for symmetric nuclear matter and increases
with both density and temperature.  

The mechanical instability is known to occur at intermediate densities,
i.e., in the mixed phase, between the gas and liquid phases at subcritical 
temperatures.  	This is clearly demonstrated by plotting
$\left({\partial P}/{\partial \rho}\right)_{T,\delta}$ as a function of
$\rho$. It is interesting to examine this quantity in the limit of low
densities. From the Gibbs-Duhem relation of Eq.\ (\ref{gibbs}) and the
expression for $\mu_q$ of Eq.\ (\ref{muq}), one has
\begin{equation}
{\rm lim}_{\rho\rightarrow 0}\left(\frac{\partial P}
{\partial \rho}\right)_{T,\delta}
={\rm lim}_{\rho\rightarrow 0}(T+a\frac{\rho}{\rho_0})=T,
\end{equation}
as expected for an ideal gas.

We note from Eq. (\ref{trans}) that the quantity $\left({\partial\mu_n}/
{\partial \delta}\right)_{T,P}$ is singular along the boundary for mechanical
instability, where $\left({\partial P}/{\partial \rho}\right)_{T,\delta}=0$.
This singularity is indicated by the spikes in Fig.\ \ref{chem1}
and Fig.\ \ref{chem2}. However, their heights at the boundary of
the gas and mixed phases is very small because the last derivative in Eq.
(\ref{trans}) is close to zero, i.e., 
\begin{equation}
{\rm lim}_{\rho\rightarrow 0}\left(\frac{\partial P}
{\partial \delta}\right)_{T,\rho}=\frac{2}{\delta}P_{\rm asy}=0.
\end{equation}
Furthermore, in the ideal gas limit the nuclear matter 
is also chemically stable, i.e.,
\begin{equation}
{\rm lim}_{\rho\rightarrow 0}\left(\frac{\partial \mu_n}{\partial \delta}
\right)_{P,T}=\frac{T}{1+\delta}>0.
\end{equation}
\begin{figure}[htp]
%\vspace{13cm}
\vspace{-5.0truecm}
\setlength{\epsfxsize=10truecm}
\centerline{\epsffile{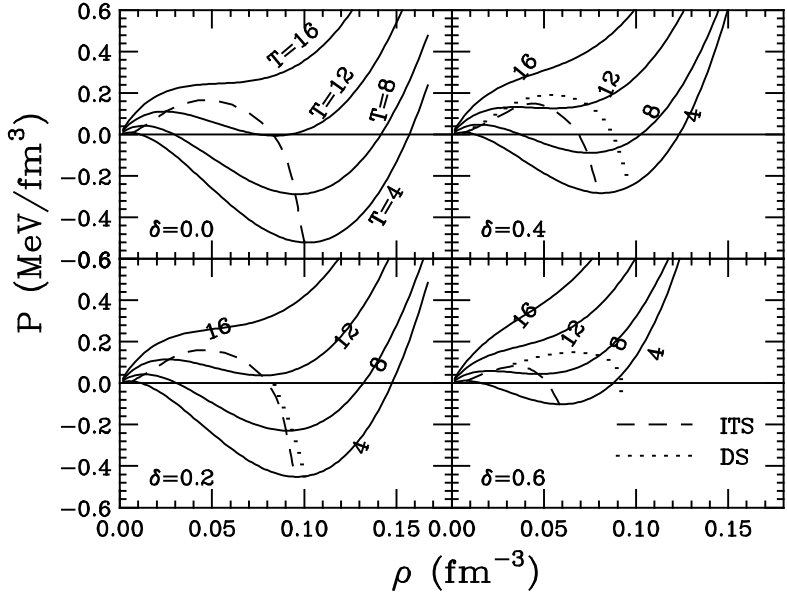}}
\caption{Pressure as a function of density at 
fixed temperatures $T=$4, 8, 12 and 16 MeV for $\delta=$0.0, 0.2, 0.4 and 
0.6. The isothermal and diffusive spinodals are shown by the dashed and 
dotted lines, respectively. Taken from Ref.\ \protect\cite{liko97}.}
\label{pre} 
\end{figure}  

As the system becomes neutron-rich, the chemical instability starts to
develop, so the boundary for mechanical instability 
shrinks while that for chemical instability expands as $\delta$ increases.
For example, in contrast to the variation of the chemical instability
the mechanical instability at a temperature of T=10 MeV gradually disappears 
as $\delta$ increases from 0 to 0.6.  Furthermore, in the
mechanically unstable region the system is chemically stable, but at
higher densities chemical instability can appear in the mechanically
stable region. Therefore, in neutron-rich nuclear matter the chemical 
instability is a more relevant quantity in determining its properties. 
As shown in Fig. \ref{pre}, where the pressure along the chemical, ({\sc ds}),
and mechanical, ({\sc its}), spinodals are plotted as functions of density,
both the isothermal pressure and the separation between the two spinodals 
increase with increasing $\delta$, so not only the chemical but also
the mechanical instability is strongly isospin-dependent.  
\begin{figure}[htp]
%\vspace{13cm}
\vspace{-5.0truecm}
\setlength{\epsfxsize=10truecm}
\centerline{\epsffile{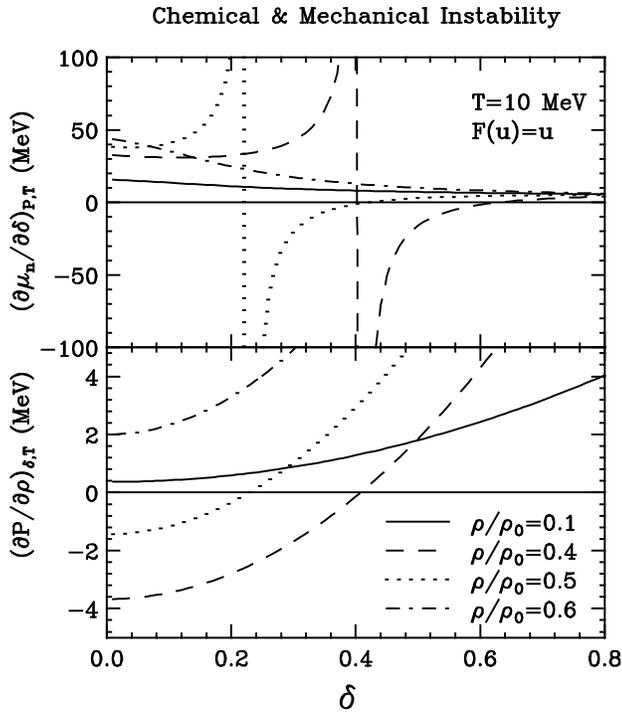}}
\caption{Chemical (upper window) and mechanical (lower window) 
instability conditions as functions of $\delta$ at a fixed 
temperature $T=10$ MeV and various densities. 
Taken from Ref.\ \protect\cite{liko97}.}
\label{chem3} 
\end{figure}  
\begin{figure}[htp]
%\vspace{12cm}
\vspace{-5.0truecm}
\setlength{\epsfxsize=10truecm}
\centerline{\epsffile{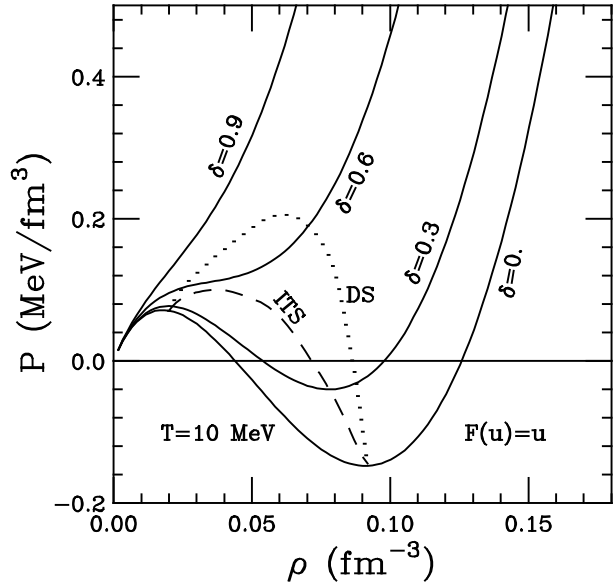}}
\caption{Pressure as a function of density at a 
fixed temperature $T=10$ MeV at various $\delta$.
The isothermal and diffusive spinodals are shown by the dashed and dotted 
lines, respectively. Taken from Ref.\ \protect\cite{liko97}.}
\label{pre1} 
\end{figure}  

The isospin-dependence of mechanical and chemical instabilities at a
fixed temperature are studied in more detail in Fig.\ \ref{chem3} and
Fig.\ \ref{pre1}.  It is seen that nuclear matter is both mechanically and
chemically stable at low and high densities. At intermediate densities
(e.g., $\rho/\rho_0=0.4,~ 0.5$) the system is mechanically unstable for
small $\delta$ (e.g., $\delta\leq 0.2$ and $\delta\leq 0.4$ for
$\rho/\rho_0=0.4$ and 0.5, respectively.), and chemically unstable for
intermediate $\delta$ (e.g.\ $0.2\leq \delta \leq 0.4$ for 
$\rho/\rho_0=0.4$ and $0.4\leq \delta \leq 0.6$ for 
$\rho/\rho_0=0.5$, respectively.) 
The corresponding boundaries in the pressure-density space are shown 
in Fig. \ref{pre1}. Again, the diffusive spinodal is more extended 
than the isothermal spinodal. The dependence of pressure on isospin is
also shown in this figure. Its strong dependence on $\delta$ is due to the
significant $P_{\rm asy}$ contribution to the total pressure.
\begin{figure}[htp]
%\vspace{15cm}
\vspace{0.0truecm}
\setlength{\epsfxsize=10truecm}
\centerline{\epsffile{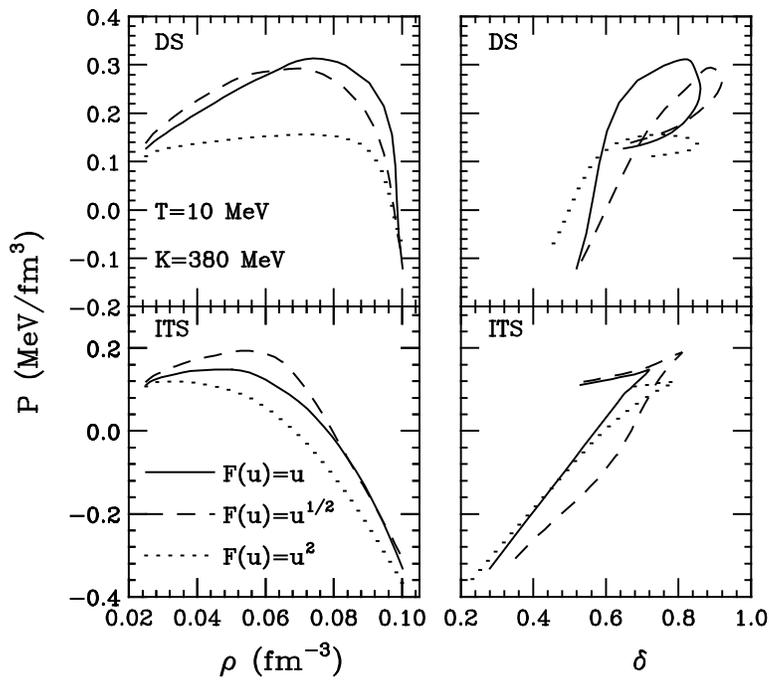}}
\caption{Pressure as a function of density (left panel) and isospin 
asymmetry (right panel) at a fixed temperature $T=10$ MeV along 
the boundary of diffusive spinodals (upper windows) and isothermal 
spinodals (lower windows) by using the three forms of $F(u)$ and a 
compressibility of $K=380$ MeV.  
Taken from Ref.\ \protect\cite{liko97}.}
\label{k380} 
\end{figure}  
\begin{figure}[htp]
%\vspace{15cm}
\vspace{0.0truecm}
\setlength{\epsfxsize=10truecm}
\centerline{\epsffile{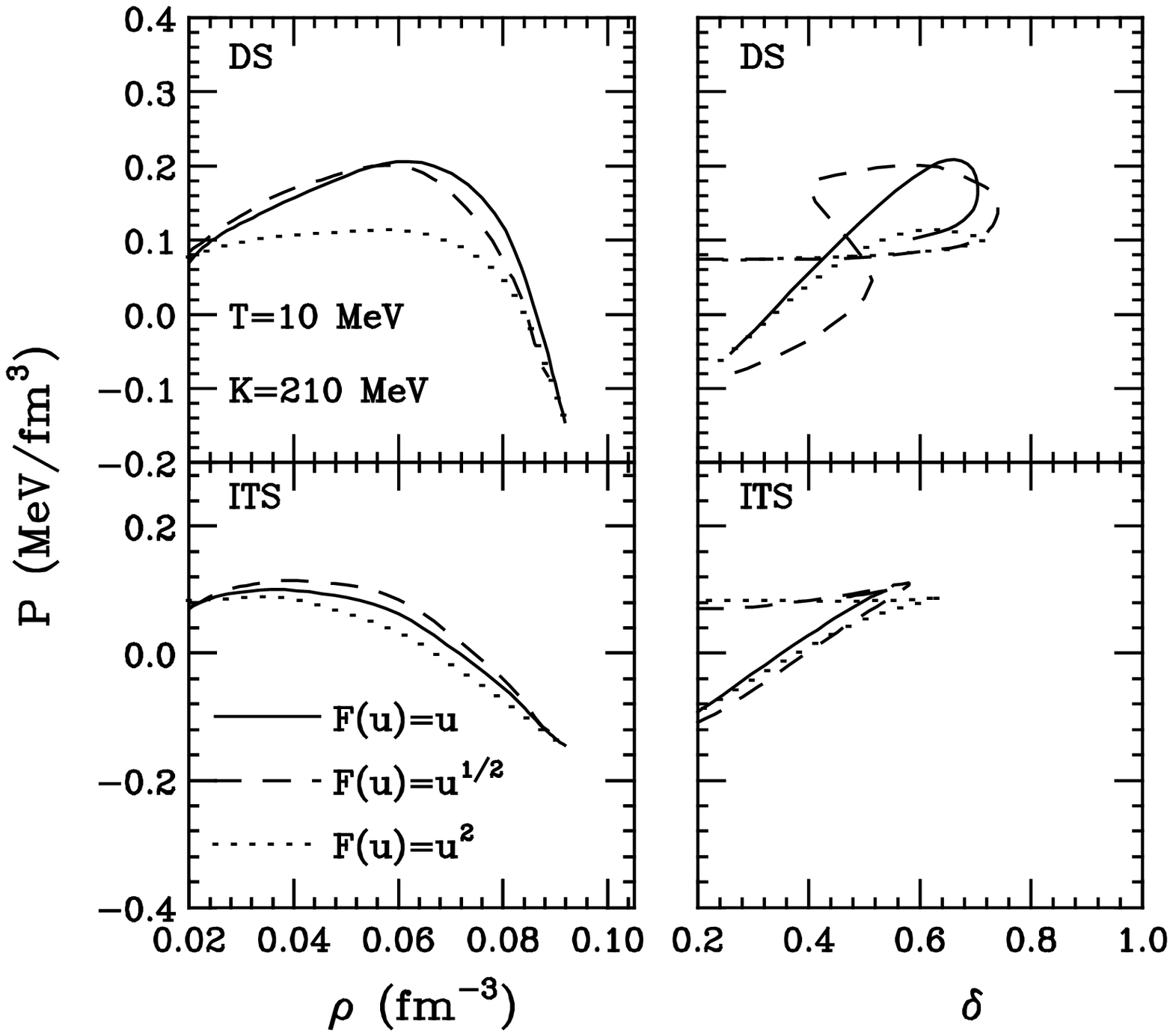}}
\caption{Same as Fig. 2.29 for a compressibility of $K=200$ MeV.  
Taken from Ref.\ \protect\cite{liko97}.}
\label{k200} 
\end{figure}  

The above analysis thus indicates a strong isospin-dependence of both 
chemical and mechanical instabilities. It is therefore of interest to know
how this dependence might be affected by the isospin-dependent and 
-independent parts of the nuclear equation of state. This is
illustrated in Fig.\ \ref{k380} and Fig.\ \ref{k200}, where the pressure is 
shown as a function of density (left panels) and $\delta$ (right panels)
along the diffusive spinodals (upper windows) and isothermal spinodals
(lower windows) at a constant temperature of $T=10$ MeV by using the
three forms of $F(u)$ and a compressibility of 380 and 200 MeV, respectively. 
Several interesting observations can be made from these two figures. 
First, the diffusive spinodals are always more 
extended in the configuration space than the isothermal spinodals. Second,
both {\sc ds} and {\sc its} 
depend on the form of $F(u)$; and this dependence is
stronger for the stiff equation of state with $K=$ 380 MeV. Third, from the
right panels of these two figures one sees that it is both chemically 
and mechanically favorable for nuclear matter to be less asymmetric 
(smaller $\delta$) in the liquid phase than in the gas phase. 
This result can also be obtained from energy considerations. Since
the equation of state for asymmetric nuclear matter contains a
$e_{\rm sym}(\rho)\delta^2$ term, it is therefore energetically favorable
for asymmetric matter to separate into a liquid phase that is less
asymmetric and a gas phase that is more asymmetric, rather than into two
phases with equal asymmetry. Finally, there
is a maximum asymmetry for both {\sc ds} and {\sc its}. 
As the density decreases from the liquid phase 
the asymmetry increases along both {\sc ds} and {\sc its} to a 
maximum value.  The asymmetry then decreases towards the gas phase. As 
shown in these
figures, the maximum asymmetry is also sensitive to both the
isospin-dependent and -independent parts of the nuclear equation of state.      

The results discussed in this section are more than
just of academic interest. Significant 
differences have been observed in the multifragmentation of 
isospin-symmetric and isospin-asymmetric nuclear matter formed in
heavy ion collisions at intermediate 
energies \cite{dem96,kunde96,toke96,sob97}. {\sc buu} calculations
suggest that the abundance of neutron rich isotopes observed in experiments
is primarily the result of the coalescence of the clusters with neutrons
from the gas phase \cite{sob97}.  That these neutron exist in abundance
is a direct consequence of the large isospin-asymmetry in the gas phase.
It is interesting, by the way, that the percolation model has difficulties
in explaining the fragmentation data for neutron-rich isotopes \cite{Kor97},
despite its success in matching a large number of other oberservables in
nuclear fragmentation.

One popular explanation for
nuclear multifragmentation is that the hot nuclear matter formed in 
the reaction expands adiabatically into the mechanical instability region 
where it fragments into clusters and nucleons due to the growth of
density fluctuations \cite{bsiemens}.  In describing this process
great efforts have been devoted to incorporating effects of density
fluctuations into dynamical models during the last
decade \cite{ayik1,randrup1,randrup2,colonna,ayik2}. To describe
multifragmentation in asymmetric nuclear collisions using these models, 
it will be necessary to extend them to include the isospin 
degree of freedom and its fluctuations. This will then allow one to
compare the strengths of both density and isospin fluctuations
and to study the relative time scales of chemical and mechanical
instabilities in asymmetric nuclear matter. Indeed some interesting 
new features about the collective motions new the n/p drip lines 
and the onset of chemical and mechanical instabilities have been 
found in recent studies using RPA\cite{hama96} and Landau dispersion 
relation approaches\cite{ditoro}. 

\section{New features of liquid-gas phase transition in asymmetric 
nuclear matter}\label{liquid}
The liquid-gas phase transition in asymmetric nuclear matter is not only more
complex than in symmetric matter but also has new distinct 
features. This is because the nature of phase transitions in a matter
is strongly influenced by its dimensionality. The inclusion of 
the isospin degree of freedom in studying the phase transitions 
in nuclear matter has thus attracted much attention recently
\cite{muller,pet95,lat78,bar80,lam81,lat85,gle92,kuo96}. These studies
have shown that there are indeed new features associated with the 
liquid-gas phase transition in asymmetric nuclear matter.
\begin{figure}[htp]
%\vspace{12cm}
\vspace{-5.0truecm}
\setlength{\epsfxsize=10truecm}
\centerline{\epsffile{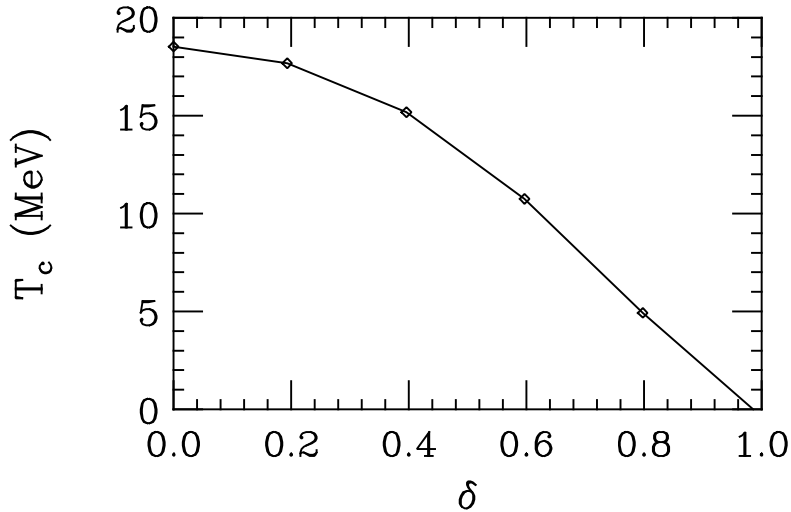}}
\caption{Critical temperature as a function of the asymmetry parameter
$\delta$ from the isospin lattice gas model.
Calculations taken from Ref.\ \protect\cite{kuo96}.}
\label{tc} 
\end{figure}  

\begin{figure}[htp]
%\vspace{12cm}
\vspace{-5.0truecm}
\setlength{\epsfxsize=10truecm}
\centerline{\epsffile{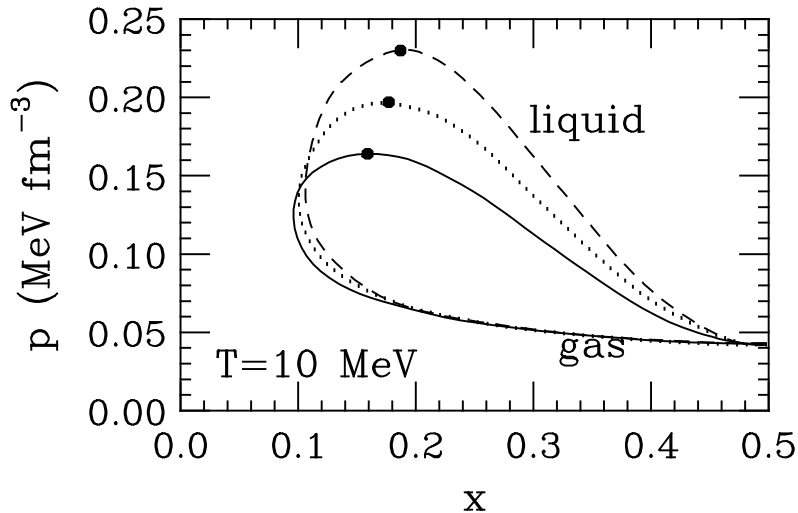}}
\caption{Binodal sections at $T=10$ MeV and $e_{\rm sym}(\rho_0)=$30, 
35 and 40 MeV shown by the solid, dotted and dashed curves, respectively.
Taken from Ref.\ \protect\cite{muller}.}
\label{cp} 
\end{figure}

First of all, the critical temperature for the liquid-gas phase transition
has been predicted to decrease with increasing neutron excess.
This can be understood qualitatively in terms of the increasing contribution
from the asymmetric pressure $P_{\rm asy}$. Fig.\ \ref{tc} shows the critical 
temperature as a function of neutron-excess $\delta$ predicted by 
the isospin lattice gas model of Kuo et al. \cite{kuo96}. This prediction
is in qualitative agreement with that based on the Skyrme interaction
\cite{lat78} and {\sc rmf} theory \cite{muller}. Using the latter theory,
M\"uller and Serot have studied the sensitivity of the critical 
point to the variation of the symmetry energy $e_{\rm sym}(\rho_0)$ and
the bulk compressibility $K$. In Fig.\ \ref{cp} the binodal sections
predicted by the {\sc rmf} theory at $T=$ 10 MeV for $e_{\rm sym}(\rho_0)=$30,
35 and 40 MeV are shown by the solid, dotted and dashed curves, respectively.   
As one would expect from our discussions in the previous section,
the minimum proton fraction at which the phase separation occurs
increases only by 10\%, and the critical temperature decreases by less
than 1\% as the symmetry energy increases from $e_{\rm sym}(\rho_0)=$30
to 45 MeV. On the other hand, the critical pressure increases by about       
40\%. However, the shape of binodal surfaces are rather similar for different
values of symmetry energy. As we have stressed earlier, in the {\sc rmf} 
theory the symmetry energy has a characteristic linear density dependence.
Other models with different density dependence in the symmetry energy
naturally may predict critical points at different temperature,
pressure and proton fraction. It is well-known that the critical point of 
liquid-gas phase transition is sensitive to the bulk compressibility $K$.
M\"uller and Serot have found that the critical temperature changes by about
13\% when $K$ varies from 200 to 300 MeV using the same symmetry energy 
$e_{\rm sym}(\rho_0)=$ 35 MeV. Although this variation is larger than that due 
to the variation of the symmetry energy, it is unlikely to have a significant
effect on the liquid-gas phase transition.  
\begin{figure}[htp]
%\vspace{12cm}
\vspace{-5.0truecm}
\setlength{\epsfxsize=10truecm}
\centerline{\epsffile{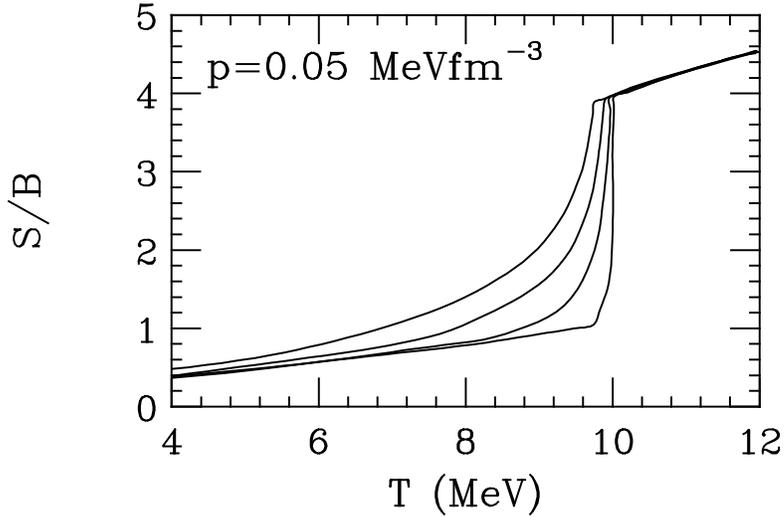}}
\caption{Specific entropy as a function of temperature at a constant pressure
for proton fractions of 0.5, 0.45, 0.4 and 0.35, respectively.
Taken from Ref.\ \protect\cite{muller}.}
\label{entropy} 
\end{figure}  

An important question about the liquid-gas phase transition in 
asymmetric nuclear matter is its order. Glendenning \cite{gle92} 
as well as M\"uller 
and Serot \cite{muller} have recently stressed that the isospin 
degree of freedom plays an important role in phase transitions.
In the liquid-gas phase transition, not only the total baryon density 
but also the total charge density is conserved. For an asymmetric matter,
the latter constraint leads to a smoother variation of the thermodynamical
quantities, such as the temperature, specific entropy, and heat capacity,
during the phase transition. The order of phase transition in asymmetric
matter is then expected to be higher than that in symmetric nuclear matter.
In Fig.\ \ref{entropy}, the variation of temperature and specific
entropy for proton fractions of 0.5, 0.45, 0.4 and 0.3 are shown.
In the symmetric case, the specific entropy increases by 3 units in a
temperature interval of 1.5 MeV, which increases to about 5 MeV as the 
proton fraction decreases to 0.35. Thus, both the specific entropy 
and temperature change continuously during the phase transition, so
the heat transferred to the system is used not only to convert the liquid
to vapor but also to heat the matter. The first-order liquid-gas phase
transition in symmetric nuclear matter therefore changes to a second-order 
one in asymmetric nuclear matter.

Another important question about the liquid-gas phase transition in
asymmetric matter is the relative proton fraction in the two phases.
Since the  {\sc eos} of asymmetric matter contains a term $e_{\rm sym}\delta^2$,
it is energetically favorable for the system to separate into a
neutron-rich liquid phase and a proton-rich gas phase rather than two
phases with equal proton concentration. The same conclusion has been 
reached from the study of chemical and dynamical instabilities in the
previous section. However, one should be cautious in
comparing this conclusion directly with experimental observations on the
charge/mass ratio of heavy fragments and single particles or light clusters
because the liquid-gas phase transition may occur in heavy residues 
during the later, expansion phase of nuclear reactions. In the latter
case, the charge/mass ratio of the heavy-residue depends strongly on the
symmetry potential during the early stage of the reaction. We shall discuss
this in more detail in section \ref{npratio}.
 
To confirm experimentally these new features of liquid-gas phase
transition in asymmetric matter is challenging. One possible way is to
study the dependence of nuclear multifragmentations on the charge/mass
ratio of the reaction system.

\chapter{Isospin dependence of in-medium nucleon-nucleon 
cross sections}\label{xnn}
In extracting information about the structure
of radioactive nuclei, such as the radii and the distributions of the
constituent neutrons and protons, the model used most 
is Glauber's.  In the optical limit, the total interaction
cross section is determined by a transmission function, which is a
convolution of the nucleon-nucleon cross section and the density
distribution of nucleons from both the target and projectile in the
overlapping region. Knowing the proton density distribution (which
can be determined from other means such as electron scattering) 
and the nucleon-nucleon cross sections,
the neutron density distribution can then be determined.
Usually, only the isospin averaged free nucleon-nucleon cross section is 
used as input. The effects of including the isospin-dependence of 
the in-medium nucleon-nucleon cross section needs to be studied. 
Also, the isospin-dependent in-medium nucleon-nucleon
cross sections are needed in transport model to extract
the isospin-dependent  {\sc eos} and to understand the 
isospin-dependent phenomena found in heavy-ion collisions. In this chapter we
will first discuss the isospin dependence of free-space nucleon-nucleon
cross sections which are normally used as inputs in the isospin-dependent
transport models. Then we shall review recent theoretical studies on 
the in-medium nucleon-nucleon cross sections based on the many-body theories
discussed in the previous Chapter.

\section{Isospin dependence of free-space NN cross sections}
\begin{figure}[htp]
%\vspace{9cm}
\vspace{-5.0truecm}
\setlength{\epsfxsize=10truecm}
\centerline{\epsffile{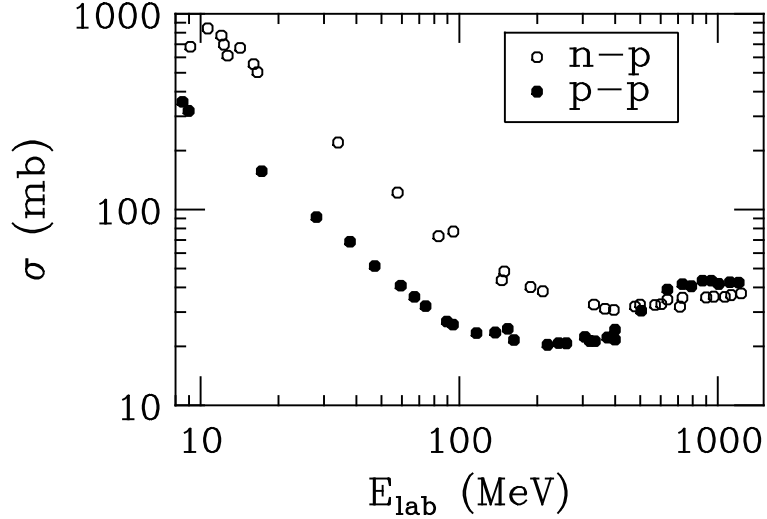}}
\caption{Cross sections of neutron-proton and proton-proton scatterings as 
functions of bombarding energy. Taken from Ref.\ \protect\cite{nndata}.}
\label{freesigma} 
\end{figure}

It is well-known that the scattering cross section between two nucleons 
depends on their isospin. Fig.\ \ref{freesigma} compares the free-space
cross sections for neutron-proton and proton-proton or neutron-neutron
scattering as functions of bombarding energy. The data in the energy
range of 10 MeV $\leq E_{\rm lab} \leq 1000 $ MeV can be parameterized 
by \cite{cha90}
\begin{eqnarray}
\sigma_{np}^{\rm free}&=&-70.67-18.18\beta^{-1}+25.26\beta^{-2}+113.85
\beta~({\rm mb}),\\
\sigma_{pp}^{\rm free}&=&13.73-15.04\beta^{-1}+8.76\beta^{-2}+68.67
\beta^4~({\rm mb}),
\end{eqnarray}
where $\beta\equiv v/c$ is the velocity of the projectile nucleon.
It is seen that below about 500 MeV the neutron-proton cross section
is about a factor of 2 to 3 higher than the proton-proton or 
neutron-neutron cross section, indicating that the $T$ matrix for the
total isospin $I=0$ channel is much stronger than that of the $I=1$ 
channel. In nuclear medium, one expects that the strength of $G$ matrix,
which differs from the free space $T$ matrix because of the Pauli effects
and many-body effects, for these two channels are also different.
How their relative strengths may change at finite density and temperature 
is a question of current interest. Since the reaction dynamics of heavy-ion
collisions is governed by both the nuclear  {\sc eos} and the in-medium 
nucleon-nucleon cross sections, both are important and have to be studied
simultaneously. In fact, the main uncertainty in the extracted nuclear 
{\sc eos} 
from heavy-ion collisions is due to our poor knowledge of the nucleon-nucleon
cross sections in medium. It is thus desirable to find experimental
observables that are only sensitive to either the  {\sc eos} 
or the cross sections. 
This is particularly important for studying the symmetry potential since
it is quite weak compared to the nuclear  {\sc eos} 
of symmetric nuclear matter.  
As we shall discuss in the next chapter, transport models are very
useful for such studies. 

\section{Isospin dependence of in-medium NN cross sections}
The study of in-medium nucleon-nucleon cross sections have been 
mostly based on the Bethe-Goldstone ({\sc bg}) 
equation \cite{har87,cug87,ber88,boh89,fas89,koh91,gqli93,alm1,alm2}. 
Medium effects appear in the {\sc bg} 
equation mainly through the Pauli blocking 
factor for intermediate states and the self-energies of the two nucleons 
in the denominator of the propagator. 
However, results from these studies differ appreciably, with some models
predicting a decrease of nucleon-nucleon cross sections while others an
increase. 

In the Dirac-Brueckner approach of Refs.\ \cite{har87,gqli93}, 
model parameters are fixed by fitting free-space nucleon-nucleon scattering
data and deuteron properties. Nucleon-nucleon cross sections in nuclear medium
at zero temperature have been predicted to decrease with increasing density.
For example, at normal nuclear matter density and a bombarding energy 
of 50 MeV, both $\sigma_{np}$ and $\sigma_{pp}$ are reduced by about 
a factor of two. Results of the calculations in Ref.\ \cite{gqli93} can be 
parameterized by
\begin{eqnarray}
\sigma_{np}^{\rm medium}&=&\left[31.5+0.092abs(20.2-E_{\rm lab}^{0.53})^{2.9}
\right]\cdot
\frac{1.0+0.0034E_{\rm lab}^{1.51}\rho^2}{1.0+21.55\rho^{1.34}}~({\rm mb}),\\
\sigma_{pp}^{\rm medium}&=&\left[23.5+0.0256(18.2-E_{\rm lab}^{0.5})^{4}
\right]\cdot
\frac{1.0+0.1667E_{\rm lab}^{1.05}\rho^3}{1.0+9.704\rho^{1.2}}~({\rm mb}).\\
\end{eqnarray}
\begin{figure}[htp]
%\vspace{11cm}
\vspace{-5.0truecm}
\setlength{\epsfxsize=10truecm}
\centerline{\epsffile{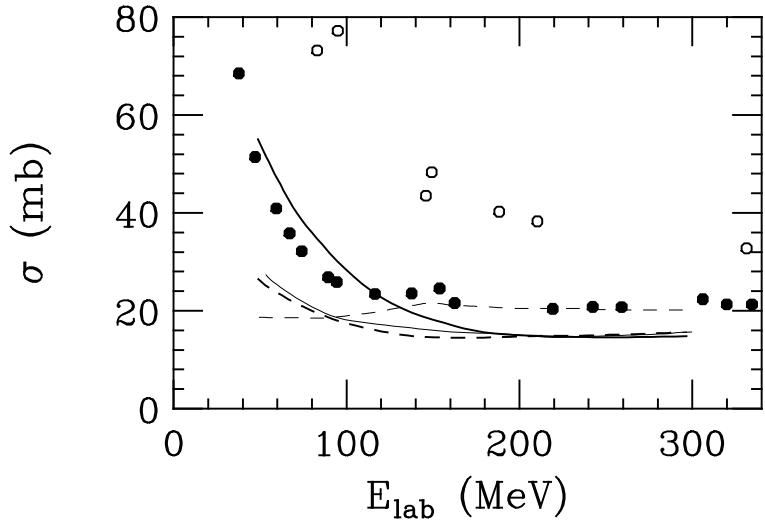}}
\caption{The neutron-proton (solid lines) and proton-proton (dashed lines)
cross sections as 
functions of bombarding energy at normal (thick lines) and twice normal
density (thin lines). The calculations are taken from Ref.\ 
\protect\cite{gqli93} and are compared to the free cross sections of Fig.\
3.1 (solid and open circles).}
\label{gqli} 
\end{figure}  

Fig.\ \ref{gqli} shows the neutron-proton and proton-proton cross sections
at normal and twice normal nuclear matter density.
One sees that the strong isospin dependence of nucleon-nucleon 
cross sections in free-space disappears gradually as the bombarding energy 
and density increase. Furthermore, at densities larger than about twice
the normal density, the proton-proton cross section even 
becomes higher than the neutron-proton cross section.  The respective 
effects of the Pauli blocking and selfenergy corrections 
have not been studied in Ref.\ \cite{gqli93}. 
Bohnet {\it et al.} \cite{boh89} have also studied the in-medium 
nucleon-nucleon cross sections during the collisions of two slabs of 
nuclear matter at zero temperature. The Pauli blocking factor has been
estimated using two Fermi spheres separated by
the beam momentum, and it is found that the in-medium cross section generally
increases with density.  
\begin{figure}[htp]
%\vspace{8cm}
\vspace{-5.0truecm}
\setlength{\epsfxsize=10truecm}
\centerline{\epsffile{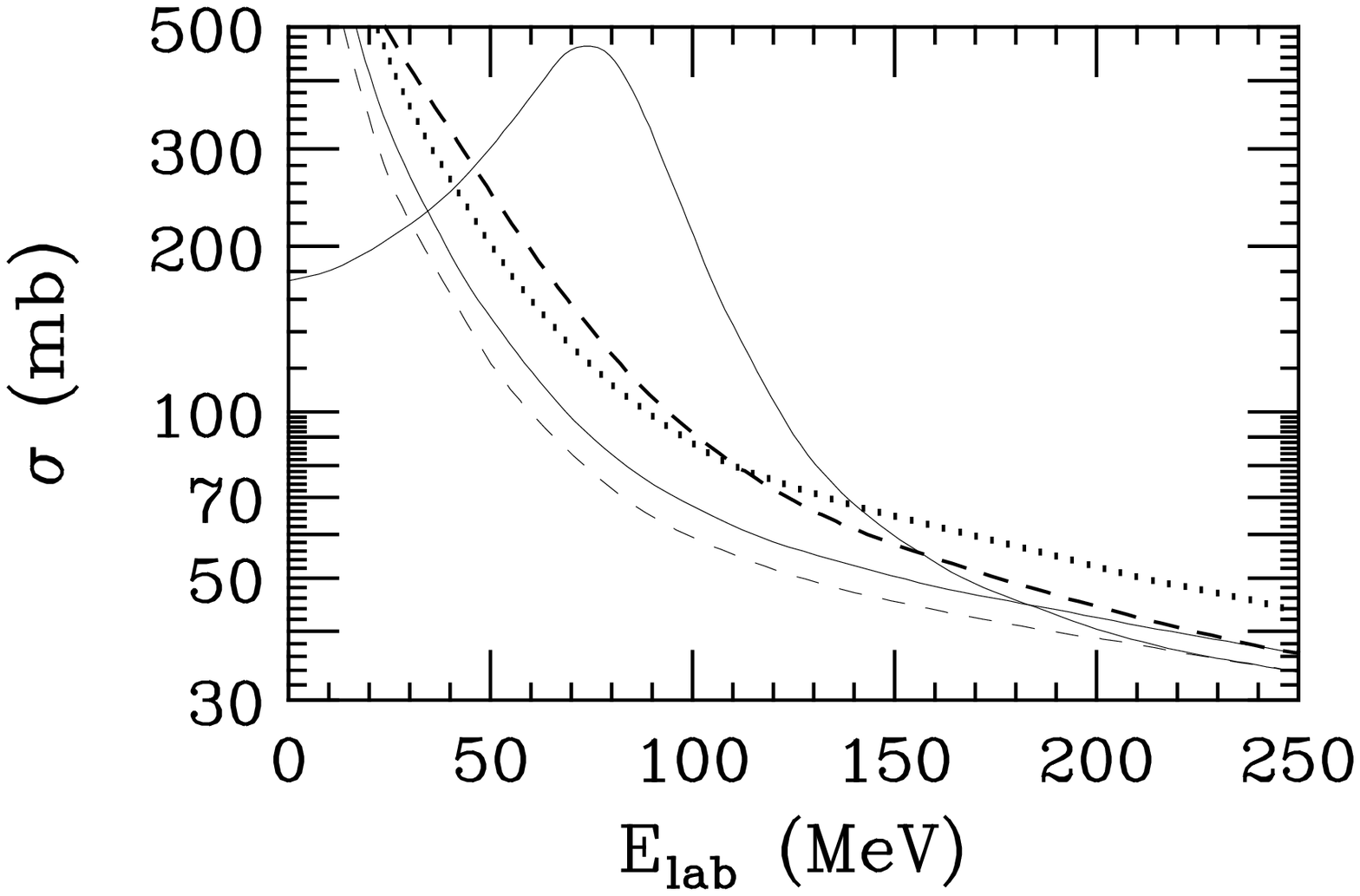}}
\caption{Isospin averaged nucleon-nucleon cross section as 
a function of bombarding energy at a density of $0.5\rho_0$.
Taken from Ref.\ \protect\cite{alm2}. The dotted line represents the free
cross section, the solid lines represent the in-medium cross sections at 
temperatures of 10 (thin line) and 35 MeV (thick line).  The dashed lines
are the corresponding cases without Pauli-blocking.}
\label{alm} 
\end{figure}  

In the work by Alm {\it et al.} \cite{alm1,alm2}, the cross section in a
hot nuclear matter has been evaluated by including also the hole-hole 
collisions in the Pauli blocking operator.
Fig.\ \ref{alm} shows their results at 
temperatures of 10 MeV and 35 MeV and a density of $0.5\rho_0$ as a function 
of bombarding energy. Effects of the Pauli blocking
and selfenergy corrections are separated by comparing full calculations with 
those by setting the Pauli blocking operator equals to 1. First, it is 
seen that at both temperatures the selfenergy correction suppresses the 
cross section, while the Pauli blocking operator for intermediate states 
enhances the cross section. Second, at energies above about 200 MeV predictions 
for different temperatures converge to values smaller than the
free-space cross section. It is also seen that at lower temperatures the 
cross section has a strong peak above the free-space cross section.  This 
has been interpreted as a precursor effect of the superfluid phase transition 
in nuclear matter \cite{alm1,alm2}.    

In summary of this section, our current theoretical understanding about 
the in-medium nucleon-nucleon cross section, especially its dependence 
on isospin, density and temperature, is very limited. Most calculations
are for nuclear matter only. Ideally, for heavy-ion collisions one should 
evaluate the in-medium cross sections selfconsistently at densities and
temperatures determined by the reaction dynamics, and this requires 
much more work in the future. It is worth mentioning that recent 
studies on collective flow in heavy-ion collisions at intermediate 
energies have shown an indication of reduced in-medium 
nucleon-nucleon cross sections \cite{wes93,kla93,hun96}. 
An empirical in-medium nucleon-nucleon cross section \cite{kla93}
\begin{equation}\label{msigma}
\sigma_{NN}^{\rm medium}=(1+\alpha\frac{\rho}{\rho_0})\sigma_{NN}^{\rm free} 
\end{equation}
with the parameter $\alpha\approx 0.2$ has been found to better reproduce the
flow data \cite{wes93,hun96}. It is thus very promising that heavy-ion 
collisions may shed some light on the in-medium cross sections and 
their isospin dependence.

\chapter{Isospin-dependent phenomenology in heavy-ion collisions at 
intermediate energies}
Several new and
interesting phenomena related to the isospin have been observed in
heavy-ion collisions with radioactive beams \cite{yen96}. 
To understand such phenomena many existing models developed for 
heavy-ion collisions, such as the isospin-dependent 
percolation model \cite{Kor97}, isospin-dependent lattice 
gas model \cite{kuo96} and isospin-dependent transport models,
have been extended to include the isospin degree of freedom.
We shall discuss in this chapter some of these phenomena and their 
explanations. Transport models are particularly
useful for studying the isospin-dependent  {\sc eos} 
and the in-medium nucleon-nucleon
cross sections, and we shall therefore also discuss in this chapter the 
role of isospin in these models and their predictions. 

\section{Isospin-dependent transport models for heavy-ion collisions}\label{buu}
A powerful framework to test various predictions on the isospin dependence 
of nuclear  {\sc eos} and in-medium nucleon-nucleon cross sections is 
the nuclear transport theory. The isospin degree of freedom has been included
in transport models at various levels of complexity 
\cite{far91,sob94,pawel,lir93,remaud,hart88,betty,libauer1,libauer2,lis95,li96}.    
These models were developed from the well-known Boltzmann-Uehling-Uhlenbeck 
({\sc buu}) transport theory \cite{greiner,bert88,cas90,bauer92} or 
the Quantum Molecular Dynamics ({\sc qmd}) theory \cite{aich91}. by 
evolving separately the phase space distribution functions for protons and 
neutrons.  In these models, the initial phase space distribution
of nucleons in the target and projectile, nucleon-nucleon cross sections, 
mean-field potentials, and Pauli blocking are all isospin-dependent.
For the isospin-dependent nucleon-nucleon cross section, the 
free-space cross sections shown in Fig.\ \ref{freesigma} have usually
been used. 

\begin{figure}[htp]
%\vspace{15cm}
\vspace{-1.0truecm}
\setlength{\epsfxsize=10truecm}
\centerline{\epsffile{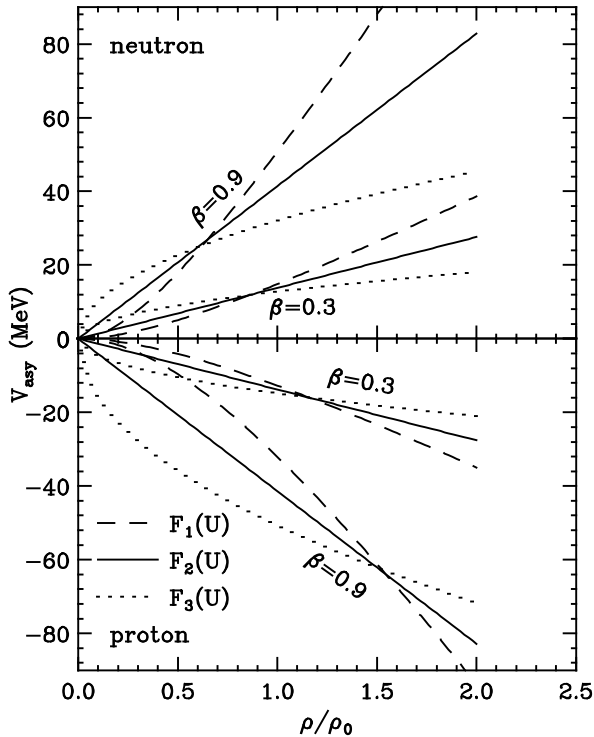}}
\caption{Symmetry potentials for neutrons and protons corresponding to
the three forms of $F(u)$ (see text). Taken from Ref.\ \protect\cite{li96}.}
\label{spotential} 
\end{figure}

For the isospin-dependent mean-field potential, 
equations of state based on nuclear many-body theories discussed in section 
\ref{eoslaw} have been used. For illustration we discuss here 
the properties of symmetry potential given in Eq.\ (\ref{vasy}). 
For the simplest form of $F(u)$, i.e., $F(u)=F_2(u)=u$, one has 
\begin{equation}\label{vasy1}
V_{\rm asy}^{n(p)}=\pm 2e_a u\delta=\pm 2e_a\frac{\rho_n-\rho_p}{\rho_0}.
\end{equation} 
This is the asymmetric part of nuclear mean-field potential used in 
Refs.\ \cite{pawel,betty,lis95,xu,li93}.
The asymmetric energy density of Eq. (\ref{simple}) used 
in Refs. \cite{far91,jou95} leads to the same 
mean-field potential at $\rho\approx \rho_0$ as Eq.\ (\ref{vasy1}). 
To illustrate the magnitude and variation of the symmetry potential, 
we show in Fig.\ \ref{spotential} $V_{\rm asy}^{n(p)}(\rho,\delta)$ 
using the three forms of $F(u)$ given in Eq. (\ref{fu}) 
and $e_{\rm sym}(\rho_0)=32$ MeV.
It is seen that the repulsive (attractive) mean-field potential for neutrons 
(protons) depends sensitively on the form of $F(u)$, the neutron excess 
$\delta$, and the baryon density $\rho$. In collisions of neutron-rich 
nuclei at intermediate energies, both $\delta$ and $\rho$ can be appreciable
in a large space-time volume where the isospin-dependent mean-field
potentials, which are opposite in sign for neutrons and protons, are strong. 
This will affect 
differently the reaction dynamics of neutrons and protons, leading to 
possible differences in their yields and energy spectra. 

In modeling heavy-ion collisions the nuclear mean-field potential
should also include a Coulomb term $V_C^p$ for protons, i.e., 
\begin{equation}\label{pot}
V^{n(p)}(\rho,\delta) = a (\rho/\rho_0) + b (\rho/\rho_0)^{\sigma}
	+V^p_{c}+V_{\rm asy}^{n(p)}(\rho,\delta).
\end{equation}

To extract 
$e_{\rm sym}(\rho)$ from the experimental data requires observables that 
are sensitive to the asymmetric part but not the symmetric part of 
the nuclear {\sc eos}. 
This is necessary because the magnitude of $V_{\rm asy}^{n(p)}$ is rather small 
compared to the symmetric part in Eq.\ (\ref{pot}). 
In addition, these observables 
should not depend strongly on other factors that affect the reaction 
dynamics, such as the in-medium nucleon-nucleon cross sections. 
In this respect, it is interesting to mention that the possibility
of extracting the isospin-dependent  {\sc eos} from the elastic scatterings of 
neutron-rich nuclei has been examined recently by using a 
double-folding model \cite{kho96,chr95}. Both the isovector and
isoscalar parts of the interaction potential between two neutron-rich nuclei
are evaluated. 
\begin{figure}[htp]
%\vspace{12cm}
\vspace{-5.0truecm}
\setlength{\epsfxsize=10truecm}
\centerline{\epsffile{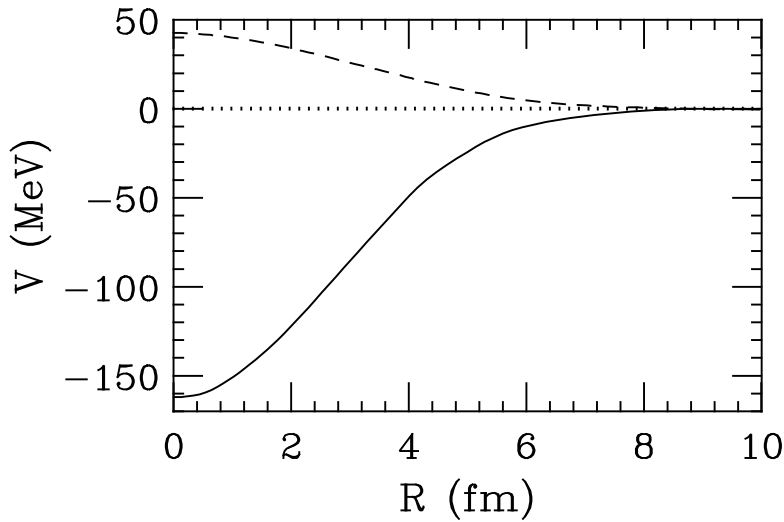}}
\caption{The symmetry and isoscalar potential for 
the scattering of $^{11}{\rm Li}+^{14}{\rm C}$ at a beam energy 
of 60 MeV/nucleon.  Shown is the calculation utilizing the BDM3Y1-Paris 
potential.  The dashed line represents the symmetry potential, and the
solid line is the isoscalar potential.
The results are taken from Ref.\ \protect\cite{kho96}.}
\label{v0v1} 
\end{figure}  

\begin{figure}[htp]
%\vspace{12cm}
\vspace{-5.0truecm}
\setlength{\epsfxsize=10truecm}
\centerline{\epsffile{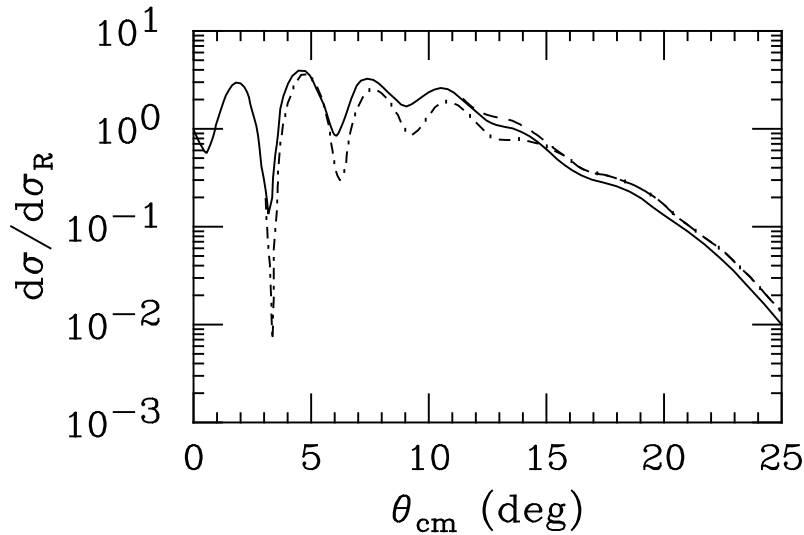}}
\caption{Elastic scattering cross section for $^{11}{\rm Li}+^{14}{\rm C}$ 
at a beam energy of 60 MeV/nucleon. Left window: the solid line is the 
result of a full calculation; the dashed line is obtained without 
the symmetry potential; and the dash-dotted line is obtained using the 
real potential folded with unrealistic compact density distribution 
for neutrons in $^{11}{\rm Li}$.
Taken from Ref.\ \protect\cite{kho96}.}
\label{escatter} 
\end{figure}

Shown in Fig.\ \ref{v0v1} are the isovector and isoscalar potentials 
for the scattering of $^{11}{\rm Li}+^{14}{\rm C}$ at a beam energy of
60 MeV/nucleon. The strength of the symmetry potential (isovector) 
is much smaller than that of the isoscalar potential, and consequently 
the optical potential and thus the elastic scattering cross section 
is dominated by the isoscalar potential as shown in Fig.\ \ref{escatter}.
As a result, it is difficult to extract the symmetry potential from 
elastic scattering data. 

These studies further indicate the importance of selecting appropriate 
observables in order to learn the isospin-dependent part of the 
nuclear  {\sc eos}. As in charge-exchange reactions, which are 
useful for extracting the isospin-dependent part of the nucleon-nucleon 
interaction \cite{hod63,len89}, the ratio of preequilibrium 
neutrons to protons in heavy-ion collisions at intermediate energies
have been found to be suitable for extracting the isospin-dependent  {\sc eos}
\cite{lik96}.
We shall discuss this ratio in detail in the next section using
the isospin-dependent transport model of 
Refs.\ \cite{lir93,libauer1,libauer2,lis95,li96}.

To describe properly reactions induced by neutron-rich nuclei, 
it is important to use as realistic as possible initial neutron and proton 
distributions in the transport model.  This can be achieved by
initializing neutrons and protons randomly 
in a sphere with a radius corresponding to the rms radius of a heavy 
nucleus, as we have discussed in section \ref{radio}. Then the {\sc buu/lv} 
model 
with a proper symmetry potential can describe the gross properties 
of the nucleus after evolving it in time,
despite the fact that
the model does not contain any information on its microscopic
structure.  Since the binding energy as well as density distributions 
of neutrons and protons depend on both the strength and form of 
symmetry potential, i.e., one obtains different density profiles using 
different symmetry potentials, it is thus possible to learn
about the asymmetric  {\sc eos} from studying saturation properties of 
neutron-rich nuclei. However, the information inferred 
from these studies is only around the saturation density. What is relevant
for astrophysics, but poorly known, is the 
asymmetric  {\sc eos} at other densities.  These can only be reached in
heavy-ion collisions. The effects of 
different symmetry potentials on particle emission in heavy ion collisions
might be stronger than on equilibrium properties of neutron-rich nuclei. 
To observe these effects the same initial states for both the target and 
projectile have to be used in calculations using different symmetry potentials. 
In Refs.\ \cite{lis95,li96}, this has been achieved 
by using initial nucleon distributions given by the {\sc rmf} theory.

\section{Isospin dependence of preequilibrium nucleon emission}\label{npratio}
In this section, we will first discuss the ratio of 
preequilibrium neutrons to protons $(R_{n/p}(E_{\rm kin})\equiv 
dN_n/dN_p)$ from collisions of neutron-rich nuclei at intermediate energies by
using the isospin-dependent {\sc buu} model.  Then we will
discuss experimental observations.
\begin{figure}[htp]
%\vspace{12cm}
\vspace{-5.0truecm}
\setlength{\epsfxsize=12truecm}
\centerline{\epsffile{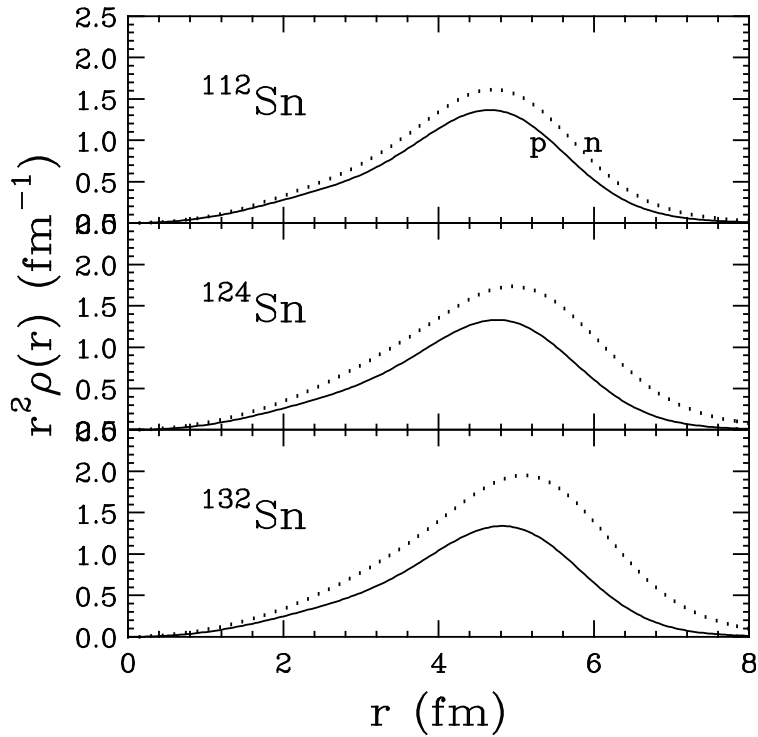}}
\caption{Density distributions of neutrons (dotted) and protons (solid) in 
${\rm Sn}$ isotopes calculated using the {\sc rmf} theory.}
\label{densitysn} 
\end{figure}  

As an example, collisions of $^{112}{\rm Sn}+^{112}{\rm Sn}$, 
$^{124}{\rm Sn}+^{124}{\rm Sn}$ and $^{132}{\rm Sn}+^{132}{\rm Sn}$ 
at a beam energy of 40 MeV/nucleon are studied. The first two reactions 
have been recently investigated experimentally at {\sc nscl/msu} by the 
MSU-Rochester-Washington-Wisconsin collaboration \cite{dem96,kunde96}. 
Preequilibrium particles have been measured in these experiments and are 
now being analyzed \cite{udo}. The last system 
is included for discussions and comparisons. 
The density distributions of neutrons and protons in the ${\rm Sn}$ isotopes
are calculated using the {\sc rmf} theory \cite{ren} and are shown in 
Fig.\ \ref{densitysn}, and the local Thomas-Fermi approximation is 
then used to generate the nucleon momenta. To identify free nucleons, 
a phase-space coalescence method has been used at 200 fm/c after the 
initial contact of the two nuclei, when the quadrupole moment of 
the nucleon momentum distribution in the heavy residue is almost zero, 
indicating the approach of thermal equilibrium.
A nucleon is considered as free if it is not 
correlated with other nucleons within a spatial distance of 
$\triangle r= 3$ fm and a momentum distance of $\triangle p = 300$ MeV/c.
Otherwise, it is bounded in a cluster. These results are not sensitive to 
these parameters if they are varied by less than 30\% around these values.

\begin{figure}[htp]
%\vspace{12cm}
\vspace{-5.0truecm}
\setlength{\epsfxsize=10truecm}
\centerline{\epsffile{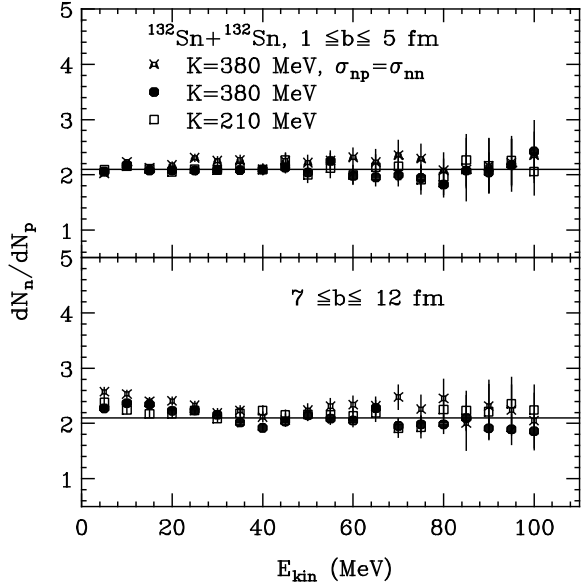}}
\caption{The ratio of preequilibrium neutrons to protons
as a function of kinetic energy in central (upper window) and peripheral 
(lower window) reactions of $^{132}{\rm Sn}+^{132}{\rm Sn}$ at a beam 
energy of 40 MeV/nucleon without the Coulomb and symmetry potentials. 
Taken from \protect\cite{lik96}.}
\label{rekin} 
\end{figure}

\begin{figure}[htp]
%\vspace{12cm}
\vspace{-1.0truecm}
\setlength{\epsfxsize=10truecm}
\centerline{\epsffile{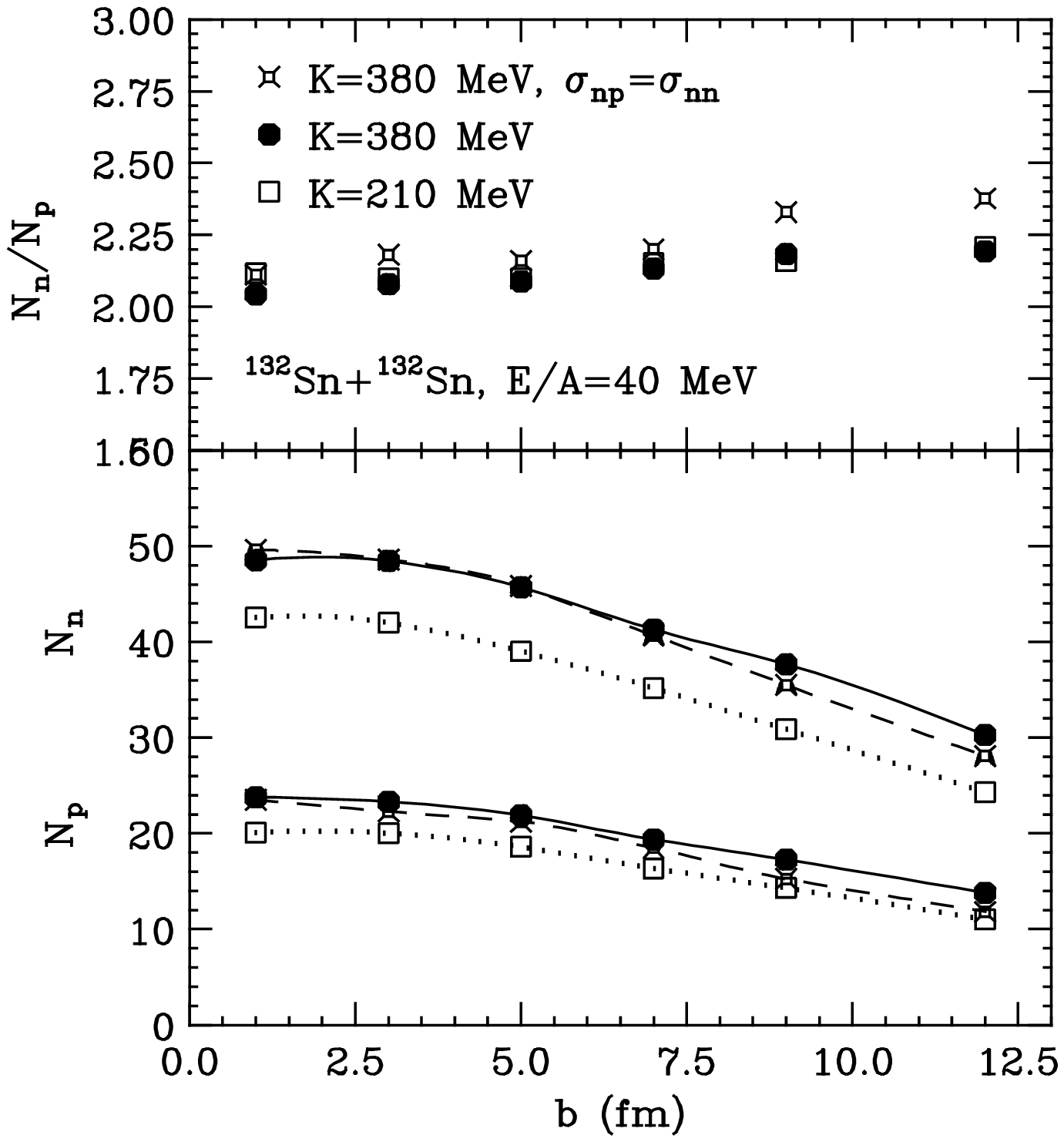}}
\caption{The total numbers of preequilibrium neutrons and protons 
(lower window) and their ratio (upper window) as functions of 
impact parameter for the reaction of $^{132}{\rm Sn}+^{132}{\rm Sn}$ at a 
beam energy of 40 MeV/nucleon.} 
\label{rtotal} 
\end{figure}

Effects of the compressibility $K$ of symmetric 
nuclear matter and the in-medium nucleon-nucleon cross sections on 
the ratio $R_{n/p}(E_{\rm kin})$ can be studied by neglecting both the 
Coulomb and symmetry potentials in the {\sc buu} 
model. In Fig.\ \ref{rekin} this 
ratio is shown as a function of nucleon kinetic energy 
for central (upper window) and peripheral (lower window) collisions of 
$^{132}{\rm Sn}+^{132}{\rm Sn}$ at a beam energy of 40 MeV/nucleon. 
The total number of preequilibrium neutrons and protons and their ratios 
for the same reaction are shown in Fig.\ \ref{rtotal}.  
When varying the compressibility $K$ from 210 MeV (open squares) to 
380 MeV (filled circles), it is seen that although the yields of both 
protons and neutrons increase, their ratio remains similar
for all impact parameters. This is simply because the effects of an
isospin-symmetric {\sc eos} on neutrons and protons are identical. 

The experimental neutron-proton cross section
is about three times the neutron-neutron (proton-proton) cross section
in the energy range studied here. Setting the two cross 
sections equal (fancy squares), it has been found that the proton and
neutron yields and 
their ratios change by less than 10\% even in peripheral collisions
of $^{132}{\rm Sn}+^{132}{\rm Sn}$. This result is also easy to understand 
since both colliding nucleons have the same probability to gain enough 
energy to become unbound \cite{lir93}. Thus, the in-medium, isospin-dependent 
nucleon-nucleon cross sections do not affect much the ratio 
$R_{n/p}(E_{\rm kin})$.
In the absence of Coulomb and symmetry potentials their ratio is thus 
almost independent of the nucleon kinetic energy and has a constant value  
of about $2.1\pm 0.3$ in both central and peripheral collisions of
$^{132}{\rm Sn}+^{132}{\rm Sn}$. 

\begin{figure}[htp]
%\vspace{15cm}
\vspace{-5.0truecm}
\setlength{\epsfxsize=10truecm}
\centerline{\epsffile{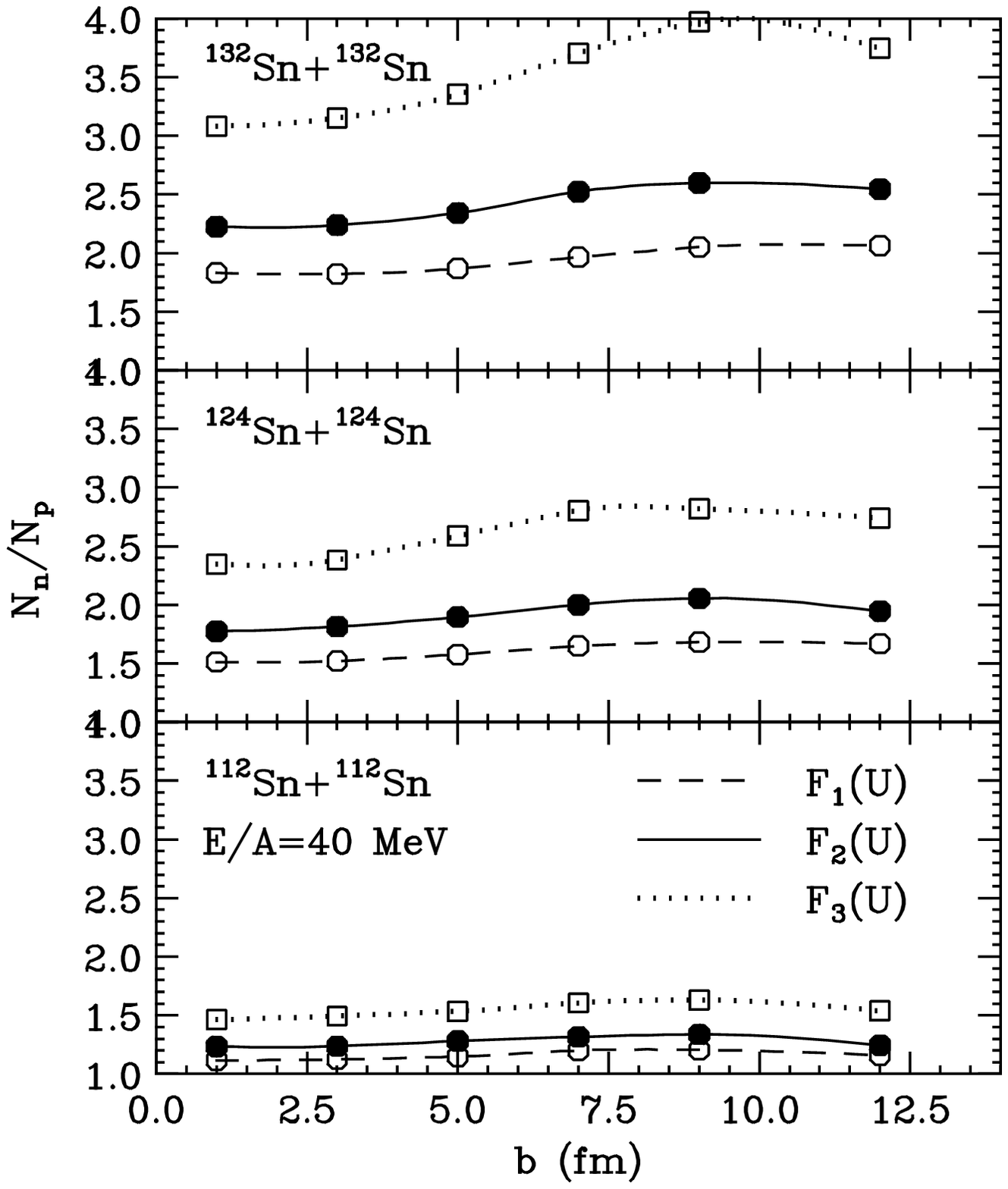}}
\caption{The ratio of preequilibrium neutrons to 
protons as a function of impact parameter using the three forms of $F(u)$.} 
\label{ratio2} 
\end{figure}

Including the Coulomb and the asymmetric term of the 
 {\sc eos} in Eq.\ (\ref{pot}), one can then study the effects of the symmetry
energy, $e_{\rm sym}(\rho)$, since the Coulomb effect is well-known. 
The symmetry potential has the following effects on 
preequilibrium nucleons. First, it
tends to make more neutrons than protons unbound.
One therefore expects that a stronger symmetry potential leads to 
a larger ratio of free neutrons to protons. 
Second, if both neutrons and protons are already free, the symmetry 
potential makes neutrons more energetic than protons. 
\begin{figure}[htp]
%\vspace{15cm}
\vspace{-5.0truecm}
\setlength{\epsfxsize=14truecm}
\centerline{\epsffile{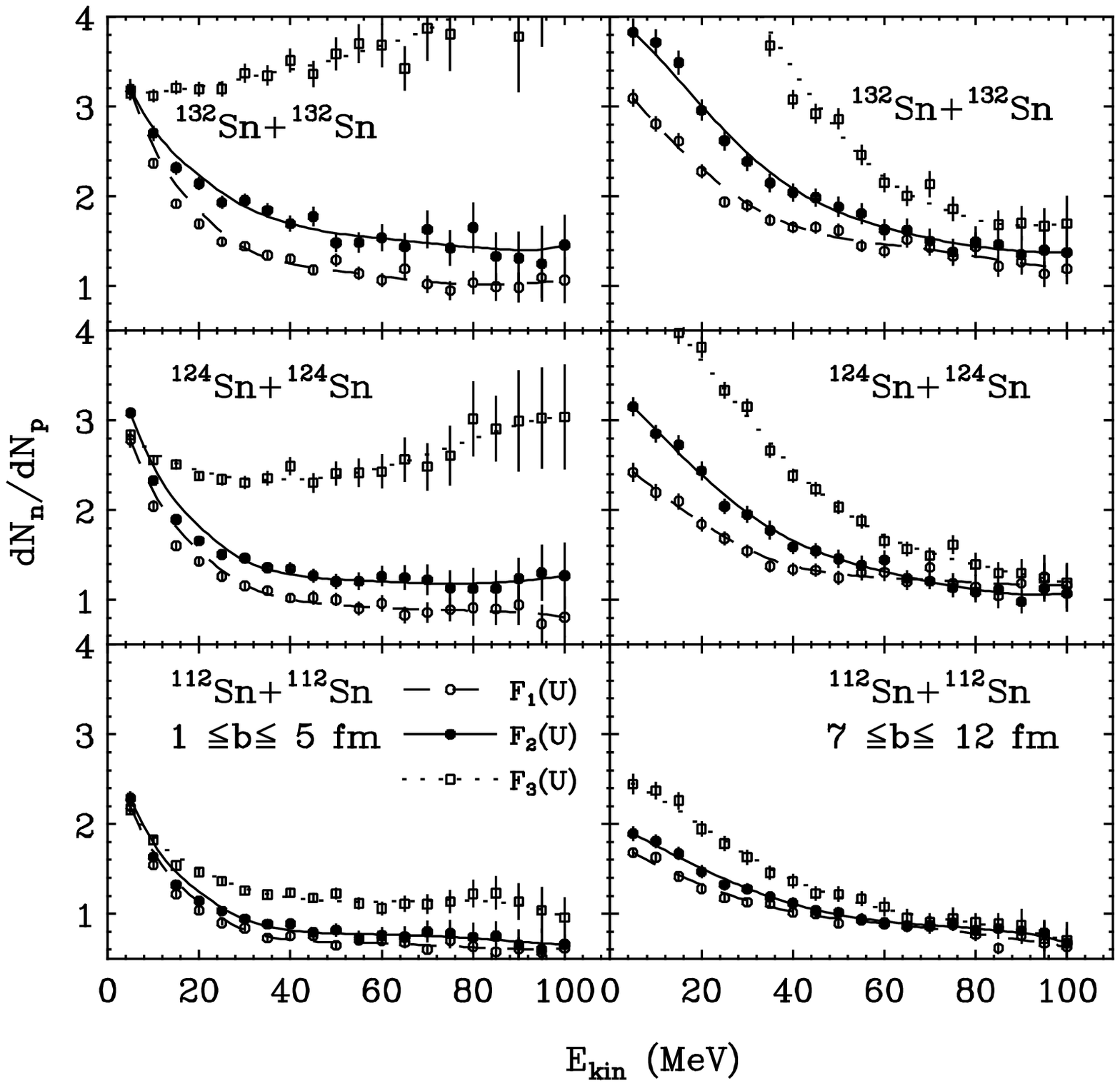}}
\caption{The ratio of preequilibrium neutrons to protons
as a function of kinetic energy in central (left windows) and peripheral 
(right windows) reactions using the three forms of $F(u)$. Taken 
from \protect\cite{lik96}.}
\label{ratio1} 
\end{figure}  
These effects are clearly demonstrated in Fig.\ \ref{ratio1} where the ratios 
$R_{n/p}(E_{kin})$ are shown as a function of kinetic energy. 
These results are obtained by using the three forms of $F(u)$ for 
typical central (left windows) and peripheral (right windows) collisions 
of $^{112}{\rm Sn}+^{112}{\rm Sn}$, $^{124}{\rm Sn}+^{124}{\rm Sn}$ and 
$^{132}{\rm Sn}+^{132}{\rm Sn}$, 
respectively. The increase of the ratios at lower kinetic energies 
in all cases is due to Coulomb repulsion which shifts protons from 
lower to higher kinetic energies. On the other hand, the different 
ratios calculated using different $F(u)$'s reflect clearly the 
effect mentioned above, i.e., with a stronger symmetry 
potential the ratio of preequilibrium neutrons to protons 
becomes larger for more neutron rich systems. 
\begin{figure}[htp]
%\vspace{10cm}
\vspace{-5.0truecm}
\setlength{\epsfxsize=10truecm}
\centerline{\epsffile{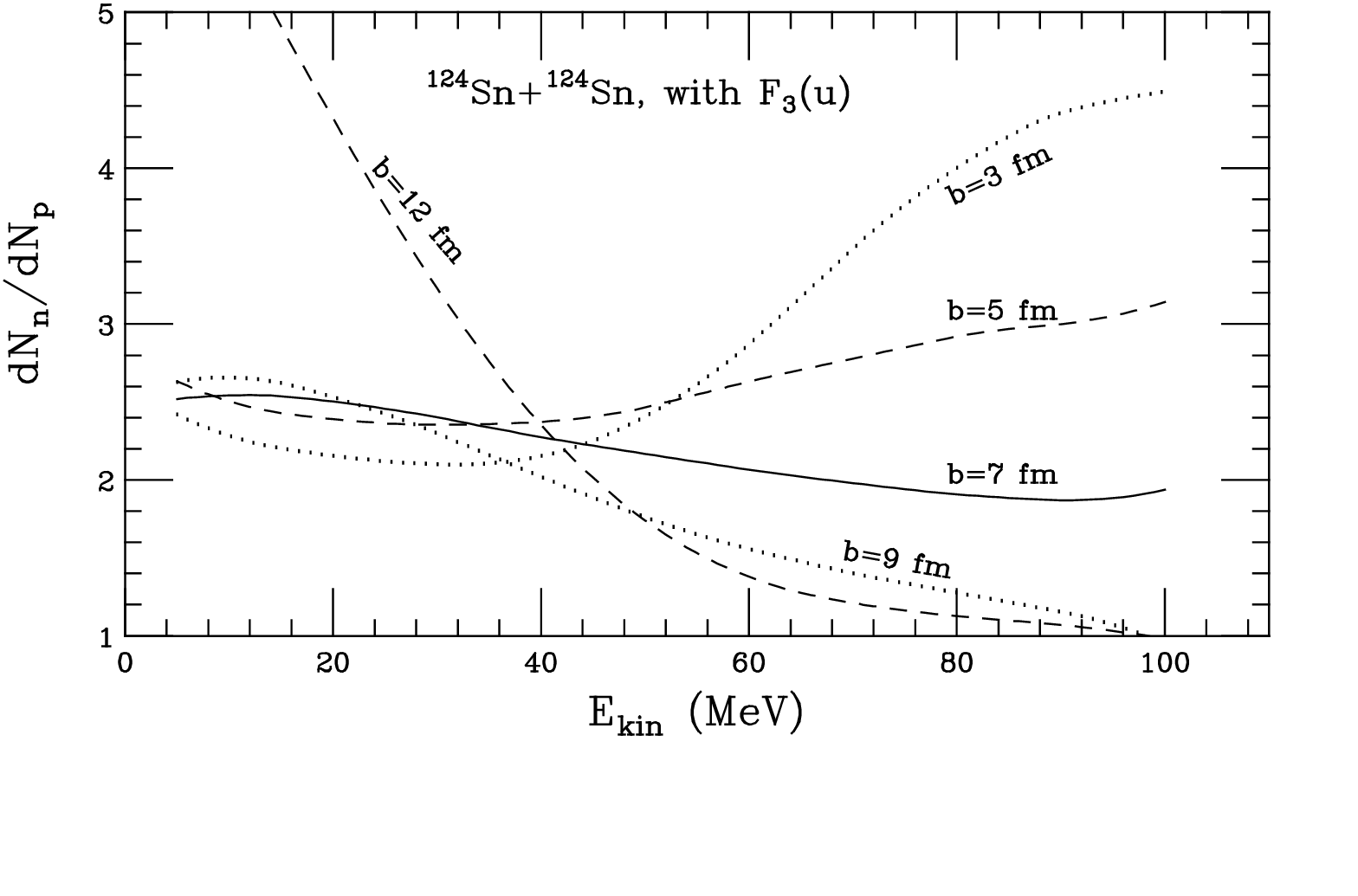}}
\caption{Impact parameter dependence of the ratio $R_{n/p}(E_{\rm kin})$ 
as a function of kinetic energy for the reaction of $^{124}{\rm Sn}
+^{124}{\rm Sn}$ using $F_3(u)$.} 
\label{ratiob} 
\end{figure}  
Between the central and peripheral collisions the ratios vary 
smoothly as a function of impact parameter. 
This is shown in Fig. \ref{ratiob} where the energy-dependence of the 
ratio $R_{n/p}(E_{\rm kin})$ is shown for different impact parameters for
the reaction of $^{124}{\rm Sn}+^{124}{\rm Sn}$ using $F_3(u)$.
The impact parameter dependence of the ratio of the total yield of 
neutrons to that of protons shown in Fig.\ \ref{ratio2} clearly 
demonstrates effects of the neutrons skins in more neutron-rich 
system at larger impact parameters.

It is interesting to note that effects due to 
different symmetry potentials are seen in different kinetic energy 
regions for central and peripheral collisions. 
This can be understood from the impact parameter dependence 
of $R_{n/p}(E_{\rm kin})$, shown in Fig.\ \ref{ratiob}, and the 
reaction dynamics. In central collisions, effects of the symmetry 
potential are most prominent at higher kinetic energies. This is 
because most of finally observed free neutrons 
and protons are already unbound in the early stage of the reaction as 
a result of violent nucleon-nucleon collisions. 
The symmetry potential thus mainly affects the nucleon energy spectra 
by shifting more neutrons to higher kinetic energies with respect to protons. 
In peripheral collisions, however, there are fewer 
nucleon-nucleon collisions; whether a nucleon can become unbound
depends strongly on its potential energy. With a stronger symmetry potential 
more neutrons (protons) become unbound (bound) 
as a result of a stronger symmetry potential, but
they generally have smaller kinetic energies. Therefore, in peripheral 
collisions effects of the symmetry potential show up chiefly at lower 
kinetic energies. For the two systems with more neutrons the effects 
of the symmetry potential are so strong that in central (peripheral) 
collisions different forms of $F(u)$ can be clearly distinguished from 
the ratio of preequilibrium neutrons to protons at higher (lower) kinetic 
energies. However, because of energy thresholds in detectors, it is 
difficult to measure low energy nucleons, especially neutrons. Furthermore, 
the low energy spectrum also has appreciable contribution from 
emissions at the later stage when the system is in equilibrium. Therefore, 
the measurement of the ratio $R_{n/p}(E_{\rm kin})$ 
in central heavy-ion collisions for nucleons with energies higher than 
about 20 MeV is more suitable for extracting the  {\sc eos} of asymmetric 
nuclear matter.
\begin{figure}[htp]
%\vspace{12cm}
\vspace{-8.0truecm}
\setlength{\epsfxsize=12truecm}
\centerline{\epsffile{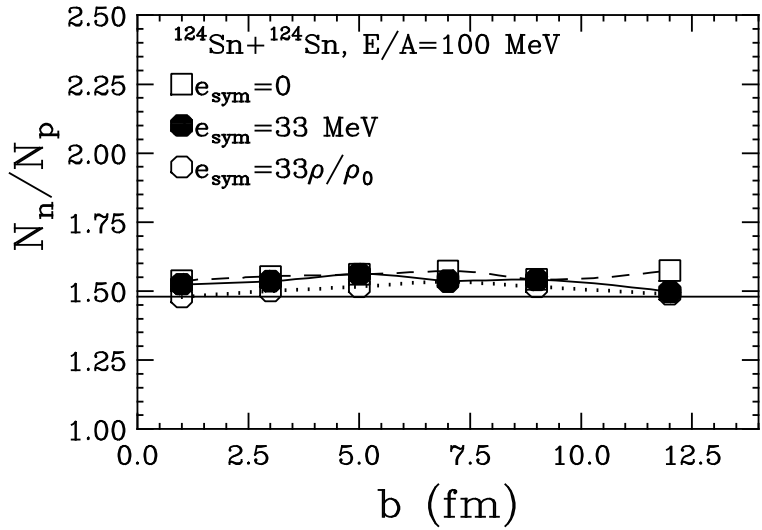}}
\caption{The ratio of preequilibrium neutrons to 
protons as a function of impact parameter for the reaction 
of $^{124}{\rm Sn}+^{124}{\rm Sn}$ at a beam energy of 100 MeV/nucleon.}
\label{r100} 
\end{figure}  
The beam energy range where the symmetry potential is relevant for
heavy-ion collisions depends on the isospin asymmetry of the reaction 
system and the observables to be studied. To observe effects of a weak 
mean-field potential, such as the symmetry potential, one needs to use 
relatively 
low beam energies so that the dynamics is not dominated by nucleon-nucleon 
collisions. On the other hand, to study the density dependence and to 
reach a stronger mean-field potential the reactions should be energetic 
enough to achieve sufficient compression. Thus, it is necessary to study 
theoretically the beam energy dependence of the isospin effects on 
preequilibrium nucleons. Preliminary studies on the neutron/proton ratio 
in the reaction of $^{124}{\rm Sn}+^{124}{\rm Sn}$ indicate that the 
beam energy should not be greater than about 100 MeV. To illustrate this, 
we show in Fig. \ref{r100} the $R_{n/p}(E_{\rm kin})$ 
ratio as a function of impact parameter for this 
reaction at a beam energy of 100 MeV. Calculations corresponding to 
three typical cases of the symmetry energy, i.e., $e_{\rm sym}=0$, 
33 MeV, and $33\rho/\rho_0$ MeV, have been carried out.
It is seen that differences among results from these calculations 
are rather small and are all close to the 
N/Z ratio of the reaction system, i.e., 1.48.

There seems already experimental evidence in heavy-ion collisions for 
the existence of a strong symmetry potential. For example, in experiments 
which measure preequilibrium nucleons, a number of puzzles 
have been observed, and they can be explained at least qualitatively 
by effects of the symmetry potential. In the following, 
several of these puzzles are discussed. 
In experiments of heavy-ion collisions around the Fermi 
energy \cite{hils87,hils88,hils92,hils95}, 
Hilscher {\it et al.} have found that the neutron/proton
ratio, $((N/Z)_{\rm free})$, of preequilibrium nucleons is consistently 
higher than that of the projectile-target system,
$(N_{P}+N_{t})/(Z_{p}+Z_{t})$, and cannot be explained by the Coulomb 
effect. For example, in the reaction of
$^{12}{\rm C}+^{165}{\rm Ho}$ at a beam energy of 32 MeV/nucleon, they have
measured the neutron and proton spectra at $14^{\circ}$ and energies between 70 
and 100 MeV, and found that the multiplicity of neutrons is larger 
than that of protons by a factor of $1.4\pm 0.2, 1.7\pm 0.3$ and $2.4\pm 0.3$
for linear-momentum transfers of 52\%, 73\% and 93\%, 
respectively.  Therefore, the neutron to proton ratio is much higher than
that of the reaction system $(N/Z)_{\rm cs}$=1.42 
in central collisions corresponding to higher linear-momentum transfer,
This result cannot be
explained by the standard Fermi jet model for preequilibrium nucleon 
emission \cite{hils87}. On the other hand, it is in agreement 
with the {\sc buu} predictions discussed above. 
From Fig.\ \ref{ratio1} and Fig.\ \ref{ratio2}, it is seen that
with a large symmetry potential the n/p ratio for 
energetic nucleons can be much larger than that of the colliding system.
Quantitative comparisons with the experiments will be very useful for
extracting the isospin-dependent  {\sc eos}. 
\begin{figure}[htp]
%\vspace{15cm}
\vspace{-5.0truecm}
\setlength{\epsfxsize=10truecm}
\centerline{\epsffile{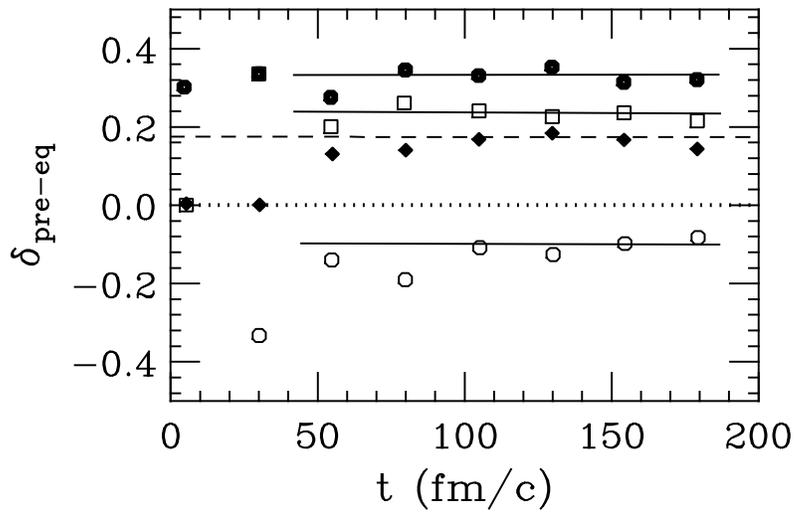}}
\caption{The ratio of preequilibrium neutrons to 
protons as a function of time for the reaction of $^{12}{\rm C}
+^{165}{\rm Ho}$ at a 
beam energy of 35 MeV/nucleon and an impact parameter of 1.0 fm. 
Solid circles: $c=46$ MeV with Coulomb; open squares: $c=32$ MeV with
Coulomb; open circles: $c=0$ MeV with Coulomb; solid diamonds: $c=0$ MeV 
without Coulomb.
Taken from \protect\cite{lir93}.}
\label{lirandrup} 
\end{figure}  
Semi-quantitative comparisons between theoretical results and some of the data 
have already been made using a simple symmetry potential energy density 
of Eq.\ (\ref{simple}) within the {\sc buu/lv} model \cite{lir93,jou95}. In 
Fig.\ \ref{lirandrup} the relative neutron excess of preequilibrium nucleons
$\delta_{\rm pre-eq} = (N-Z)/(N+Z)$ is shown as a function of time for
the reaction of $^{12}{\rm C}+^{165}{\rm Ho}$ at a beam energy of 35 
MeV/nucleon and an
impact parameter of 1.0 fm. In the early stage of the reaction, the 
neutron excess is seen to fluctuate due to violent collisions, but  
reaches a constant value
after about 100 fm/c. The dashed line at 0.17 is the neutron excess of 
the reaction system. Without the symmetry 
and Coulomb potential the final neutron excess of preequilibrium 
nucleons is about the same as that of the reaction system. 
In the case of including the Coulomb but not the symmetry potential 
($c=0$) more protons are emitted in the preequilibrium stage of the reaction 
due to the Coulomb repulsion. This effect has also 
been observed in other {\sc buu} calculations \cite{far91,sob94,pawel}. 
Furthermore, the ratio $(N/Z)_{\rm free}$ increases monotonically as
strength of the symmetry energy increases . To have a ratio larger 
than the N/Z of the reaction system one needs to use a strength parameter 
$c\geq 20$ MeV corresponding to $e_{\rm sym}(0)\geq 32$ MeV. More detailed 
comparisons with data are required to see what form and strength of the 
symmetry potential are needed to reproduce the data. 

\begin{figure}[htp]
%\vspace{10cm}
\vspace{-5.0truecm}
\setlength{\epsfxsize=10truecm}
\centerline{\epsffile{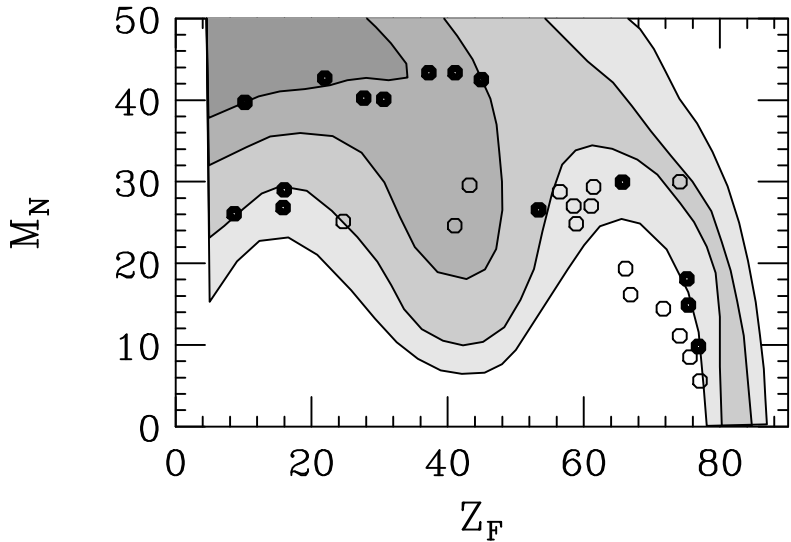}}
\caption{Experimental neutron multiplicity versus fragment 
charge for peripheral reactions of Pb+Au at a beam energy 
of 29 MeV/nucleon. {\sc lv} 
model calculations with (without) the symmetry potential
are given by filled (open) symbols. 
Taken from \protect \cite{jou95}.}
\label{ganil} 
\end{figure}

A more quantitative 
comparison between the recent data from Ganil and the {\sc LV} model 
calculation \cite{jou95} has been carried out. In Fig.\ \ref{ganil}, 
the experimentally measured 
correlations between neutron multiplicities, $M_N$, in the forward direction 
($6^\circ \le \theta \le 20^\circ$)
and fragment charges, $Z_f$, in the reaction of Pb+Au at 
$E_{\rm beam}/A$= 29 MeV 
are shown as contours. The contour lines separating the areas in different 
shades (from white to dark gray) have the values 0.6, 1, 2, and 3, 
respectively. Calculations using the {\sc lv} model without the 
symmetry potential $(c=0)$ are shown as open symbols and are seen to
significantly underpredict the neutron multiplicities, especially in 
central collisions. 
In calculations with a symmetry energy strength parameter $c=20$ MeV 
(filled circles)
the emission of neutrons is enhanced, so the data are better described.

\begin{figure}[htp]
%\vspace{10cm}
\vspace{-5.0truecm}
\setlength{\epsfxsize=10truecm}
\centerline{\epsffile{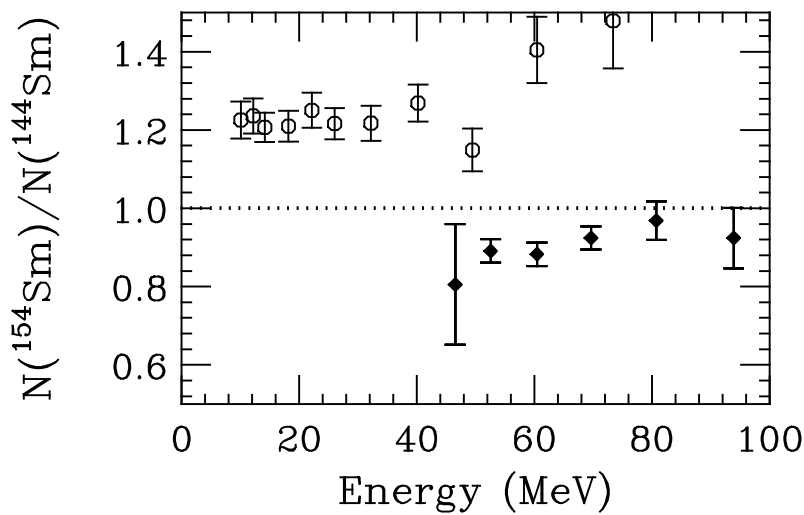}}
\caption{The ratio of free neutrons (open circles) and protons (solid diamond)
in central reactions of $^{32}{\rm S}+^{144, 154}{\rm Sm}$ as a function
of the kinetic energy of the emitted particle, at a 
beam energy of 26 MeV/nucleon. Taken from
\protect\cite{hils88}.}
\label{hilscher} 
\end{figure}

Another interesting experimental observation 
is the ratio of free neutrons and protons from central reactions of 
$^{32}{\rm S}+^{144,154}{\rm Sm}\rightarrow F,F+n,p$,
where $F$ denotes nuclear fragments, at $E_{beam}/A=26$ MeV 
as shown in Fig.\ \ref{hilscher}. We note that protons emitted at 
velocities higher than the projectile velocity are mainly 
preequilibrium particles. Several interesting observations can be 
made from these data. First, the emission of protons 
in the neutron-richer system $({\rm S}+^{154}{\rm Sm})$ is suppressed although 
both systems have the same number of protons to begin with. As we have 
discussed in detail in the previous section, protons feel an attractive 
symmetry potential, and the emission of high energy protons is thus suppressed 
with respect to neutrons. Correspondingly, the emission of high energy 
neutrons is enhanced, and this is indeed consistent with the neutron data 
at higher kinetic energies. Second, even for neutrons with lower 
kinetic energies the ratio of free neutrons from the two systems 
is about 1.23 and is much larger than the ratio (1.12) of neutrons of 
the two systems. More recently, Schr\"oder {\it et al.} have 
studied systematically the spectra of preequilibrium neutrons and
protons in both isospin symmetric and asymmetric collisions \cite{udo2}.
It is found that the n/p ratio of preequilibrium nucleons does not scale with
the N/Z ratio of the combined system. In particular, a preliminary analysis 
of the experimental data indicates that the ratio 
$(n/p)_{^{48}{\rm Ca}+^{112}{\rm Sn}}/(n/p)_{^{40}{\rm Ca}+^{112}{\rm Sn}}$ 
at a beam energy of 35 MeV/nucleon is about 4 to 12 depending on the impact 
parameter.  A similar observation has been made by
Hilscher {\it et al.}. Both results are thus consistent with 
the existence of a strong symmetry potential.

Enhanced production of free neutrons than that predicted by 
the standard statistical model
have also been reported in Refs.\ \cite{dem96,gon90}.
In Ref.\ \cite{han95} it was pointed out that within the experimental and 
model uncertainties the production of neutron-deficient
residues is much higher than the statistical model predictions.
Because of charge conservation, this result 
is thus equivalent to an enhanced neutron emission as shown in
Refs.\ \cite{kunde96,gon90}. This is again consistent 
with the existence of a strong symmetry potential. To extract quantitative
information about the nuclear symmetry potential requires, however, more
detailed studies using the transport model.

To summarize, we have shown that in the isospin-dependent {\sc buu} model 
for heavy-ion collisions at intermediate energies
the calculated ratio of preequilibrium neutrons to protons is very 
sensitive to the symmetry potential. Longstanding puzzles observed in
preequilibrium neutrons and protons can be understood qualitatively
by considering effects of the symmetry potential. More quantitative comparisons
with data are still needed to constrain the form and strength of the symmetry
potential.
 
\section{Isospin equilibrium as a probe of nuclear stopping power}
\begin{figure}[htp]
%\vspace{12cm}
\vspace{.0truecm}
\setlength{\epsfxsize=12truecm}
\centerline{\epsffile{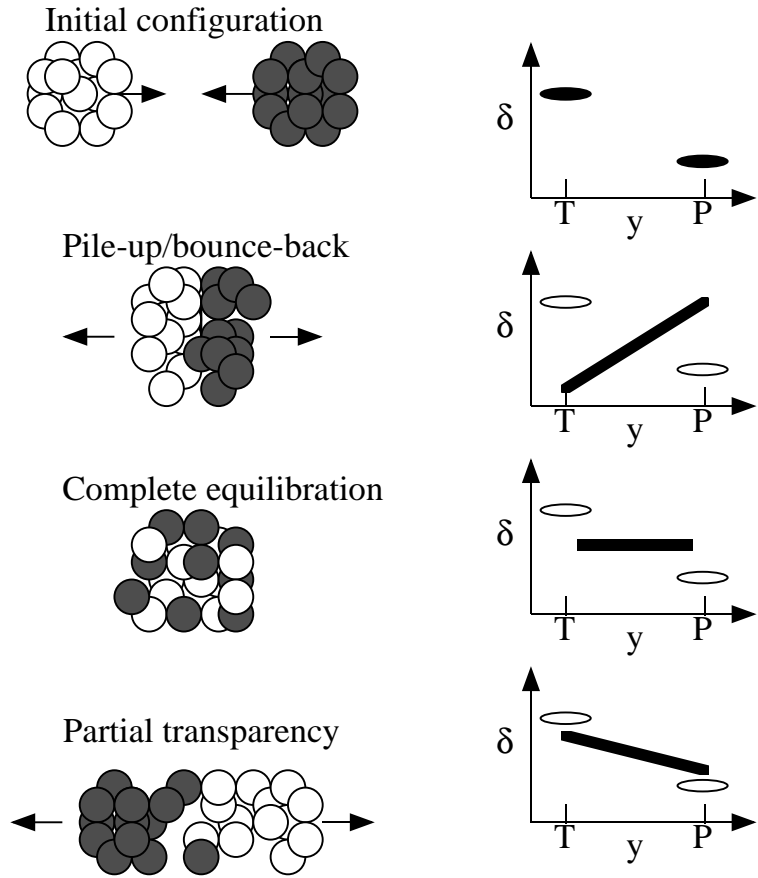}}
\caption{Illustration of using the rapidity distribution of 
neutron/proton ratio as a probe of nuclear stopping power. (A similar 
illustration can be found in Ref.\ \protect{\cite{bass94}}.) }
\label{bass} 
\end{figure}

The degree of stopping in heavy-ion collisions is an important
quantity in determining the outcome of a reaction.
It is usually studied by measuring the rapidity distribution or the asymmetry
of the nucleon momentum distribution. 
If we were able to tag the nucleons from projectile and target, then this
task would be much easier.  By using nuclei with different N/Z ratios, such
a tag can be provided.
Recently, it has thus been suggested
that the degree of equilibration in the isospin degree of freedom
also provides a means to probe nuclear stopping in heavy ion 
collisions \cite{bass94}, as illustrated in Fig.\ \ref{bass}. 

The pictures on the left show the nucleon distributions
in coordinate space, while that on the right are the rapidity distributions
for the three scenarios of pile-up and bounce-back, 
stopping and mixing (isospin equilibrium), and 
transparency. It is seen that for a projectile and a target with 
very different N/Z ratios, a comparison of the rapidity distribution of N/Z
before (shown in top picture on the right) and after the collision can 
give direct information about the
degree of stopping between the target and projectile \cite{lis95,bass94}.
This method thus allows one to study whether there is a transition 
from full stopping to 
transparency as the beam energy varies. Also, if isospin equilibrium can be
reached in the collisions, it is then possible to determine
its time scale relative to that for thermal equilibrium.
Moreover, one can study the dependence of isospin equilibrium 
on the nuclear  {\sc eos} and the in-medium nucleon-nucleon cross sections.
We note that this method is restricted to beam energies that are below 
the pion production threshold; so the total nucleon charge is 
conserved.
\begin{figure}[htp]
%\vspace{12cm}
\vspace{-5.0truecm}
\setlength{\epsfxsize=10truecm}
\centerline{\epsffile{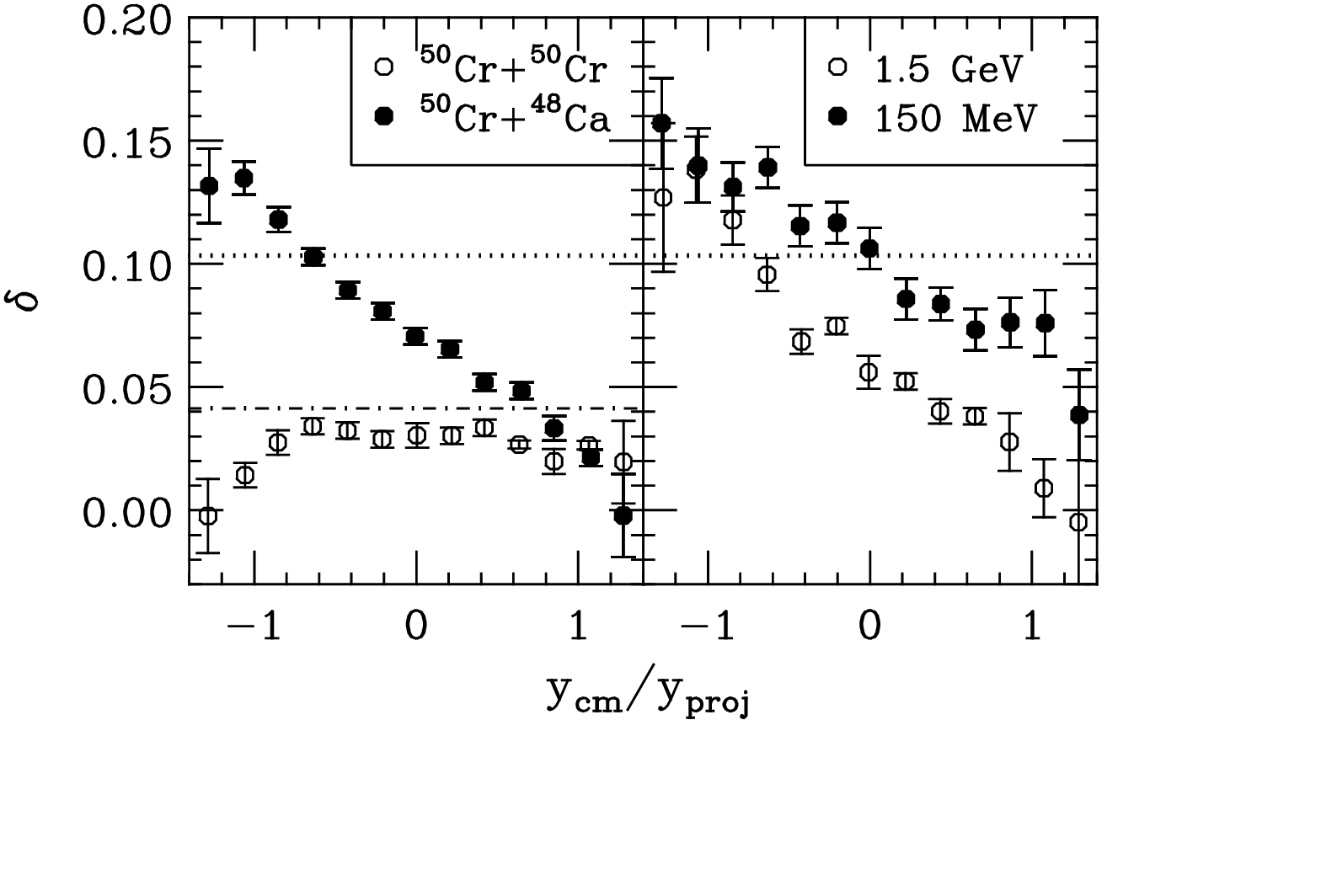}}
\caption{The neutron to proton asymmetry versus rapidity predicted by
the {\sc qmd} model. Taken from Ref. \protect\cite{bass94}.}
\label{qmd} 
\end{figure} 

These ideas have been studied in the isospin-dependent {\sc qmd} model 
under various conditions as shown in Fig.\ \ref{qmd}. In the left window
central collisions of $^{50}{\rm Cr}+^{48}{\rm Ca}$ and $^{50}{\rm Cr}+
^{50}{\rm Cr}$ at $E_{\rm beam}/A=1.0$ GeV are compared, and it shows 
clearly that there is a transparency in the asymmetric system.
In the right window two calculations for the asymmetric system
for different beam energies are shown, and it is seen that even 
at 150 MeV/nucleon the asymmetric system
shows transparency.  Bass et al.\ further show that the signature for 
transparency seen in the N/Z ratio is not altered by cluster formation.
On the other hand, the stopping power is affected by the magnitude of
the nucleon-nucleon cross section.  As expected, increasing the 
in-medium nucleon-nucleon cross section by a factor of 5 
results in a transition from transparency to full stopping.    
\begin{figure}[htp]
%\vspace{14cm}
\vspace{-5.0truecm}
\setlength{\epsfxsize=10truecm}
\centerline{\epsffile{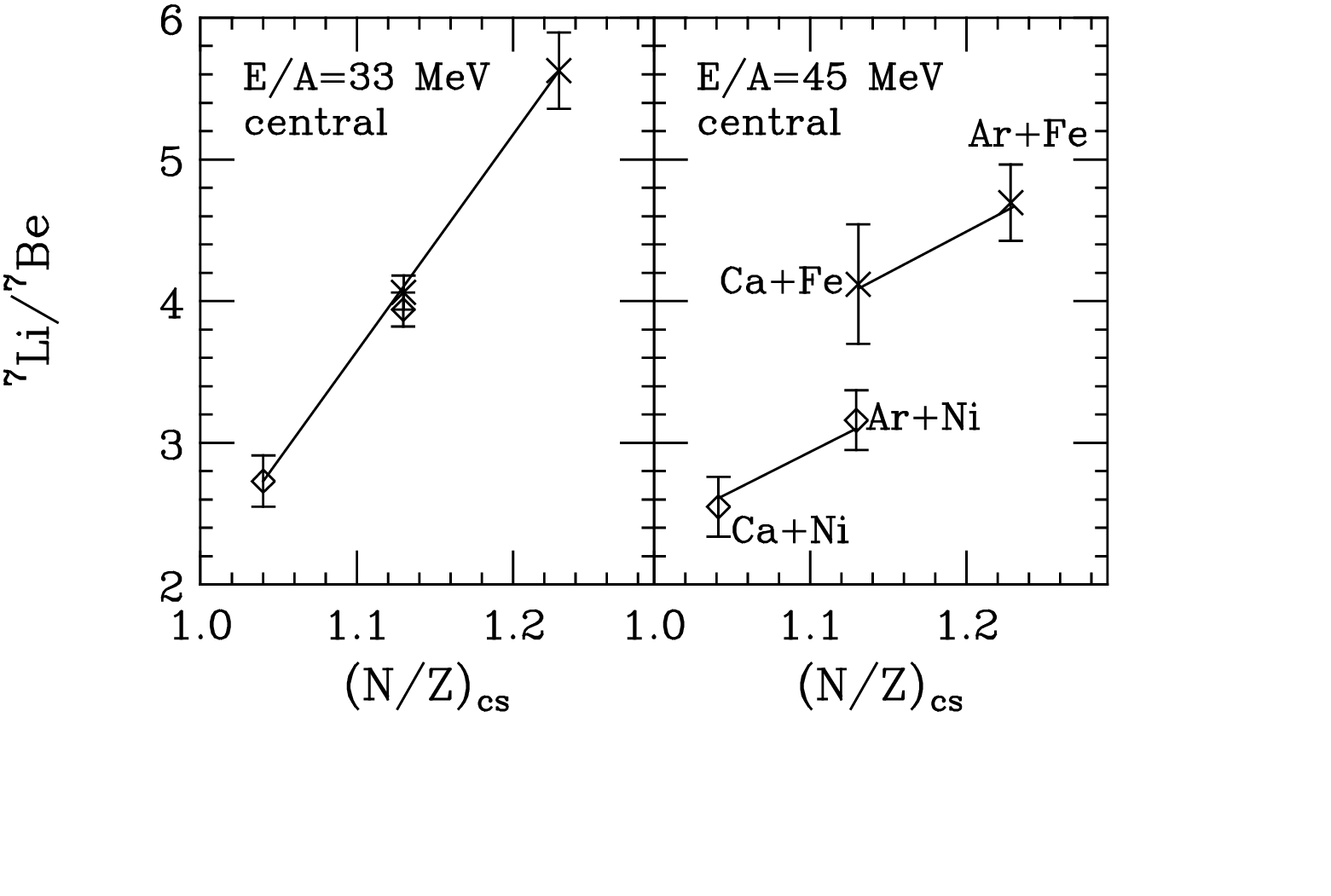}}
\caption{Isobaric ratios from central collisions plotted as a function 
of N/Z ratio of the combined target and projectile system at 
$E_{\rm beam}/A$=35 and 45 MeV. Taken from Ref. \protect\cite{sherry3}.}
\label{sherry} 
\end{figure} 

The idea of using isospin to study the nature of stopping in nuclear 
collisions has been introduced earlier in Ref.\ \cite{lis95} to explain 
the angular distribution of isotope ratios observed in 
experiments by Yennello {\it et al.} \cite{sherry1,sherry2,sherry3}.
In these experiments the isotopic composition of intermediate mass 
fragments from central collisions of $^{40}{\rm Cl},~ ^{40}{\rm Ar}$ 
and $^{40}{\rm Ca}$ with $^{58}{\rm Fe}$ and $^{58}{\rm Ni}$ at 
$E_{\rm beam}/A$=25, 35, 45 and 53 MeV was studied at {\sc nscl/msu} and 
{\sc tamu}. 
It was found that at $E_{\rm beam}/A$=25 and 33 MeV the isotopic 
ratios $^9{\rm Be}/^7{\rm Be}, ~^{11}{\rm B}/^{10}{\rm B}$ and 
$^{13}{\rm C}/^{12}{\rm C}$ increase 
linearly with increasing $(N/Z)_{\rm cs}$ ratio of the combined 
target and projectile system, but are independent of the N/Z ratio 
of the target or projectile. Shown in the left window of 
Fig.\ \ref{sherry} are typical results of reactions at 33 MeV/nucleon,
which indicate that the isospin is equilibrated 
in the composite system formed in these reactions 
before the emission of fragments.

The most striking and unexpected feature was 
observed from the isotopic ratios in central collisions at 
$E_{\rm beam}/A=$ 45 and 53 MeV. A typical result at $E_{beam}/A=$45 MeV
is shown in the right window of Fig.\ \ref{sherry}. 
It is seen that the isotopic ratios depend 
on the N/Z ratio of the target and projectile in reactions with 
target-projectile combinations having the same $(N/Z)_{CS}$ ratio, 
such as $^{40}{\rm Ca}+^{58}{\rm Fe}$ and $^{40}{\rm Ar}+^{58}{\rm Ni}$. 
In ref.\ \cite{sherry3}, similar results are also shown for other isotope
ratios.

Moreover, data at very forward and backward angles
show that the isotope ratios do not simply depend on $(N/Z)_{\rm cs}$.
Instead, light fragments at backward angles are seen to have a much stronger
dependence on $(N/Z)_{\rm target}$, while at forward angles they depend
more on $(N/Z)_{\rm projectile}$. These results demonstrate that
the isospin degree of freedom in reactions at
$E_{\rm beam}/A=$ 45 and 53 MeV is not equilibrated during the time when 
fragments are emitted.
Therefore, a transition from isospin equilibration to non-equilibration 
is observed as the beam energy increases from below to above 
the Fermi energy.

The above observation has profound implications on the reaction mechanism 
leading to multifragmentation. It not only establishes the relative time scale
for multifragmentation in these reactions but also indicates that 
the assumption of isospin equilibrium taken for granted in various 
statistical models for nuclear multifragmentation at intermediate 
energies is not valid. Indeed, a statistical model study 
was made and failed to show any entrance channel effect \cite{sherry1}. 
Calculations using an intranuclear cascade code 
{\sc isabel} \cite{yariv} show that the N/Z ratio of the residue is very close 
to that of the initial combined system \cite{sherry1} and 
thus also fails	 to reproduce those features observed at 
$E_{\rm beam}/A=$ 45 and 53 MeV. 
Although the exact origin of this failure is not clear,
one expects that the reaction dynamics at these relatively low 
energies cannot be described by the nucleon-nucleon cascade alone, and 
should include also nuclear mean-field potential.
\begin{figure}[htp]
%\vspace{15cm}
\vspace{-3.0truecm}
\setlength{\epsfxsize=10truecm}
\centerline{\epsffile{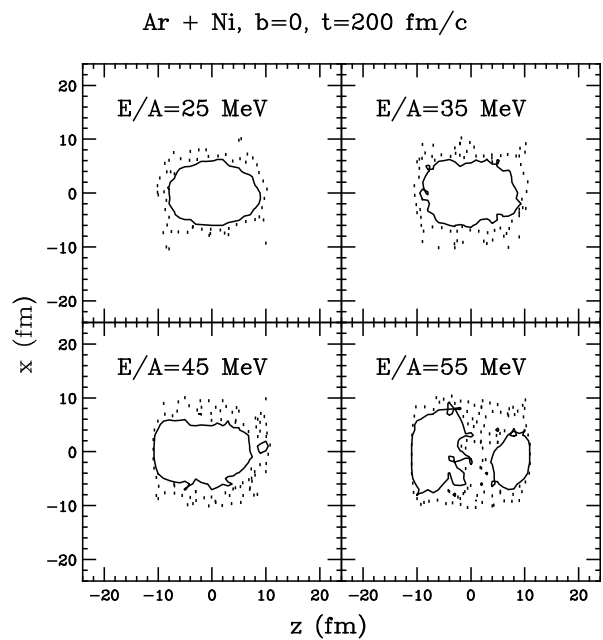}}
\caption{Density contours in the reaction plane at t= 200 fm/c
in head-on collisions of Ar+Ni at $E_{\rm beam}/A=$ 25, 35, 45 and 55 MeV.
Taken from Ref. \protect\cite{lis95}.}
\label{lish1} 
\end{figure} 

The experimental observation discussed above can be explained by
the isospin-dependent {\sc buu} model outlined in section \ref{buu}.
Calculations based on this model have been performed over a range of impact 
parameters.  For peripheral collisions it shows a memory of the initial
target and projectile, which is, however, gradually lost as the collisions 
become more central. A calculation at b=0 thus gives the most 
interesting test of any non-equilibrium effect. In Fig.\ \ref{lish1}, we
show the energy dependence of the density contours in the reaction plane 
at t= 200 fm/c in head-on collisions of Ar+Ni at $E_{\rm beam}/A=$ 25, 
35, 45 and 55 MeV.  The solid contours with $\rho=\rho_0/8$ correspond 
essentially to bound composite systems or heavy residues formed in 
the reactions. The dotted contours with $\rho=0.05~ \rho_0$ are 
free nucleons emitted mainly before thermal equilibrium is
reached.  The most interesting feature in Fig.\ \ref{lish1} is the 
formation of two heavy residues in the reaction at $E_{\rm beam}/A=55$ MeV. 
For reactions at $E_{\rm beam}/A=$ 25, 35 and 45 MeV,
these heavy residues do not break up even after 300 fm/c. 
\begin{figure}[htp]
%\vspace{15cm}
\vspace{0.0truecm}
\setlength{\epsfxsize=10truecm}
\centerline{\epsffile{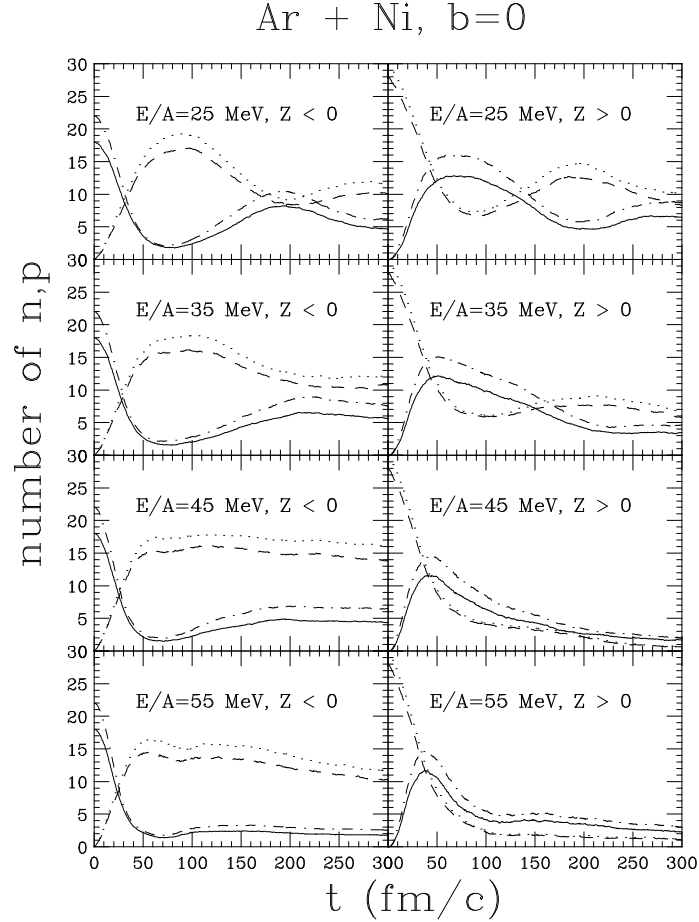}}
\caption{The neutron and proton numbers in the residues on the left 
($Z < 0$) and right ($Z\geq 0$) side of the origin.
Solid lines are the proton number from the projectile, 
while dot-dashed lines are the neutron number from 
the projectile which moves towards the right.
Dashed lines are the proton number from 
the target, while dotted lines are the neutron number from 
the target which moves towards the left.
Taken from Ref. \protect\cite{lis95}.}
\label{lish2} 
\end{figure} 

Studies on similar systems at much lower energies show
that the isospin degree of freedom is one of the fastest to 
equilibrate \cite{gatty}.
To see if the heavy residues observed above is also in isospin equilibrium,
we show in Fig.\ \ref{lish2} the neutron and proton numbers in the 
residues on the left 
($Z < 0$) and right ($Z\geq 0$) side of the origin in head-on 
reactions of Ar+Ni at $E_{\rm beam}/A=$25, 35, 45 and 55 MeV. 
The solid and dot-dashed lines are, respectively, the proton and 
neutron numbers from the projectile, which is incident from the left. 
The dashed and dotted lines are, respectively, the proton and neutron numbers
from the target, which moves from the right. It is seen 
that the neutron and proton numbers on both sides not only decrease
but also fluctuate.  The decreases is mainly due to nucleon-nucleon 
collisions and particle emissions, 
while the fluctuation is due to both the restoring force from the 
mean-field potential and nucleon-nucleon collisions. 
At $E_{\rm beam}/A=$ 25 MeV the neutron and proton numbers on the two sides 
become very close to each other and the amplitude of oscillation 
is rather small by the time of 300 fm/c. This indicates 
that the heavy residue is very close to isospin equilibrium, i.e., 
the proton and neutron distributions are independent of space-time.
The damping of the oscillation is faster at $E_{\rm beam}/A=35$ MeV so the
particle distribution reaches isospin equilibrium also sooner. The isotopic 
composition of fragments emitted from the residues in these low energy 
reactions after about 300 fm/c would therefore essentially reflect the 
$(N/Z)_{\rm cs}$ ratio of the initial composite system, and there is little
forward-backward asymmetry. These features are in good agreement with
those found in the data of $E_{\rm beam}/A=$25 and 35 
MeV \cite{sherry2,sherry3,sherry4}.

At higher energies, such as $E_{\rm beam}/A=$ 45 and 55 MeV, 
there is little oscillation in the overlapping region between the
target and projectile. This is mainly because 
the incoming momenta of projectile-nucleons and target-nucleons 
are very large so the mean-field potential 
cannot reverse the directions of motion of many nucleons during a relatively 
short reaction time. As a result, there exists a large isospin asymmetry or 
non-equilibration at these two energies.
In particular, on the left side of the origin the N/Z ratio of the residue 
is more affected by that of the target while on the right side it is more
affected by that of the projectile. However, the
N/Z ratios on both sides are not simply those of the target and projectile
but a combination of the two depending on the complicated reaction dynamics.
In the case of $E_{\rm beam}/A=$ 55 MeV, at the time of about 200 fm/c 
the heavy residue has broken up into two pieces with some longitudinal
collectivity. The forward moving residue has 
an average N/Z ratio of about 6/5 and an excitation energy of 
about 8.6 MeV/nucleon, while the backward moving residue has an 
average N/Z ratio of about 16.5/14 and an excitation energy of 
about 6.8 MeV/nucleon. Both residues are thus in approximate
thermal equilibrium in their own center of mass frame \cite{lis95}
but not in thermal equilibrium with each other.

\begin{figure}[htp]
%\vspace{14cm}
\vspace{-5.0truecm}
\setlength{\epsfxsize=10truecm}
\centerline{\epsffile{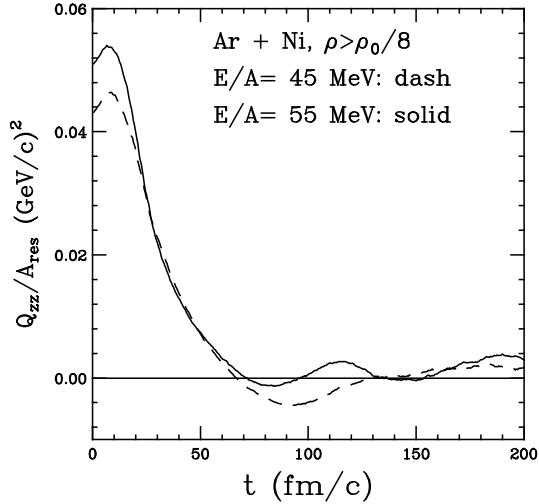}}
\caption{The scaled quadrupole moment of heavy residues formed 
in head-on collisions of Ar+Ni at $E_{\rm beam}/A=$ 45 and 55 MeV.
Taken from Ref. \protect\cite{lis95}.}
\label{qzz} 
\end{figure}

The signature of nuclear transparency from studying the isotopic
ratios is consistent with that from studying the kinematic 
observables, such as the scaled quadrupole moment $Q_{ZZ}/A_{\rm res}$ 
defined as 
\begin{equation}
Q_{zz}/A_{\rm res}(t)=1/A_{\rm res}\int\frac{d\vec{r}d\vec{p}}
{(2\pi)^3}(2p^2_z-p^2_x-p^2_y)f(\vec{r},\vec{p},t),
\end{equation}
where $f(\vec{r},\vec{p},t)$ is the Wigner function from the {\sc buu} model 
calculations and $A_{\rm res}$ is the mass of the residue. 
A zero value of $Q_{zz}$ is a necessary but not a sufficient condition for 
thermal equilibrium, and a positive value of $Q_{ZZ}$ signals transparency.
The quadrupole moments of heavy residues formed in the reactions 
at $E_{beam}/A=$ 25 and 35 MeV are about zero by the time 
of about 300 fm/c, indicating the establishment of thermal equilibrium. 
Since the heavy residue formed in the reaction at $E_{\rm beam}/A=$ 
55 MeV has already broken up into two pieces at about 200 fm/c, it is 
interesting to examine more closely the evolution of $Q_{zz}/A_{\rm res}$ 
in this reaction, which is shown in Fig.\ \ref{qzz} 
for particles at local densities larger than $\rho_0/8$ 
for the reactions at E/A= 45 and 55 MeV.
It is seen that $Q_{zz}$ has a small positive value when the system breaks 
up, indicating thus a transparency in the collisions. This finding is also 
in good agreement with the kinematic signature of transparency found recently 
at Ganil for the same reaction system at similar beam energies using the 
{\sc indra}
detector \cite{borderie}. We note that in recent
calculations using the Antisymmetrized Molecular Dynamics ({\sc amd}) 
transparency 
was found at 35 MeV/nucleon for head-on collisions of 
$^{40}{\rm Ca}+^{40}{\rm Ca}$ \cite{amd}. Moreover, the signatures 
of transparency 
from both the isotopic ratios and the quadrupole moments are also 
in good agreement with earlier {\sc buu} predictions on the systematics 
of nuclear stopping power \cite{bauer88}. 
  
\begin{table}
\caption{Values of N-Z for the quasitarget(QT) and quasiprojectile (QP)}
\label{tableofc}
\medskip
\centerline{
\begin{tabular}{ccccccccccccccc}
\hline\\
%\hline\hline\\
\multicolumn{1}{c}{Reaction/N-Z} &\multicolumn{1}{c}{c=0}
&\multicolumn{1}{c}{c=20}&\multicolumn{1}{c}{c=28}\\\\
\hline\\
\multicolumn{1}{c}{$^{56}{\rm Ca}+^{40}{\rm Ca}$~~QT} &\multicolumn{1}{c}{3.74}
&\multicolumn{1}{c}{4.64}&\multicolumn{1}{c}{5.02}\\\\
\multicolumn{1}{c}{~~~~~~~~~~~~~~~~~~~QP} &\multicolumn{1}{c}{8.76}
&\multicolumn{1}{c}{5.82}&\multicolumn{1}{c}{4.92}\\\\
\multicolumn{1}{c}{$^{48}{\rm Ca}+^{48}{\rm Ca}$~~QT} &\multicolumn{1}{c}{6.52}
&\multicolumn{1}{c}{5.78}&\multicolumn{1}{c}{5.52}\\\\
\multicolumn{1}{c}{~~~~~~~~~~~~~~~~~~~QP} &\multicolumn{1}{c}{6.96}
&\multicolumn{1}{c}{5.84}&\multicolumn{1}{c}{5.12}\\\\
\hline\\
%\hline\hline
\end{tabular}
}
\end{table}

Effects of the symmetry potential on isospin equilibrium have 
been studied at lower beam energies where isospin equilibrium is usually 
reached. Within the {\sc lv} model using the simple symmetry potential 
energy density of Eq.\ (\ref{simple}) \cite{far91}, it has been found that 
a stronger symmetry potential enhances the degree of 
isospin equilibrium. Shown in Table \ref{tableofc} are the values of 
$N-Z$ in the quasitarget (QT) and quasiprojectile (QP) formed in the 
reaction of $^{56}{\rm Ca}+^{40}{\rm Ca}$ at an impact parameter 
of 7 fm and a beam energy of 15 MeV/nucleon. 
For comparisons, results from reactions of a symmetric system
$^{40}{\rm Ca}+^{48}{\rm Ca}$ 
are also listed in the table. In both reactions the total mass and 
charge numbers are the same. The difference in the values of $N-Z$ for the
symmetric system is completely due to the numerical fluctuations of the 
calculations. In the asymmetric reaction the initial value of $N-Z$ is 
16 and 0 for the projectile and target, respectively. It is seen that 
significant stopping is achieved even in the case of no symmetry 
potential. However, without the symmetry potential isospin equilibrium 
cannot be reached, and the degree of isospin equilibrium increases with
the strength of the symmetry potential.

In summary, isospin equilibration in heavy-ion collisions can be used to 
study the nature of stopping. The degree of isospin equilibrium 
depends sensitively on both the in-medium nucleon-nucleon cross section and the
{\sc eos} of asymmetric nuclear matter. In both experiments and {\sc buu} model 
calculations a transition from isospin equilibrium at lower energies 
to nonequilibrium at higher energies has been observed. The isospin 
nonequilibrium signals the onset of nuclear transparency. 
For head-on collisions of the system $A_{\rm pro}+A_{\rm tar}\approx 
40+58$ the transparency starts at a beam energy as low as 45 MeV/nucleon. 

\section{Isospin dependence of nuclear collective flow}
Nuclear collective flow in heavy-ion collisions at intermediate energies 
has been a subject of intensive theoretical and experimental studies 
during the last decade; for a general introduction and overview 
see Ref. \cite{gary}. The study of the dependence of 
collective flow on entrance channel parameters, such as the beam energy, 
mass number and impact parameter, have revealed much interesting physics about 
the properties and origin of collective flow. In particular, 
from studying the beam energy dependence it has been found that the 
transverse collective flow changes from a negative one to a positive one at an 
energy $E_{\rm bal}$, defined as the balance energy, due to the competition 
between the attractive nuclear mean-field potential and the repulsive
nucleon-nucleon collisions \cite{exp1,exp2,exp3,exp4,exp5,exp6,exp7,exp8,exp9}. 
The balance energy has been found to depend sensitively on the mass number, 
impact parameter and properties of the colliding nuclei, such as the 
thickness of their surfaces \cite{kla93}. Furthermore, detailed theoretical
studies using transport models (for a review see 
e.g. \cite{greiner,bert87,bauer92}) have shown that both the strength 
of transverse flow and the balance energy can be used to extract 
information about the nuclear equation of state and in-medium 
nucleon-nucleon cross sections 
\cite{gale87,zhang,betty,xu,li93,dani85,moli1,moli2,bert87,dani89,vd,pan}.
\begin{figure}[htp]
%\vspace{10cm}
\vspace{-5.0truecm}
\setlength{\epsfxsize=10truecm}
\centerline{\epsffile{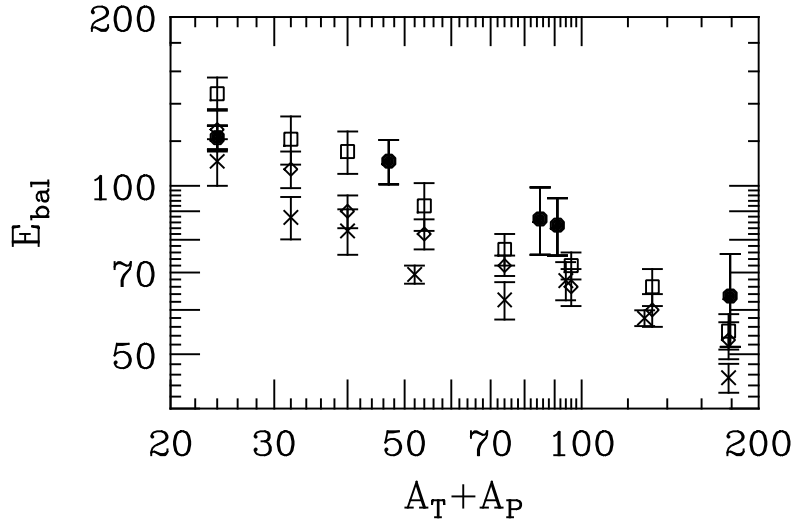}}
\caption{The balance energy as a function of the combined mass. The
experimental data are compared with {\sc buu} 
calculations using a soft 
 {\sc eos} and
a reduction of in-medium nucleon-nucleon cross section by 0\% (crosses),
10\% (diamonds) and 20\% (squares) are compared with the data (filled circles). 
Taken from Ref. \protect\cite{wes93}.}
\label{westfall} 
\end{figure} 

Shown in Fig.\ \ref{westfall} is the mass dependence of the balance 
energy measured by Westfall {\it et al.} in comparison with the predictions of 
Klakow {\it et al.} using the {\sc buu} model \cite{kla93}. It is seen that 
{\sc buu}
calculations reproduce the measured scaling of balance energy 
with mass, but underpredict the values of balance energy. Using an in-medium 
cross section of Eq.\ (\ref{msigma}) with a parameter of $\alpha=-0.2$ the 
{\sc buu}
calculations can better reproduce the data. This results thus indicate 
that the collective flow and balance energy depend strongly on the in-medium 
nucleon-nucleon cross section.

Nuclear transverse flow is expected to be isospin-dependent for several 
reasons. First, it is well-known that nucleon-nucleon 
collisions cause repulsive collective flow, and this effect is roughly
proportional to the number of collisions in the overlapping region. 
While the number of particles in this region might not be so strongly 
isospin-dependent, the number of collisions during the reaction of two 
neutron-rich nuclei is smaller. This is due to both the smaller 
neutron-neutron cross sections (see Chapter \ref{xnn}) and the larger Pauli 
blocking of neutron-neutron collisions in these reactions.
This effect is stronger in peripheral collisions 
of neutron-rich nuclei where two thick neutron skins overlap during 
the reaction. Second, the Coulomb potential also causes repulsive 
scatterings, and this effect is obviously weaker in the neutron-richer system. 
Third, the isospin-independent part of the nuclear mean-field potential
is attractive at densities up to about $2\rho_0$ and $4\rho_0$, 
respectively, for the stiff ($K=380$ MeV) and soft ($K=200$ MeV)  {\sc eos}. 
Since the mean field effect is approximately proportional 
to the total surface area of the colliding system \cite{kla93},
collisions of neutron-rich nuclei are expected to be influenced by stronger
attractive mean-field potentials due to their more extended surfaces.
Finally, the generally repulsive symmetry potential is stronger for 
collisions of neutron-rich nuclei. However, since the magnitude of the symmetry 
potential is smaller than the isospin-independent mean-field potential, 
effects of the
symmetry potential on the collective flow are expected to be small. 
Although a more quantitative study on the relative importance of the above
mechanisms remains to be worked out, one expects that the overall isospin 
effect is that the neutron-richer system will have a stronger negative flow 
below the balance energy and a weaker positive flow at higher energies.   
Moreover, the relative importance of different contributions to the collective 
flow depends on the beam energy. As the beam energy increases the 
repulsive nucleon-nucleon collisions become dominant and effects 
of the neutron skin is thus less important. Also, the isospin dependence 
of the nucleon-nucleon cross sections becomes weaker at higher 
energies as discussed in Chapter \ref{xnn}. It is therefore 
understandable that the isospin effects on the collective 
flow decrease as the beam energy increases.
These discussions thus show that 
both the strength of flow and the balance energy are 
expected to be isospin-dependent. 

\begin{figure}[htp]
%\vspace{15cm}
\vspace{-1.0truecm}
\setlength{\epsfxsize=10truecm}
\centerline{\epsffile{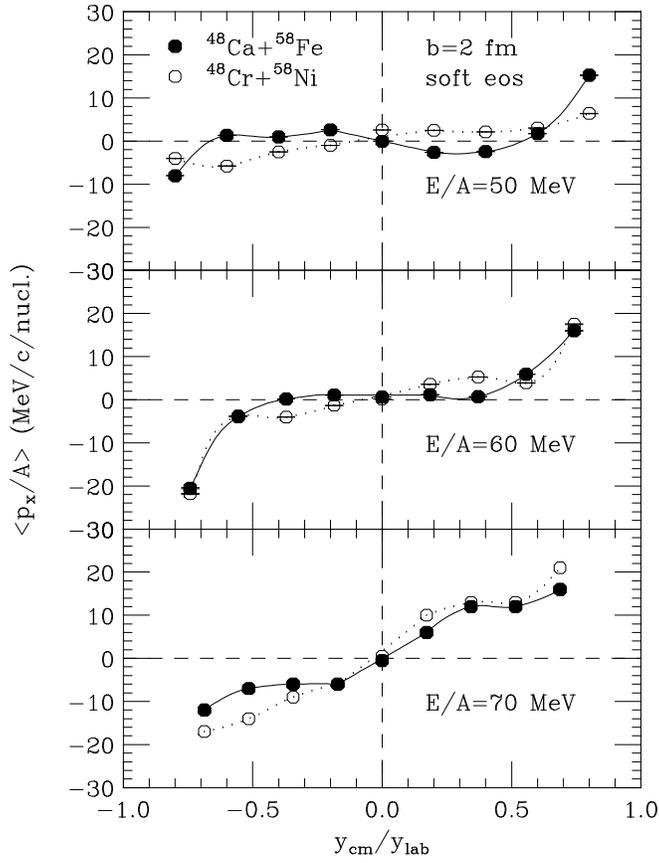}}
\caption{The transverse momentum distribution in the reactions of
$^{48}{\rm Cr}+^{58}{\rm Ni}$ and $^{48}{\rm Ca}+^{58}{\rm Fe}$ at an impact 
parameter of 2 fm 
and beam energies of 50, 60 and 70 MeV/nucleon, respectively. 
Taken from Ref. \protect\cite{li96}.}
\label{flow1} 
\end{figure}

The isospin dependence of collective flow and balance energy has recently been
studied theoretically with isospin-dependent {\sc buu}
models \cite{li96,dani96} 
and experimentally by Pak {\it et al.} at {\sc nscl/msu} \cite{pak1,pak2}.  
Within the isospin-dependent {\sc buu} model the standard transverse 
momentum analysis \cite{dani85} (see also \cite{gary}) are carried out
for two reaction systems, $^{48}{\rm Cr}+^{58}{\rm Ni}$ and 
$^{48}{\rm Ca}+^{58}{\rm Fe}$, 
which have the same mass number of $48+58$ but different neutron/proton ratios 
of 1.04 and 1.30, respectively. Typical results for central 
collisions at an impact parameter of 2 fm 
and beam energies of 50, 60 and 70 MeV/nucleon are shown in 
Fig.\ \ref{flow1}. At a beam energy of 50 MeV/nucleon, 
the transverse flow in the reaction of $^{48}{\rm Ca}+^{58}{\rm Fe}$ is 
still negative 
while that in the reaction of $^{48}{\rm Cr}+^{58}{\rm Ni}$ becomes already
positive. The difference disappears at beam energies above 70 MeV/nucleon.
\begin{figure}[htp]
%\vspace{10cm}
\vspace{-7.0truecm}
\setlength{\epsfxsize=12truecm}
\centerline{\epsffile{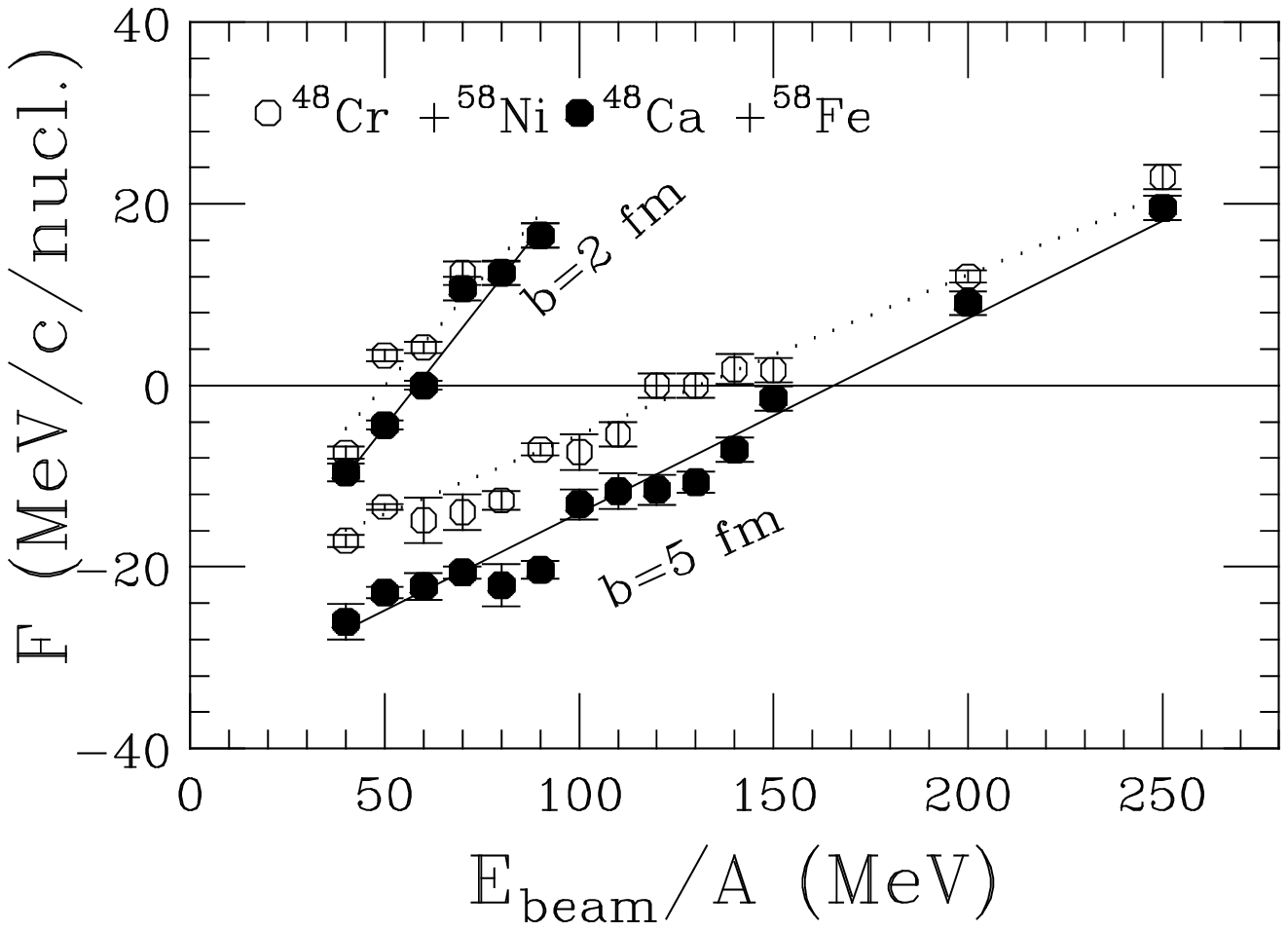}}
\caption{Flow parameters for the reactions of
$^{48}{\rm Cr}+^{58}{\rm Ni}$ and $^{48}{\rm Ca}+^{58}{\rm Fe}$ at 
impact parameters of 2 and 5 fm. 
Taken from Ref. \protect\cite{li96}.}
\label{flow2} 
\end{figure} 

More quantitatively, one can compare the flow parameter $F$ defined as the
slope of the transverse momentum distribution at the center of
mass rapidity $y_{\rm cm}$. The beam energy dependence of the flow parameter 
for the two reaction systems at impact parameters of 2 and 5 fm 
are shown in Fig.\ \ref{flow2}. The lines are the least-square 
fit to calculations using linear functions, i.e., 
$F({\rm Ca+Fe})=-32.2+0.55\,E/A$ 
and $F({\rm Cr+Ni})=-23.9+0.48\,E/A$ at $b=2$ fm; and $F({\rm Ca+Fe})=
-35.9+0.22\,E/A$ and $F({\rm Cr+Ni})=-23.2+0.18\,E/A$ at $b=5$ fm. 
It is seen that 
in both central and peripheral collisions the neutron-rich system 
$^{48}{\rm Ca}+^{58}{\rm Fe}$ shows systematically smaller 
flow parameters, indicating thus a stronger attractive interaction during the 
reaction. The effect is more appreciable in peripheral collisions -- as 
expected. Consequently, the balance energy in $^{48}{\rm Ca}+^{58}{\rm Fe}$ 
reaction is higher than that in the reaction of $^{48}{\rm Cr}+^{58}{\rm Ni}$ 
by about 10 to 30 MeV/nucleon. The difference between flow parameters in 
the two systems 
decreases as the beam energy increases and finally disappears as the 
beam energy rises far above the balance energy. 
\begin{figure}[htp]
%\vspace{15cm}
\vspace{-3.0truecm}
\setlength{\epsfxsize=10truecm}
\centerline{\epsffile{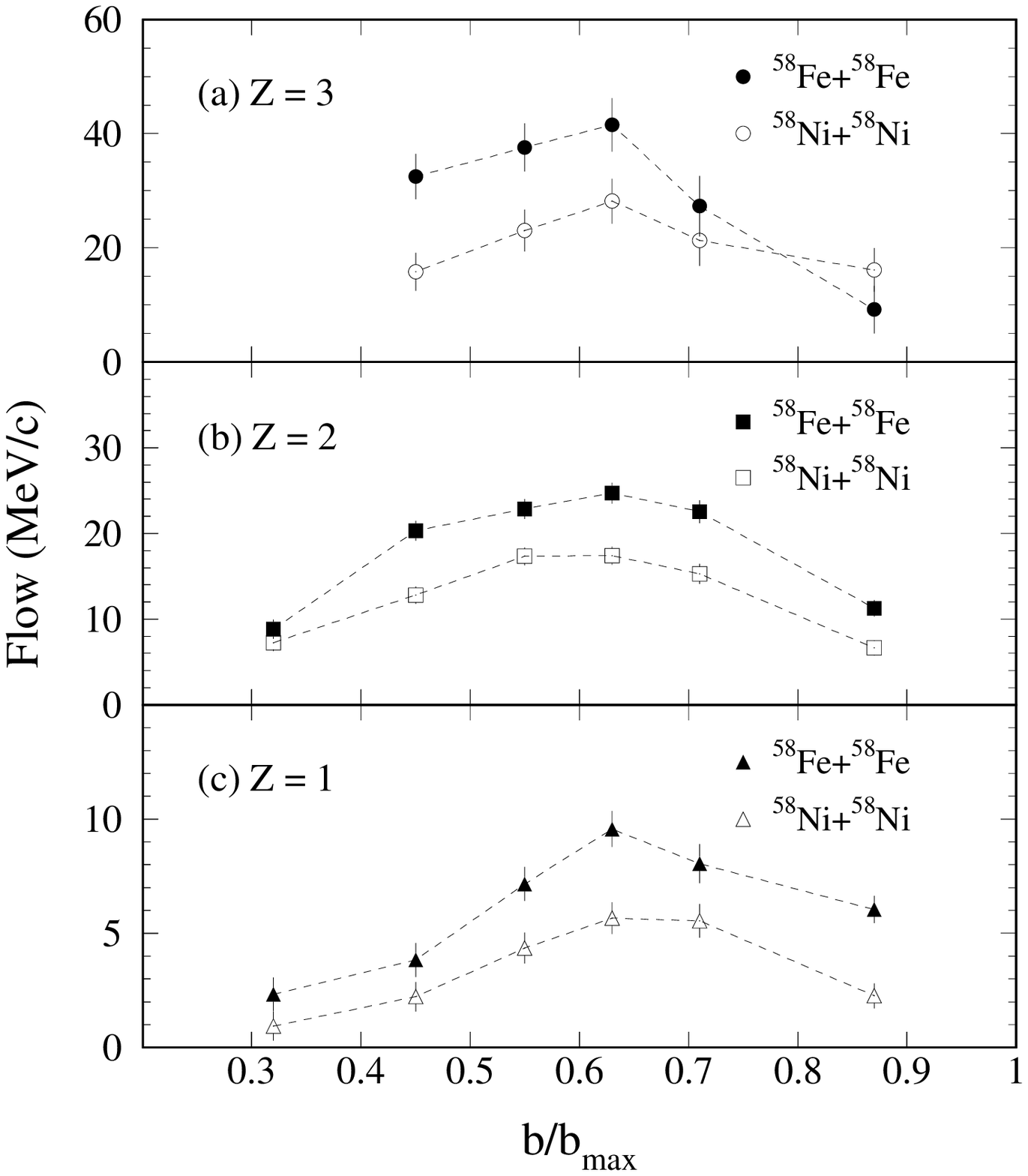}}
\caption{Flow parameters for the reactions of
$^{58}{\rm Fe}+^{58}{\rm Fe}$ and $^{58}{\rm Ni}+^{58}{\rm Ni}$ as a 
function of the reduced 
impact parameter at a beam energy of 55 MeV/nucleon.
Taken from Ref. \protect\cite{pak1}.}
\label{data1} 
\end{figure} 

\begin{figure}[htp]
%\vspace{14cm}
\vspace{-3.0truecm}
\setlength{\epsfxsize=10truecm}
\centerline{\epsffile{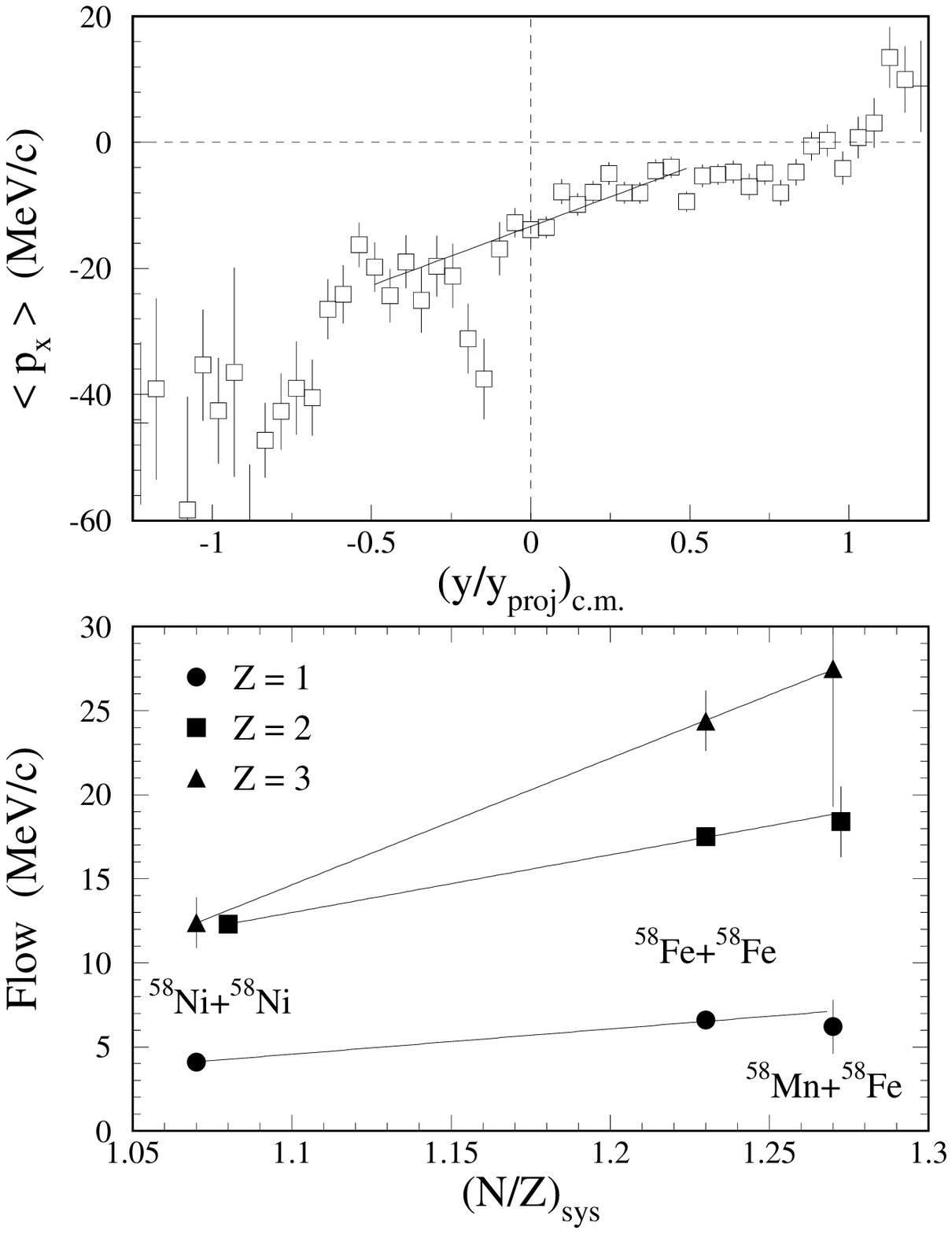}}
\caption{(Upper window) Mean transverse momentum in the reaction
plane versus the reduced c.m. rapidity for Z=2 fragments from 
impact-parameter-inclusive $^{58}+^{58}Fe$ collisions at 55 MeV/nucleon.
(Lower window) Isospin dependence of the flow parameter 
for inclusive collisions at a beam energy of 55 MeV/nucleon.
Taken from Ref. \protect\cite{pak1}.}
\label{data2} 
\end{figure} 

The experimental investigation of the isospin dependence of transverse
collective flow has been recently carried out by Pak et al. {\it et al.} 
at {\sc nscl/msu}. Since one cannot identify the direction of
flow using the transverse momentum analysis in experiments, 
the absolute value of flow parameter is usually extracted. 
Shown in Fig.\ \ref{data1} are the flow parameters of particles with 
charge $Z=1$, $Z=2$ and $Z=3$ 
as functions of reduced impact parameter $b/b_{\rm max}$ for the
reactions of $^{58}{\rm Fe}+^{58}{\rm Fe}$ and $^{58}{\rm Ni}+^{58}{\rm Ni}$ 
at a beam energy of 55 MeV/nucleon. Note that at this beam energy 
flow is still dominated by the attractive mean-field potential and is 
thus negative -- if the same convention as in the {\sc buu} calculations 
is used. It is seen that the flow parameter for the neutron-richer 
system is consistently higher and is in agreement with the {\sc buu} 
predictions. Pak et al. have also studied the flow parameter as a function
of the isotope ratio of the composite projectile plus target system for three
different fragment types from three isotopic entrance channels. Shown in the
upper window of Fig.\ \ref{data2} is the mean transverse momentum in the 
reaction plane versus the reduced c.m. rapidity for Z=2 fragments from 
impact-parameter-inclusive $^{58}Mn+^{58}Fe$ collisions at 55 MeV/nucleon.  
The flow parameter extracted for the impact-parameter-inclusive events 
is plotted in the lower window of Fig.\ \ref{data2} as a function of the ratio 
of neutrons to protons of the combined 
system $(N/Z)_{\rm cs}$. 
The flow parameter increases linearly with the
ratio $(N/Z)_{\rm cs}$ for all three types of particles. 
\begin{figure}[htp]
%\vspace{15cm}
\vspace{-1.0truecm}
\setlength{\epsfxsize=10truecm}
\centerline{\epsffile{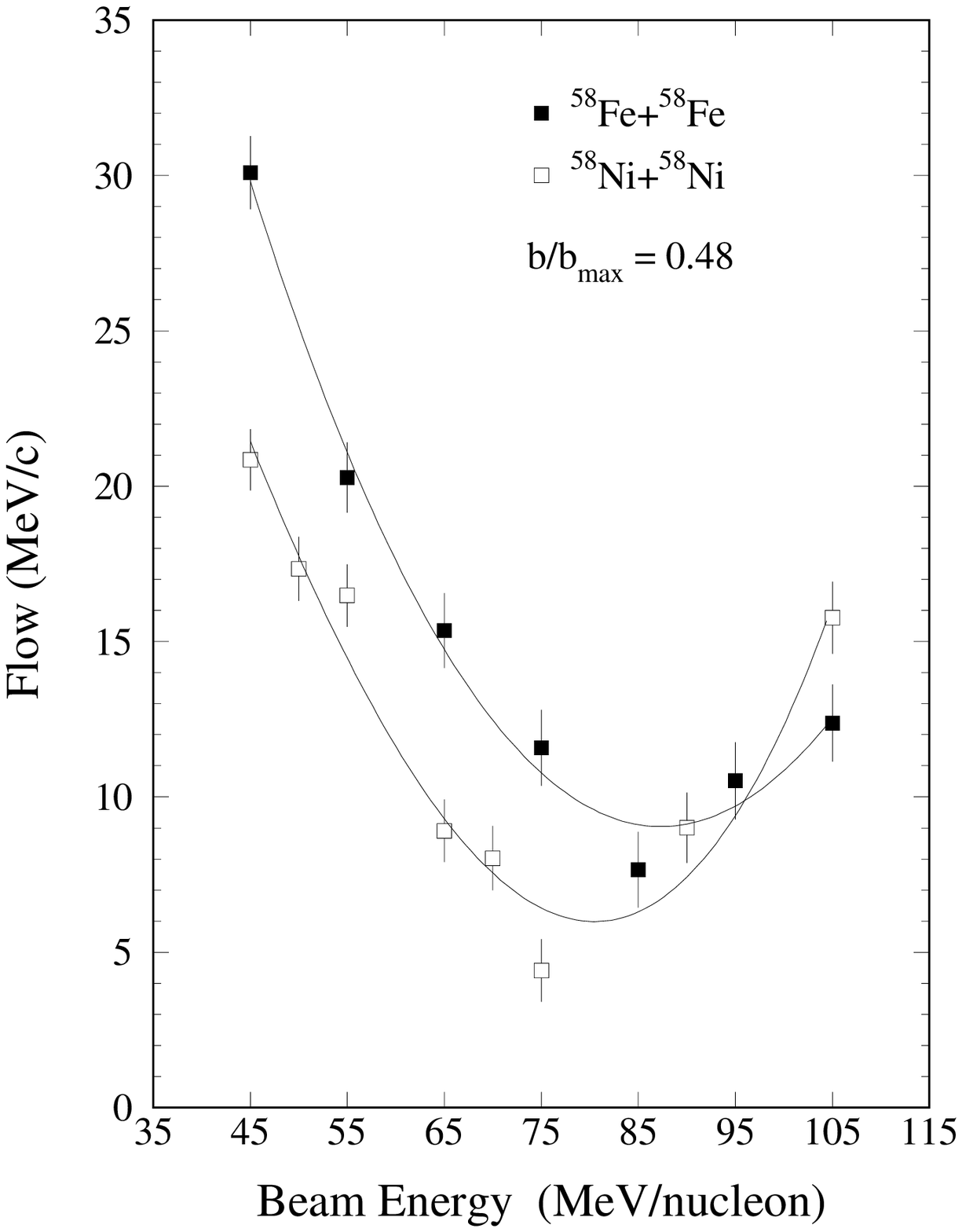}}
\caption{Energy dependence of the flow parameter 
for midcentral collisions of $^{58}{\rm Fe}+^{58}{\rm Fe}$ and 
$^{58}{\rm Ni}+^{58}{\rm Ni}$, 
respectively. Taken from Ref. \protect\cite{pak2}.}
\label{data3} 
\end{figure} 
Fig.\ \ref{data3} shows the measured energy dependence of the flow parameter
for midcentral collisions. The curves are drawn to guide the eyes. 
In the whole energy range the flow parameter shows dependence
on the isospin asymmetry of the colliding system. The minima of the curves 
correspond to the balance energies where the flow changes sign. 
Because of the isospin dependence of nucleon-nucleon cross sections, 
there are less number of nucleon-nucleon collisions in the neutron-rich system. 
Below the balance energy attractive mean-field potentials
dominate.  There fewer collisions mean less kinetic pressure and 
help generate more attractive flow in the 
neutron-rich system. Above the balance energy, where the repulsive
nucleon-nucleon collisions dominate, less collisions in the neutron-rich system
results in a smaller flow parameter. For both systems 
studied in the experiment the balance energy occurs at 
about $E_{\rm beam}/A=90$ MeV. These features in the measured energy 
dependence are also in quantitative agreement 
with those predicted in Fig.\ \ref{flow2} using the {\sc buu} model.

\begin{figure}[htp]
%\vspace{14cm}
\vspace{-2.0truecm}
\setlength{\epsfxsize=10truecm}
\centerline{\epsffile{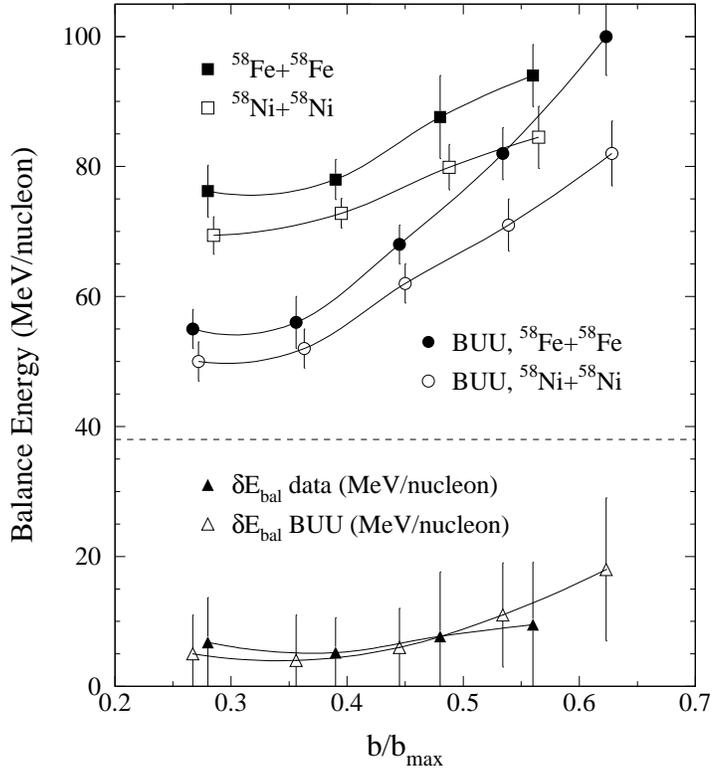}}
\caption{Upper window: comparison of the measured and calculated 
balance energies for collisions of $^{58}{\rm Fe}+^{58}{\rm Fe}$ and 
$^{58}{\rm Ni}+^{58}{\rm Ni}$ as functions of impact parameter. 
Lower window: comparison of 
the difference in balance energies of the two systems as a function of 
impact parameter. Taken from Ref. \protect\cite{pak2}.}
\label{data4} 
\end{figure}

The comparison of theoretical predictions on collective flow with
the experimental data is shown in Fig.\ \ref{data4}. In the upper window 
the measured and predicted balance energies are compared for both systems, 
and in the lower window the measured and predicted 
differences in balance energies of the two systems are
compared. One observes that the difference of balance energies, i.e., the 
$\delta E_{\rm bal}$, is well reproduced. The model, however, underpredicts
the balance energy in central collisions and overpredicts the data in
peripheral collisions. These discrepancies can be
understood as follows \cite{wes93,hun96,pak3}. We mentioned 
earlier in this section that in central collisions 
nucleon-nucleon collisions are important, and a reduction of nucleon-nucleon 
cross sections by about 20\% was needed to reproduce the measured
balance energy -- see Fig.\ \ref{westfall}.  Once one includes the
isospin dependence, we expect a different reduction of the in-medium 
nucleon-nucleon cross sections is needed to reproduce the data.  This 
remains to be studied. In the case of peripheral 
collisions, the mean-field potential has a more important effect on flow
than nucleon-nucleon collisions. In particular, the repulsive 
momentum-dependent mean-field potential \cite{gale87,gale90,csernai92,fai}, 
which is not included in the calculations presented here, is expected to 
reduce the balance energy in heavy ion collisions at large impact 
parameters \cite{pak3}.

It is encouraging that both model calculations and experiments show
that the transverse flow is sensitive to the isospin asymmetry of the
colliding systems.  
Studies using the isospin-dependent {\sc buu} model indicate, however,
that effects of the symmetry potential on the transverse flow 
and balance energy are not appreciable, because they are 
small compared to those of
the isospin-{\em in}dependent part of the nuclear  {\sc eos}. 
By neglecting the Coulomb interaction, the balance energies for both 
systems increase, but the difference remains about the same 
in central collisions and becomes larger in peripheral
collisions. Thus, the observed isospin dependence of the transverse flow and 
balance energy is mainly due to the isospin-dependent in-medium 
nucleon-nucleon cross sections. The isospin-dependent {\sc buu} 
model reproduces
the measured isospin dependence of the balance energy, but fails to reproduce
the magnitude of the balance energy. It thus indicates the need of a 
reduced in-medium nucleon-nucleon cross section and a momentum-dependent 
mean-field potential. More detailed work in refining the comparison between 
model calculations and the data remains to be done to
infer useful information about the in-medium nucleon-nucleon 
cross sections.   

\section{Isospin dependence of total reaction cross sections and radii of 
neutron-rich nuclei}
\begin{figure}[htp]
%Fig. 4.27
%\vspace{8cm}
\vspace{-5.0truecm}
\setlength{\epsfxsize=10truecm}
\centerline{\epsffile{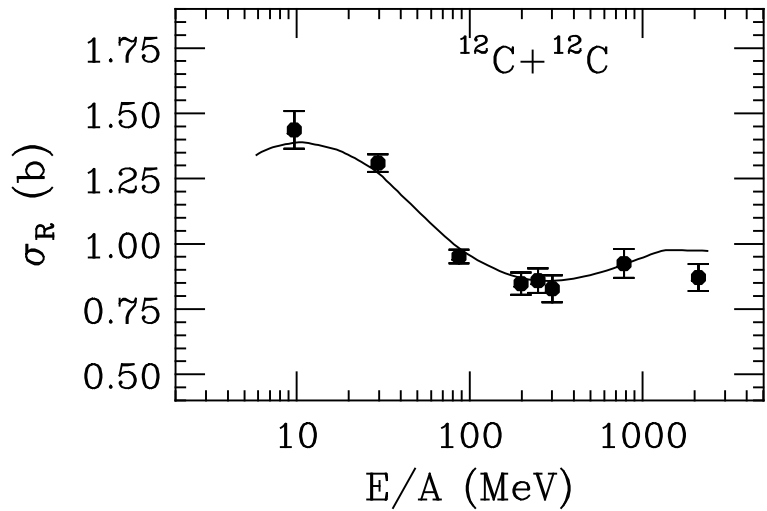}}
\caption{The energy dependence of the total reaction cross section for the
reaction of $^{12}{\rm C}+^{12}{\rm C}$. Taken from Ref. \protect\cite{kox87}.}
\label{cc12} 
\end{figure}

We discuss, in this section, the isospin dependence of total reaction cross
sections of neutron-rich nuclei.  The total reaction 
cross section provides direct information about the size of nuclei involved 
in the reaction, and has been one of the standard
techniques for studying the size and structure of radioactive 
nuclei \cite{tanihata95}. Theoretical analyses of the measured 
reaction cross sections have been done almost exclusively by using the 
Glauber model or eikonal approximation with 
various modifications. Shown in Fig.\ \ref{cc12} is the total reaction cross 
section for $^{12}{\rm C+}^{12}{\rm C}$ reactions at beam energies from about 
10 to 2000 MeV/nucleon \cite{kox87}. The model works surprisingly well 
in the whole energy range, although it should not be an appropriate model
at low energies.  Indeed, 
recent Glauber model calculations underpredict significantly the interaction 
cross sections for $^{11}{\rm Li+C}$ reactions at beam energies below 
about 100 MeV/nucleon \cite{yab92,oga92}. 
Furthermore, one expects that both 
the nuclear  {\sc eos} and the in-medium nucleon-nucleon cross sections, which
one has learned from intermediate energy heavy-ion collisions, should have 
effects on the total reaction cross section.  In the following,
we shall first review the
observed isospin dependence of total reaction cross sections, and then 
comment on the feasibility of applying isospin-dependent transport models 
to study the total reaction cross section. As we shall show that results 
from these studies have already demonstrated the possibility of 
learning the nuclear  {\sc eos}, in-medium nucleon-nucleon cross section, and 
the size of nuclei from the total reaction cross sections.
  
Several empirical formulae for describing the systematics of 
the total reaction cross sections have been developed based on 
the Glauber model \cite{kox87,kar75,shen89}. For example, 
Kox's parameterization for the total reaction cross section can
be written as \cite{kox87} 
\begin{equation}
\sigma_R(E)=\pi r_0^2 f(E),
\end{equation}
where $f(E)$ is an empirical expression describing the degree of transparency,
and depends on the masses of the colliding nuclei and the beam energy.
The strong-absorption radius parameter $r_0$ contains information about 
the size of both the target and projectile. A constant value of 
$r_0\approx 1.05$ fm independent of the beam energy and 
projectile-target combination has been obtained by analyzing a large 
amount of $\sigma_R(E)$ data for collisions between $\beta$-stable 
nuclei at energies from 33 to 2100 MeV/nucleon. Using the same 
parameterization for reactions involving neutron-rich nuclei 
$r_0^2$ has been found to increase with increasing neutron excess 
in the reaction system \cite{tanihata95,shen89,mit87}. 
\begin{figure}[htp]
%\vspace{8cm}
\vspace{-5.0truecm}
\setlength{\epsfxsize=10truecm}
\centerline{\epsffile{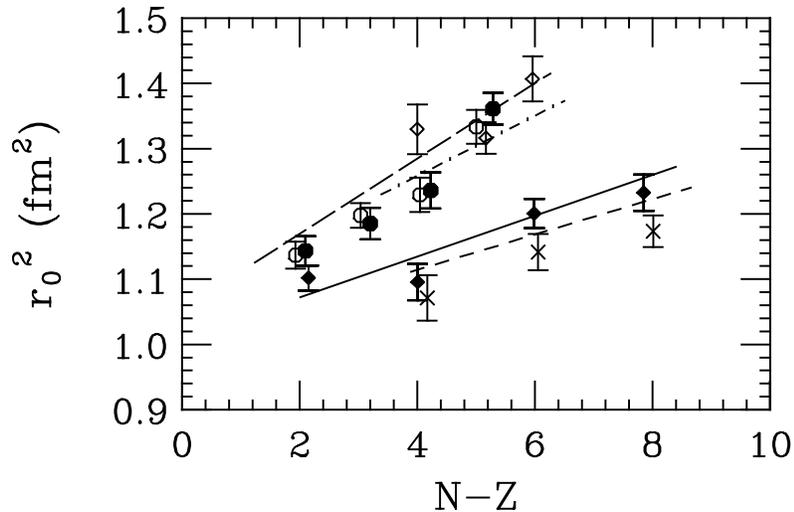}}
\caption{Square of the radius parameter of the interacting nuclei $r_0^2$ 
as a function of $N-Z$ of the reaction system. 
The open circles, filled diamonds, open diamonds, crosses and 
filled circles are for $^{16-18}{\rm N}+^{28}{\rm Si}$, $^{26-28}{\rm Na}
+^{28}{\rm Si}$, $^{12}{\rm C}+^{62-68}{\rm Zn}$ and $^{18-21}{\rm O}
+^{28}{\rm Si}$, respectively. Taken from Ref. \protect\cite{shen89}.}
\label{r02} 
\end{figure} 
Shown in Fig.\ \ref{r02} are the extracted values of $r_0^2$ as a function of 
the neutron excess $N-Z$ of the reaction system \cite{mit87}. 
The linear dependence of $r_0^2$ on $N-Z$ can be explained using an extended
Glauber model where nucleons are assumed to have a Fermi-type distribution 
with the following isospin-dependent surface diffuseness for 
neutrons \cite{shen89}
\begin{equation}
t_n=2.4+10\cdot\left(\delta-\delta_{\beta}\right)~({\rm fm}),
\end{equation}
where $\delta_{\beta}\equiv \left((N-Z)/A\right)_{\beta}$ is the relative 
neutron excess along the $\beta$ stability line.
\begin{figure}[htp]
% Fig. 4.29
%\vspace{10cm}
\vspace{-5.0truecm}
\setlength{\epsfxsize=10truecm}
\centerline{\epsffile{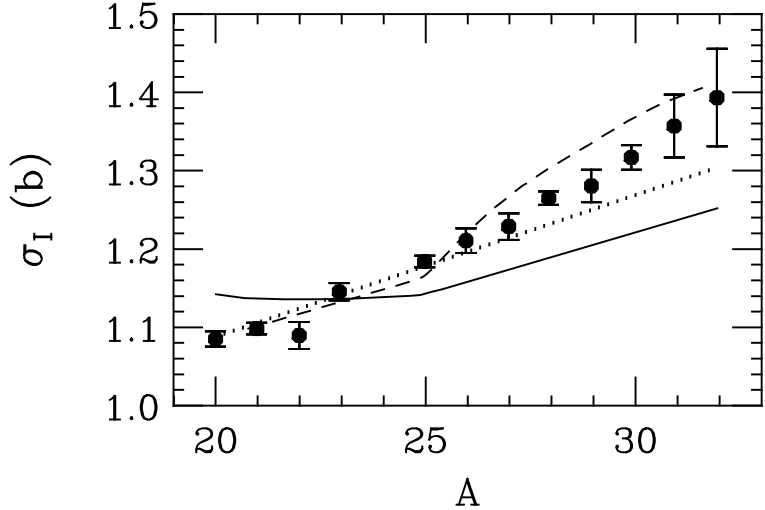}}
\caption{Interaction cross section for Na isotopes on carbon at a beam
energy of 950 MeV/nucleon. Taken from Ref. \protect\cite{suz95}.}
\label{na} 
\end{figure} 

More recently, Suzuki {\it et al.} have measured the interaction cross section 
of Na isotopes ($^{20-23, 25-32}{\rm Na}$) on a carbon target 
at a beam energy of 950 MeV/nucleon \cite{suz95}. 
In Fig.\ \ref{na} the measured cross section (filled circles) is plotted 
as a function of mass number of the Na isotopes. 
The dotted line is obtained from the expression
\begin{equation}
\sigma_I=\pi \left(R_I(C)+r_0A^{1/3}\right)^2,
\end{equation}
where $R_I(C)=2.61$ fm is the interaction radius of $^{12}{\rm C}$ and $r_0$ 
is chosen to reproduce the $^{23}{\rm Na}+^{12}{\rm C}$ data. It is 
seen that as the projectile becomes more neutron-rich this simple scaling
starts to deviate from the data.
This result is consistent with that observed in Refs.\ \cite{shen89,mit87}, 
although different formulae have been used there to describe
the reaction cross section. The results from the Glauber model are 
also shown in the figure. The solid line is obtained by assuming that 
the density 
distribution for neutrons is the same as that for protons, except a 
normalization factor (N/Z). With this assumption the model underpredicts 
the cross section for neutron-rich nuclei but overpredicts that for
the neutron-poor nuclei.  Using the neutron and proton density distributions 
predicted by the {\sc rmf} theory, the data 
can be well reproduced as shown by the dashed line. 
The increasing difference between the dashed and solid lines 
reflects the effects due to the larger neutron-skin of 
Na isotopes. In fact, a gradual increase of the neutron-skin up to 0.4 fm 
for neutron-rich Na isotopes has been deduced from fitting the total reaction
cross sections using the Glauber model \cite{suz95}. 
      
In the optical limit, the Glauber model predicts a reaction cross section
\begin{equation}
\sigma_R=2\pi\int_0^{\infty}[1-T(b)]bdb,
\end{equation}
where $T(b)$ is the transmission function at an impact parameter $b$.
In the standard Glauber model, $T(b)$ is evaluated from a convolution of the
nucleon-nucleon cross section $\sigma_{kl}$, with the density distributions 
of target $(T)$ and projectile $(P)$ nucleons in the overlapping 
region \cite{kar75}, i.e., 
\begin{equation}\label{tb}
T(b)=exp\left(-\sum_{kl}\sigma_{kl}\int \rho_{Tk}^z(\vec{s})
\rho_{Pl}^z(\vec{b}-\vec{s})d\vec{s}\right),
\end{equation}
where the indices $k, l$ are used to distinguish neutrons and protons and 
$\rho^z$ is the nucleon density distribution along the beam direction,
i.e.,
\begin{equation}
\rho^z(\vec{s})=\int_{-\infty}^{\infty}\rho(\sqrt{s^2+z^2})dz.
\end{equation}
\begin{figure}[htp]
%\vspace{12cm}
\vspace{-5.0truecm}
\setlength{\epsfxsize=10truecm}
\centerline{\epsffile{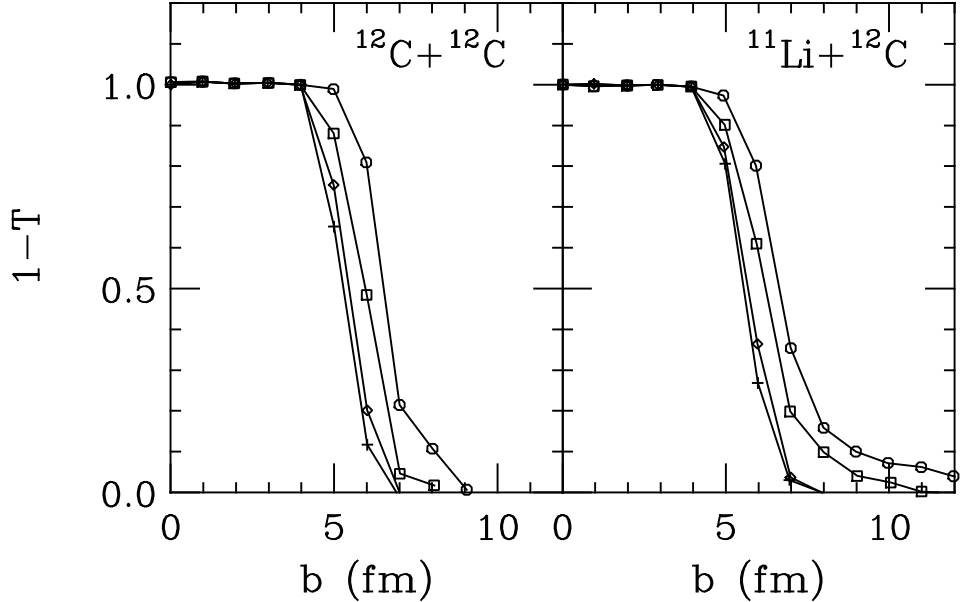}}
\caption{The impact parameter dependence of the reaction probability $1-T(b)$
for the reactions of $^{12}{\rm C}+^{12}{\rm C}$ (left window) and 
$^{11}{\rm Li}+^{12}{\rm C}$ (right window) at four beam energies.
Circles, squares, diamonds, and crosses represent the results for E/A = 
43, 75, 400, and 800 MeV, respectively.
Taken from Ref. \protect\cite{shen1}.}
\label{ma1} 
\end{figure} 

In transport models, the transmission function $T(b)$ can be calculated
directly by simply counting events in which no nucleon-nucleon collision 
have occurred.  This approach, however, is hampered by
large statistical fluctuations. 
It is thus numerically more reliable to determine $T(b)$ by using
the observation that the number of nucleon-nucleon collisions encountered by a
nucleon in heavy ion collisions at all impact parameters 
and for both low and high beam energies 
can be very well described by a Poisson distribution
\cite{bauer88,cole,har85,jac87}. Therefore, the transmission function at 
a given impact parameter $b$ can be simply expressed by the average number
of collisions per nucleon $\overline{N}(b)$, i.e.,
\begin{equation}
T(b)={\rm exp}\left(-\overline{N}(b)\right).
\end{equation}
The isospin dependence of the reaction dynamics 
is contained in $\overline{N}(b)$. However, the isospin-dependent
transport models have not been used to calculate the total reaction cross
section. Instead, the normal isospin-independent {\sc buu} model was used by
Shen {\it et al.} in calculating the total cross sections \cite{shen1}. 
In their study neutrons and protons are distributed randomly in a sharp 
sphere with a radius of $r=r_0A^{1/3}$. By varying the nuclear  {\sc eos}, 
in-medium 
nucleon-nucleon cross section and the radius parameter $r_0$ 
Shen {\it et al.} have already demonstrated the usefulness of 
transport models in studying the total reaction cross section. 
Fig.\ \ref{ma1} shows the reaction probability $1-T(b)$ as a function of impact
parameter for the reaction of $^{12}{\rm C}+^{12}{\rm C}$ and 
$^{11}{\rm Li}+^{12}{\rm C}$ at the four
beam energies from 43 to 800 MeV/nucleon. It is seen that in central 
collisions these nuclei behave as black spheres. 
The observed energy dependence of the total reaction cross section is due to 
the energy dependence of partial transparency in peripheral collisions.

\begin{figure}[htp]
%\vspace{15cm}
\vspace{-5.0truecm}
\setlength{\epsfxsize=10truecm}
\centerline{\epsffile{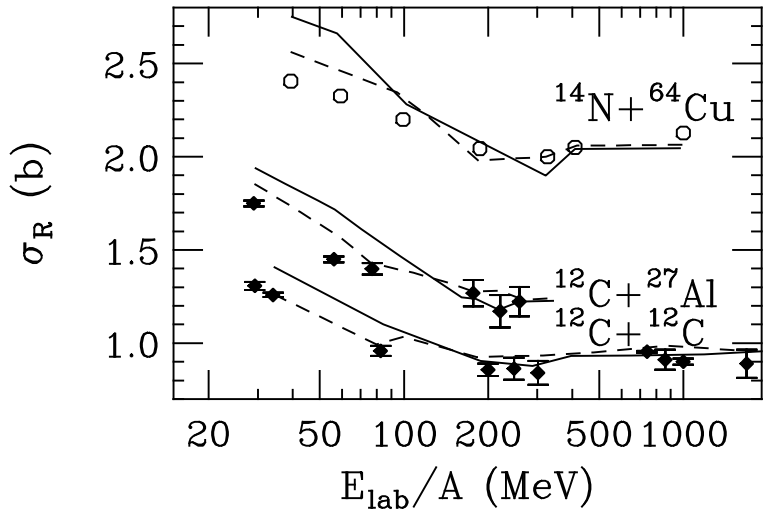}}
\caption{The energy dependence of the total reaction cross section 
for three different reactions.
The data are shown by filled circles while the solid (dashed) 
lines are calculations with the stiff (soft) 
nuclear  {\sc eos}. Taken from Ref. \protect\cite{shen1}.}
\label{ma2} 
\end{figure} 
The calculated and measured energy dependence 
of the total reaction cross section are compared in Fig.\ \ref{ma2}.
The filled circles are the experimental data, while the solid (dashed) 
lines are the {\sc buu} calculations with a stiff (soft) nuclear  {\sc eos}. 
Both calculations can reproduce the data at beam energies above about 
100 MeV/nucleon. The dependence of the total cross section to the nuclear 
 {\sc eos} appears at lower beam energies, but this is exactly where the 
Glauber-type models underpredict the
total reaction cross sections \cite{yab92,oga92}.   

\begin{figure}[htp]
%\vspace{8cm}
\vspace{-5.0truecm}
\setlength{\epsfxsize=10truecm}
\centerline{\epsffile{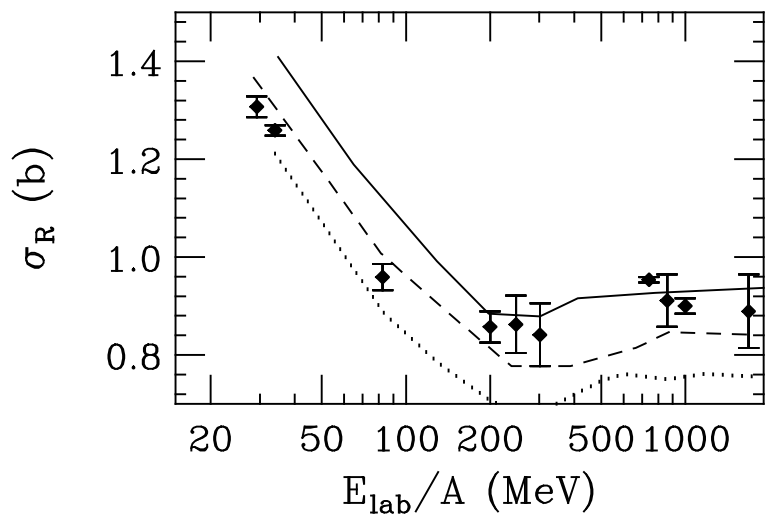}}
\caption{Effects of in-medium nucleon-nucleon cross sections on 
the energy dependence of total reaction cross section 
for the reaction of $^{12}{\rm C}+^{12}{\rm C}$. 
The data are plotted with filled circles. 
All calculations are carried out with the stiff 
nuclear  {\sc eos}. The solid line is obtained using the 
free-space $pp$ cross section; while the dashed and dotted
lines are calculated with a reduction of the cross section by 
20\% and 40\%, respectively. 
Taken from Ref. \protect\cite{shen1}.}
\label{ma3} 
\end{figure}

{\sc buu} calculations generally overpredict the measured
cross sections at beam energies below 100 MeV/nucleon. This might be related
to the in-medium nucleon-nucleon cross sections used in the model.
In Ref.\ \cite{shen1} a parameterization which fits 
the proton-proton data and has a cutoff of 55 $mb$ at lower beam energies 
has been used for all nucleon-nucleon collisions. If the nucleon-nucleon cross
section is modified in medium, then the total reaction cross section is
expected to be affected as well. This effect is shown in Fig.\ \ref{ma3}, 
where it is seen that for all incident energies a 20\% variation of the 
nucleon-nucleon cross section results in about 15\% variation of the total
reaction cross section. This variation appears to be larger than that found
in Ref. \cite{bertsch89} using the Glauber model where a 7\% reduction 
has been found by reducing the effective, energy-independent nucleon-nucleon 
cross section 40 mb by 20\%. Furthermore, a 20\% 
reduction the {\sc buu} model can better reproduce the data. This seems to 
be consistent with that from studying the balance energy \cite{wes93} 
and the impact parameter dependence of the flow parameter \cite{hun96}.
However, one should note that the 20\% reduction of the 
proton-proton cross section used in Ref.\ \cite{shen1} for all nucleon-nucleon 
collisions corresponds to a larger reduction of the in-medium nucleon-nucleon  
cross section if its isospin dependence is taken into account.

\begin{figure}[htp]
%\vspace{8cm}
\vspace{-6.0truecm}
\setlength{\epsfxsize=10truecm}
\centerline{\epsffile{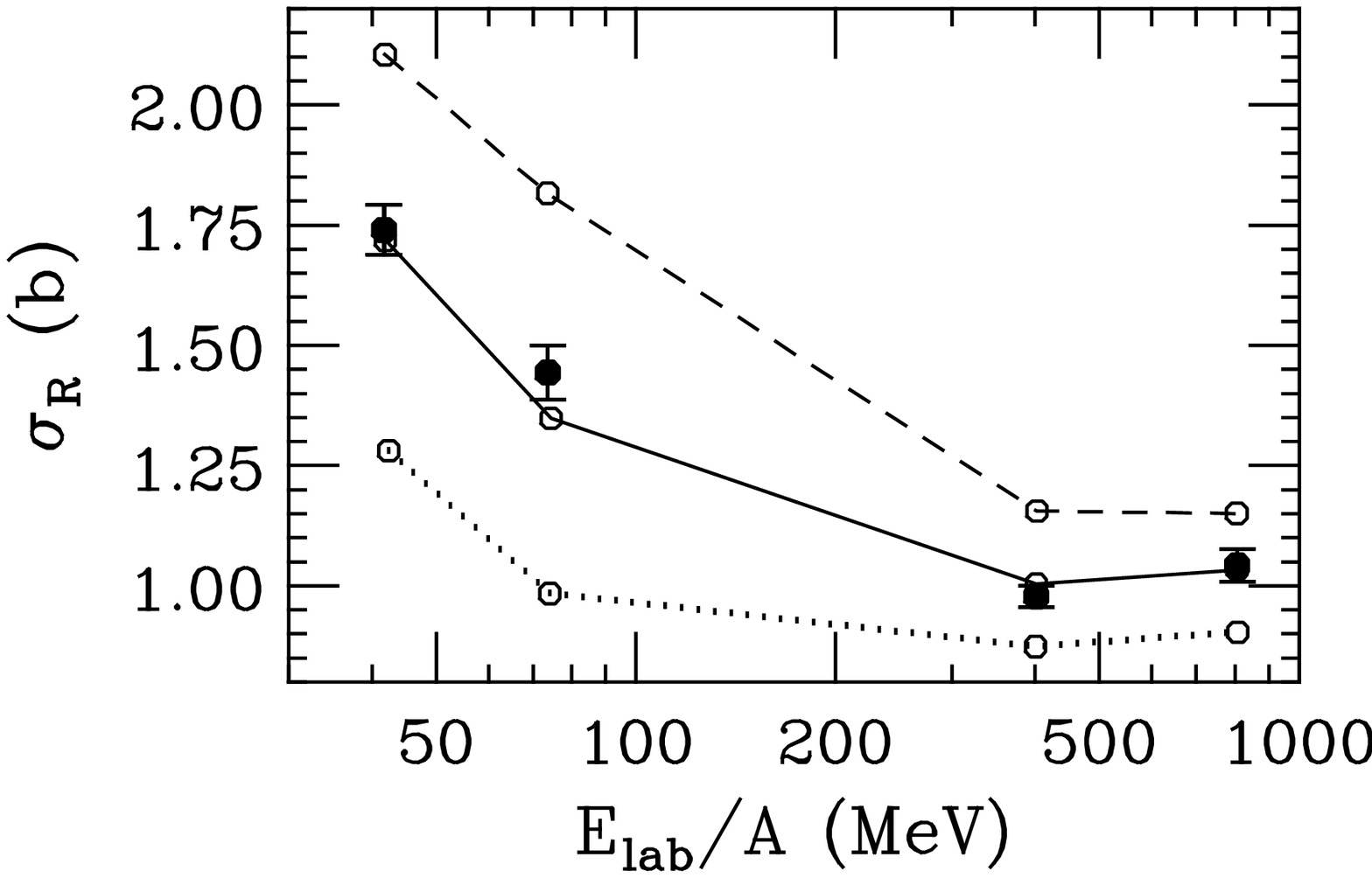}}
\caption{Effects of radius parameter $r_0$ on the 
energy dependence of total reaction cross section 
for the reaction of $^{11}{\rm Li}+^{12}{\rm C}$. 
The data are plotted with filled circles. 
The calculations are represented by the open symbols and connected
by the dashed, solid, and dotted lines, representing the choices of $r_0=1.8$,
1.6, and 1.33 fm, respectively.
All calculations are carried out with the stiff 
nuclear  {\sc eos}. Taken from Ref. \protect\cite{shen1}.}
\label{ma4} 
\end{figure}

Transport models may also be useful for studying the isospin-dependence of
the radius parameter $r_0$. Keeping the $r_0$ parameter fixed at 1.33 fm 
for $^{12}{\rm C}$ the calculated total reaction cross section 
for $^{11}{\rm Li}+^{12}{\rm C}$ has been
found to be very sensitive to the $r_0$ parameter for $^{11}{\rm Li}$.
Shown in Fig.\ \ref{ma4} are the $r_0$ dependence of the total reaction cross
section. As one expects, the total reaction cross section increases 
significantly with increasing $r_0$ for $^{11}{\rm Li}$. A radius parameter 
$r_0=1.60$ fm is needed to fit the cross section data.

From the above discussion it is clear that transport models are useful for
studying the total reaction cross section, especially at low beam energies
where Glauber-type models are not valid. So far only the isospin-independent 
{\sc buu} model has been used to study the total reaction cross section and its 
dependence on the nuclear  {\sc eos}, in-medium nucleon-nucleon 
cross section, and the
size of neutron-rich nuclei. It will be interesting to repeat such  
studies using the isospin-dependent transport models.

\section{Isospin dependence of subthreshold pion production}
The multiplicity ratio $\pi^{-}/\pi^{+}$ of charged pions in heavy-ion 
collisions is closely related to the ratio $N/Z$ in the participant region.
A systematic study of the relation between the two ratios are useful not
only for extracting information about the nucleon distributions  
but also for studying the production mechanism of subthreshold pions.

\begin{figure}[htp]
%\vspace{10cm}
\vspace{-5.0truecm}
\setlength{\epsfxsize=10truecm}
\centerline{\epsffile{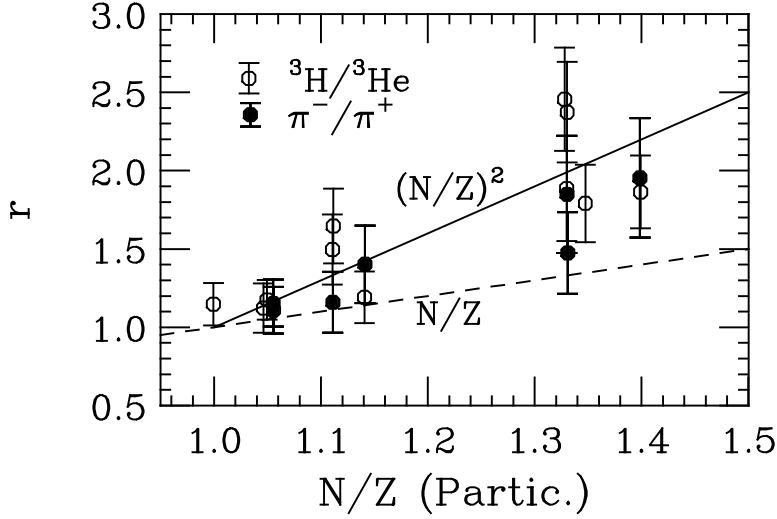}}
\caption{The ratio of $\pi^-/\pi^+$ as a function of $N/Z$ ratio 
of the participants in the reaction of Ar+KCl. 
Taken from Ref. \protect\cite{nag81}.}
\label{pion} 
\end{figure}

Assuming that pions are produced through the intermediate 
$\Delta(1232)$ resonance, i.e., $NN\rightarrow N\Delta(1232)$ 
and $\Delta\rightarrow N+\pi$, the ratio $\pi^-/\pi^+$ can be
estimated \cite{stock86,li95} and is given by 
\begin{equation}\label{pionr}
\frac{\pi^{-}}{\pi^{+}}=\frac{5N^{2}+NZ}{5Z^{2}+NZ}
\approx \left(\frac{N}{Z}\right)^2.
\end{equation}
This simple scaling formula is supported by data from relativistic heavy-ion 
experiments at the {\sc bevalac} \cite{stock86,nag81}. Fig.\ \ref{pion} shows 
the $\pi^-/\pi^+$ ratio as a function of $N/Z$ ratio of the participants 
in the reaction of Ar+KCl at a beam energies of 2.1 GeV/nucleon.
It is seen that the $\pi^-/\pi^+$ ratio scales well with $(N/Z)^2$ of the
participants.
At subthreshold energies, however, it has been found that the ratio 
$\pi^-/\pi^+$ scales approximately linearly with $N/Z$ from in 
reactions induced by 
$^{12}{\rm C}$ on $^{7}{\rm Li},~ ^{12}{\rm C},~ ^{116}{\rm Sn}$ and 
$^{124}{\rm Sn}$ at $E_{\rm beam}/A=60-85$ MeV \cite{nore88,ber84}. 
For example, at $E_{\rm beam}/A=$85 MeV the relative ratio for reactions with
$^{124}{\rm Sn}$ and $^{116}{\rm Sn}$ targets is
\begin{equation}
\left(\frac{\sigma_{\pi^-}}{\sigma_{\pi^+}}\right)(^{124}{\rm Sn})/  
\left(\frac{\sigma_{\pi^-}}{\sigma_{\pi^+}}\right)(^{116}{\rm Sn})
\approx 1.2\pm 0.3,
\end{equation}
which is compatible with the ratio of neutrons in the two targets.  
Such a linear dependence of $\pi^-/\pi^+$ on $N/Z$ at subthreshold 
energies is in contrast to the relation found at relativistic energies,
although the available data is rather limited at both energies.
It is also interesting to note that an anomalously 
large ratio of $\pi^-/\pi^+\approx 7 $ has been observed in the reaction of
$^{12}{\rm C}+^{208}{\rm Pb}$ at $E_{\rm beam}/A=85$ MeV \cite{ber84}. 
This ratio cannot be explained by the neutron-excess in the 
$^{208}{\rm Pb}$ target. Thus, the relation between $\pi^-/\pi^+$ and 
$N/Z$ of the participants has 
not been well established experimentally at subthreshold energies. 
The simple scaling law of Eq. (\ref{pionr}) thus needs to be tested in 
a wide range of beam energies and $N/Z$ ratios. High intensity beams of 
neutron-rich or radioactive nuclei are useful for this purpose.

The possibility of using the $\pi^-/\pi^+$ ratio to extract information about 
the distribution of neutrons and protons in neutron-rich nuclei has been
studied using the Glauber model \cite{lom88,libhu,bauLA}. 
Under the assumption that the nucleon density distribution is a 
Gaussian function, i.e., 
\begin{equation}
\rho(r)=\rho(0){\rm exp}(-\frac{r^2}{a^2}),
\end{equation}
the integration in Eq. (\ref{tb}) can be carried out analytically to yield the
following result
\begin{equation}
T(b)={\rm exp}(-\overline{N(b)}),
\end{equation}
with the average number of nucleon-nucleon collisions per nucleon in the 
reaction of $A+B$ at an impact parameter $b$ 
\begin{equation}
\overline{N(b)}=\frac{\bar{\sigma}\pi^2\rho_A(0)\rho_B(0)a_A^3a_B^3}
{a_A^2+a_B^2}\cdot {\rm exp}\left(-\frac{b^2}{a_A^2+a_B^2}\right).
\end{equation}
In the above $\bar{\sigma}$ is the isospin-averaged nucleon-nucleon cross 
section. Similarly, the average number of proton-proton, neutron-neutron and 
proton-neutron collisions can be obtained. If one assumes that 
pions are produced through $\Delta(1232)$ resonances, the 
inclusive $\pi^+$ and $\pi^-$ cross sections can be written 
as \cite{libhu,tel87}
\begin{eqnarray}
\frac{d\sigma_{\rm inc}^{\pi^+}}{d\Omega}&=&|f_{N\Delta}|^2Z_AZ_B
\frac{\pi^2\rho_{Z_A}(0)\rho_{Z_B}(0)a^3_{Z_A}a^3_{Z_B}}
{a^2_{Z_A}+a^2_{Z_B}}\nonumber\\
&\cdot& 2\pi\int_0^{\infty}bdb\cdot{\rm exp}
\left(-\frac{b^2}{a^2_{Z_A}+a^2_{Z_B}}\right)\cdot P_s,
\end{eqnarray}
\begin{eqnarray}
\frac{d\sigma_{\rm inc}^{\pi^-}}{d\Omega}&=&|f_{N\Delta}|^2N_AN_B
\frac{\pi^2\rho_{N_A}(0)\rho_{N_B}(0)a^3_{N_A}a^3_{N_B}}
{a^2_{N_A}+a^2_{N_B}}\nonumber\\
&\cdot& 2\pi\int_0^{\infty}bdb\cdot{\rm exp}
\left(-\frac{b^2}{a^2_{N_A}+a^2_{N_B}}\right)\cdot P_s,
\end{eqnarray}
where
\begin{equation}
P_s\equiv{\rm exp}\left[-\frac{\bar{\sigma}(AB-1)\rho_A(0)\rho_B(0)a_A^3a_B^3}
{a^2_A+a^2_B}\cdot{\rm exp}\left(-\frac{b^2}{a_A^2+a_B^2}\right)\right],
\end{equation}
is the elastic survival probability, $\rho_{N_i}$ and $\rho_{Z_i}$ 
are the neutron and proton densities of the
nucleus $i=A$ or $B$, and $f_{N\Delta}$ is the amplitude for the reaction 
$N+N\rightarrow N+\Delta$. Knowing the density distribution of 
neutrons and protons in both the target and projectile, one can then calculate
the ratio of $\pi^-/\pi^+$ from the above relations. 
It has been found that this ratio is rather sensitive to the difference 
between the neutron and proton density distributions. 
Shown in Table \ref{rpion} are the ratio $E(\bar{\sigma})$ defined as
\begin{equation}
E\equiv \frac{\sigma_{\rm inc}^{\pi^-}-\sigma_{\rm inc}^{\pi^+}}  
{\sigma_{\rm inc}^{\pi^-}+\sigma_{\rm inc}^{\pi^+}.}  
\end{equation}
In this calculation \cite{libhu} density distributions from the binding 
energy adjusted shell model \cite{bertsch89} are used.
Results using two isospin-averaged nucleon-nucleon cross
sections of 40 mb and 25 mb are compared. It is seen that the ratio $E$ 
is not so sensitive to the cross section $\bar{\sigma}$, but is 
sensitive to the difference between the density distributions of 
neutrons and protons.
This prediction has attracted some experimental interest \cite{she94} 
and can be tested in future experiments with high intensity 
radioactive beams.
 
\begin{table}
\caption{Normalized difference between $\pi^-$ and $\pi^+$ production cross
sections for reactions of ${\rm Li}$ isotopes on $^{12}{\rm C}$}
\label{rpion}
\medskip
\centerline{
\begin{tabular}{ccccccccccccccc}
\hline\\
%\hline\hline\\
\multicolumn{1}{c}{Ratio E($\bar{\sigma}$)} &\multicolumn{1}{c}{$^{7}Li+^{12}C$}
&\multicolumn{1}{c}{$^{8}Li+^{12}C$}&\multicolumn{1}{c}{$^{9}Li+^{12}C$}
&\multicolumn{1}{c}{$^{11}Li+^{12}C$}\\\\
\hline\\
\multicolumn{1}{c}{$\bar{\sigma}=40$ mb} &\multicolumn{1}{c}{0.1153}
&\multicolumn{1}{c}{0.2221}&\multicolumn{1}{c}{0.2955}
&\multicolumn{1}{c}{0.3951}\\\\
\multicolumn{1}{c}{$\bar{\sigma}=25$ mb} &\multicolumn{1}{c}{0.1143}
&\multicolumn{1}{c}{0.2210}&\multicolumn{1}{c}{0.2939}
&\multicolumn{1}{c}{0.3927}\\\\
\hline\\
%\hline\hline
\end{tabular}
}
\end{table}

\chapter{Overview and perspectives}
Calculations based on microscopic many-body theories have predicted many 
interesting new physics phenomena in isospin-asymmetric nuclear matter.
These theoretical studies have shown that the equation of state of 
isospin-asymmetric nuclear matter 
depends quadratically on the neutron excess $\delta$. The coefficient 
of the quadratic term in the  {\sc eos} of asymmetric nuclear matter, i.e., 
the symmetry energy, and especially its density dependence vary widely among
theoretical predictions. In particular, the symmetry energy at 
high densities suffers the most severe uncertainties among all properties 
of dense nuclear matter. 

Theoretical studies also indicate that the isospin-dependent {\sc eos} has
significant influence on the properties of neutron stars as well as 
on those of radioactive
nuclei. The softening of the equation of state in isospin-asymmetric 
nuclear matter is important for the prompt explosion of a type {\sc ii}
supernova. The proton concentration in neutron stars at $\beta$ equilibrium 
is almost completely determined by the symmetry energy. Consequently, the
cooling rate and neutrino emissions of neutron stars are affected 
significantly by the symmetry energy, especially its density dependence.
The binding energy, radii and deformation of radioactive nuclei are also shown
to depend critically on the isospin-dependence of the nuclear {\sc eos}.  
The nucleon-nucleon cross sections in free space are known to be also strongly 
isospin-dependent. Nuclear many-body theories have predicted that the 
isospin-dependence of nucleon-nucleon cross sections 
might be significantly altered by the nuclear medium.

Because of the additional degree of freedom, 
isospin, new phenomena are predicted to occur in isospin-asymmetric 
nuclear matter. In particular, the chemical instability is predicted to 
occur in a larger configuration space than the mechanical instability. 
Consequently, the spinodal decomposition in asymmetric nuclear matter 
might be triggered by fluctuations 
in the isospin asymmetry rather than baryon density.  
This may have significant effects on the mechanism for nuclear 
multifragmentations in nuclear reactions involving isospin asymmetric 
nuclear matter.
The liquid-gas phase transition in asymmetric nuclear matter 
has been predicted to be second-order rather than first-order as in 
symmetric nuclear matter. Moreover, the critical temperature of liquid-gas 
phase transition gradually decreases to zero from symmetric nuclear matter to
pure neutron matter. 

The predicted new physics phenomena in isospin asymmetric nuclear 
matter not only motivates further theoretical studies but also 
calls for dedicated experiments to test them.
Energetic collisions of stable and radioactive neutron-rich nuclei can create
transiently nuclear matter 
with appreciable isospin asymmetry and matter compression in the overlap 
region,
thus providing a testing ground for studying isospin
physics. In fact, many interesting isospin effects have already been observed
in heavy-ion collisions. For example, isospin non-equilibrium has been 
observed in the angular distribution of isotope ratios. Interesting
isospin-dependence of nuclear multifragmentation has been observed from 
comparing reactions of isospin-symmetric systems and highly 
isospin-asymmetric ones.
Abnormally preferential emission of preequilibrium neutrons have been observed
in many experiments. Very recently, the isospin-dependence of nuclear 
collective
flow and balance energy has been first predicted by using isospin-dependent 
transport models and later confirmed by experiments. 
The theoretical understanding of the observed isospin-dependent 
phenomena has been limited to the qualitative level in most cases. 
For example, isospin-dependent transport 
models, though very useful for predicting and explaining many observations 
qualitatively, are still unable to explain the data 
quantitatively. Nonetheless, these models are very useful in establishing 
connections between the experimental observables and the 
isospin-dependent equation of
state and in-medium cross sections. For instance, it has been shown that the
neutron/proton ratio of preequilibrium nucleons is very sensitive to the 
density dependence of the symmetry energy, but not sensitive to the in-medium 
nucleon-nucleon cross sections. On the other hand, the isospin-dependence 
of the collective flow and balance energy is mainly caused by the 
isospin-dependence of the in-medium nucleon-nucleon cross sections.     

The main goal of studying isospin physics in heavy-ion collisions 
is to explore the properties of nuclear matter in the transition region 
between symmetric nuclear matter 
and neutron matter and its equation of state. Dedicated radioactive 
beam facilities will provide excellent opportunities to achieve this goal. 
We anticipate that there will be substantial experimental activities in 
studying isospin physics in the coming years. To extract from these 
experiments more quantitative information about the  {\sc eos} of asymmetric 
nuclear matter and isospin-dependence of in-medium nucleon-nucleon 
cross sections, it is necessary to combine theoretical and computational 
techniques in the area of nuclear structure, reaction and astrophysics. 
With a continued dialog between theories and experiments,
we believe that this goal will be achieved.

\chapter{Acknowledgement}
We would like to thank R. Pak, J. Randrup, G.D. Westfall, S.J. Yennello,
and Zhongzhou Ren for collaboration in the study of several topics discussed 
in this review. We are also grateful to J.B. Natowitz, L.G. Sobotka, W.Q. Shen
and W.U. Schr\"oder for helpful conversations on the subject. The work of 
BAL and CMK was supported in part by the NSF Grant No. PHY-9509266 and the 
Robert A. Welch Foundation under Grant A-1358, and that of WB was supported 
in part by the NSF Grant No. PHY-9403666 and the NSF Presidential 
Faculty Fellow grant No. PHY-9253505.


\begin{thebibliography}{999}
\bibitem{iso91} North American Steering Committee for the Isospin 
      Laboratory, research opportunities with radioactive nuclear beams. 
      Report LALP91-51.

\bibitem{hus91} M.S. Hussein, R.A. Rego and C.A. Bertulani, 
      Phys. Rep. {\bf 201}, 279 (1991).

\bibitem{boyd92} R.N. Boyd and I. Tanihata, 
      Phys. Today, {\bf 45}(6), 45 (1992).

\bibitem{ms93} A.C. M\"uller and B.M. Sherrill, 
      Ann. Rev. Nucl. Part. Sci. {\bf 43}, 529 (1993).
 
\bibitem{tanihata95} I. Tanihata, Prog. of Part. and Nucl. Phys., 
      {\bf 35} (1995) 505.

\bibitem{gei95} H. Geissel, G. M\"unzenberg, and K. Riisageer,
      Ann. Rev. Nucl. Part. Sci. {\bf 45}, 163 (1995).

\bibitem{hansen95} P.G. Hansen, A.S. Jensen and B. Jonson, Ann. Rev. Nucl.
      Part. Sci. {\bf 45}, 591 (1995).

\bibitem{spiral} The SPIRAL radioactive ion beam facility, 
      Ganil R 9402, May, 1994. 

\bibitem{bck} E. Baron, J. Cooperstein and S. Kahana, 
      Phys. Rev. Lett. {\bf 55}, 126 (1985); 
      Nucl. Phys. {\bf A440}, 744 (1985).

\bibitem{kahana} S.H. Kahana, Ann. Rev. Nucl. Part. Sci., {\bf 39}, 231 (1989).

\bibitem{lat91} J.M. Lattimer, C.J. Pethick, M. Prakash and P. Haensel,
      Phys. Rev. Lett. {\bf 66}, 2701 (1991).

\bibitem{sum94} K. Sumiyoshi and H. Toki, 
      Astro. Phys. Journal, {\bf 422}, 700 (1994).

\bibitem{chlee}C-H. Lee, Phys. Rep. {\bf 275}, 255 (1996).

\bibitem{bru} K.A. Brueckner, S.A. Coon and J. Dabrowski, 
      Phys. Rev. {\bf 168}, 1184 (1967).
 
\bibitem{serot}B.D. Serot and J.D. Walecka, 
      Adv. Nucl. Phys. {\bf 16}, 1 (1986).

\bibitem{muller} H. M\"uller and B.D. Serot, 
      Phys. Rev. C {\bf 52}, 2072 (1995).
 
\bibitem{siemens70} P.J. Siemens, Nucl. Phys. {\bf A141}, 225 (1970).

\bibitem{sjo74}	O. Sj\"oberg, Nucl. Phys. {\bf A222}, 161 (1974).

\bibitem{cug87} J. Cugnon, P. Deneye and A. Lejeune, 
      Z. Phys. A {\bf 328}, 409 (1987).

\bibitem{bom91} I. Bombaci and U. Lombardo, Phys. Rev. C{\bf 44}, 1892 (1991).

\bibitem{mut87} H. M\"uther, M. Prakash and T.L. Ainsworth, 
      Phys. Lett. {\bf B199}, 469 (1987).

\bibitem{har87} B. ter Haar and R. Malfliet, Phys. Rev. C{\bf 50}, 31 (1994).

\bibitem{sumi92} K. Sumiyoshi, H. Toki and R. Brockmann, 
      Phys. Lett. {\bf B276}, 393 (1992). 

\bibitem{hub93} H. Huber, F. Weber and M.K. Weigel, 
      Phys. Lett. {\bf B317}, 485 (1993); Phys. Rev. C{\bf 50}, R1287 (1994).

\bibitem{fri81} B. Friedman and V.R. Pandharipande,
      Nucl. Phys. {\bf A361}, 502 (1981).

\bibitem{laga81} I.E. Lagaris and V.R. Pandharipande, 
      Nucl. Phys. {\bf A369}, 470 (1981)

\bibitem{wiringa88} R.B. Wiringa, V. Fiks and A. Fabrocini, 
      Phys. Rev. C{\bf 38}, 1010 (1988).

\bibitem{chin77} S.A. Chin, Ann. Phys. (N.Y.), {\bf 108}, 301 (1977).

\bibitem{horo87} C.J. Horowitz and B.D. Serot, 
      Nucl. Phys. {\bf A464}, 613 (1987);
      B.D. Serot and H. Uechi, Ann. Phys. (N.Y.) {\bf 179}, 272 (1987).

\bibitem{glen82} N.K. Glendenning, Phys. Lett. {\bf B114}, 392 (1982).

\bibitem{hir91} D. Hirata {\it et al.}, Phys. Rev. C{\bf 44}, 1467 (1991).

\bibitem{sug94} Y. Sugahara and H. Toki, Nucl. Phys. {\bf A579}, 557 (1994).

\bibitem{lopez88} M. Lopez-Quelle {\it et al.}, 
      Nucl. Phys. {\bf A483}, 479 (1988).

\bibitem{wer94} T.R. Werner {\it et al.}, Phys. Lett. {\bf B333}, 303 (1994).

\bibitem{kho96} Dao T. Khoa, W. Von Oertzen and A.A. Ogloblin, 
      Nucl. Phys. {\bf A602}, 98 (1996).

\bibitem{kar85} K. Kolehmainen {\it et al.}, 
      Nucl. Phys. {\bf A439}, 535 (1985);
      J. Treiner {\it et al.}, Ann. Phys. (N.Y.), {\bf 170}, 406 (1986).

\bibitem{band90}D. Bandyopadhyay, C. Samanta, S.K. Samaddar and J.N. De,
	Nucl. Phys. {\bf A511}, 1 (1990).

\bibitem{pra87} M. Prakash and T.L. Ainsworth, 
      Phys. Rev. C {\bf 36}, 346 (1987).

\bibitem{bro90}	R. Brockmann and R. Machleidt, 
      Phys. Rev. C{\bf 42}, 1965 (1990); 
      H. M\"uther, R. Machleidt and R. Brockmann,
      {\it ibid} C{\bf 42}, 1981 (1990).

\bibitem{tanihata96} I. Tanihata, Preprint RIKEN-AF-NP-229, July, 1996.

\bibitem{sumi93} K. Sumiyoshi, D. Hirata, H. Toki and H. Sagawa,
      Nucl. Phys. {\bf A552}, 437 (1993).

\bibitem{baym71} G. Baym, H.A. Bethe, and C.J. Pethick,
      Nucl. Phys. {\bf A175}, 225 (1971).

\bibitem{thor94} V. Thorsson, M. Prakash and J.M. Lattimer,
      Nucl. Phys. {\bf A572}, 693 (1994). 

\bibitem{prak88} M. Prakash, T.L. Ainsworth and J.M. Lattimer, 
      Phys. Rev. Lett. {\bf 61}, 2518 (1988).

\bibitem{mass} P.E. Haustein, Atomic data and nuclear data tables, 
      {\bf 39}, 185-395 (1988).

\bibitem{farine} M. Farine, J.M. Pearson and B. Rouben, 
      Nucl. Phys. {\bf A304}, 317 (1978).

\bibitem{pear} J.M. Pearson et. al., Nucl. Phys. {\bf A528}, 1 (1991).

\bibitem{rein88} P.-G. Reinhard, Z. Phys. {\bf A329}, 257 (1988).

\bibitem{rufa88} M. Rufa, P.-G. Reinhard, J. Maruhn, W. Greiner 
      and M.R. Strayer, Phys. Rev. C{\bf 38}, 390 (1988).

\bibitem{shar93} M.M. Sharma, M.A. Nagarajan and P. Ring, 
      Phys. Lett. {\bf B312}, 377 (1993).

\bibitem{scha96} J. Schaffner and I.N. Mishustin, 
      Phys. Rev. C{\bf 53}, 1416 (1996).
 
\bibitem{shl93} S. Shlomo and D.H. Youngblood, 
      Phys. Rev. C{\bf 47}, 529 (1993).

\bibitem{gle87} N.K. Glendenning, Z. Phys. {\bf A326}, 57 (1987); 
      {\it ibid}, {\bf A327}, 295 (1987).

\bibitem{kut94} M. Kutschera, Phys. Lett. {\bf B340},1; 
      Z. Phys. {\bf A348}, 263 (1994); M. Kutschera and W. W\'ojcik, 
      Phys. Lett. {\bf B223}, 11 (1989); Phys. Rev. C{\bf 47}, 1077 (1993).


\bibitem{pan72} V.R. Pandharipande and V.K. Garde, 
      Phys. Lett. {\bf B39}, 608 (1972).

\bibitem{pet95} C.J. Pethick and D.G. Ravenhall, Ann. Rev. Nucl.
      Part. Sci. {\bf 45}, 429 (1995). 

\bibitem{bethe90} H.A. Bethe, Rev. of Modern Physics, {\bf 62}, 801 (1990).

\bibitem{pra96} M. Prakash, I. Bombaci, M. Prakash, P.J. Ellis, J.M. Lattimer
      and R. Knorren, Phys. Rep. (1996) in press.

\bibitem{pra96b} M. Prakash, S. Reddy, J.M. Lattimer and P.J. Ellis,
      Heavy Ion Physics (1996) in press.

\bibitem{web96} F. Weber and N.K. Glendenning, Lecture notes at the 3rd 
      Mario Sch\"onberg School on Physics, July, 1996, to be published 
      in the Brasilian Journal of Teaching Physics.  

\bibitem{toki95} K. Sumiyoshi, H. Suzuki and H. Toki,
      Astronomy and Astrophysics, {\bf 303}, 475 (1995).

\bibitem{de97}V.S. Uma Maheswari, J.N. De and S.K. Samaddar, 
	Nucl. Phys. {\bf A615}, 516 (1997). 
 
\bibitem{don94} P. Donati, P.M. Pizzochero, P.F. Bortignon and R.A. Broglia,
      Phys. Rev. Lett. {\bf 72}, 2835 (1994).

\bibitem{dea95} D.J. Dean, S.E. Koonin, K.Langanke and P.B. Radha,
      Phys. Lett. {\bf B356}, 429 (1995).
 
\bibitem{eng94} L. Engvik {\it et al.}, 
      Phys. Rev. Lett. {\bf 73}, 2650 (1994).

\bibitem{pra88} M. Prakash, T.L. Ainsworth and J.M. Lattimer,
      Phys. Rev. Lett. {\bf 61}, 2518 (1988).

\bibitem{wer94b} T.R. Werner {\it et al.}, Phys. Lett. {\bf B335}, 259 (1994).

\bibitem{sch96} H. Scheit {\it et al.}, 
      Phys., Rev. Lett. {\bf 77}, 3967 (1996).

\bibitem{gla97} T. Glasmacher {\it et al.}, Phys.\ Lett.\ {\bf B395},
      163 (1997).

\bibitem{far91} M. Farine, T. Sami, B. Remaud and F. S\'ebille, 
      Z. Phys. {\bf A339}, 363 (1991).

\bibitem{sob94} L.G. Sobotka, Phys. Rev. C{\bf 50}, R1272 (1994).

\bibitem{abo92} Y. Aboussir {\it et al.}, Nucl. Phys. {\bf A549}, 155 (1992).

\bibitem{pawel} P. Danielewicz, Phys. Rev. C{\bf 46}, 2002 (1992).

\bibitem{lir93} B.A. Li and J. Randrup, HMI preprint (1993).

\bibitem{jou95} B. Jouault {\it et al.}, Preprint SUBATECH-95-11.

\bibitem{mye74} W.D. Myers and W.J. Swiatecki,
      Ann. of Phys. (N.Y.), {\bf 84}, 186 (1974).

\bibitem{lat78} J.M. Lattimer and D.G. Ravenhall, Astrophysical 
      Journal, {\bf 223}, 314 (1978). 

\bibitem{bar80} M. Barranco and J.R. Buchler, 
      Phys. Rev. C{\bf 22}, 1729 (1980).

\bibitem{shape1} L.G. Moretto {\it et al.},
      Phys.\ Rev.\ Lett.\ {\bf 69}, 1884 (1992).

\bibitem{shape2} W. Bauer, G.F. Bertsch, and H. Schulz,
      Phys.\ Rev.\ Lett.\ {\bf 69}, 1888 (1992); 
      L. Phair, W. Bauer, and C.K. Gelbke,
      Phys.\ Lett.\ {\bf B314}, 271 (1993).

\bibitem{shape3} D. Gross, Bao-An Li and A.R. DeAngelis, 
	Ann.\ Phys.\ {\bf 1}, 467 (1992); 
	Bao-An Li and D.H.E. Gross, Nucl. Phys. {\bf A554}, 257 (1993).

\bibitem{shape4} B. Borderie {\it et al.}, Phys.\ Lett.\ {\bf B302}, 15 (1993).

\bibitem{moretto} L.G. Moretto and G.J. Wozniak, Ann. Rev. Nucl. Part. Sci. 
      {\bf 43}, 123 (1993).

\bibitem{brack} M. Brack, C. Guet and H.-B H\'akansson, 
      Phys. Report, {\bf 123}, 275 (1985). 

\bibitem{jaqaman1} H.R. Jaqman, A.Z. Mekjian and L. Zamick, 
      Phys. ReV. C{\bf 27}, 2782 (1983); 
      {\it ibid} C{\bf 29}, 2067 (1984).  

\bibitem{jaqaman2} H.R. Jaqaman, Phys. ReV. C{\bf 39}, 169 (1988); 
      {\it ibid} C{\bf 40}, 1677 (1989).

\bibitem{liko97} B.A Li and C. M. Ko, Nucl. Phys. {\bf A618}, 498 (1997). 

\bibitem{gale87} C. Gale, G.F. Bertsch and S. Das Gupta, 
      Phys. Rev. C{\bf 35}, 1666 (1987).

\bibitem{zhang} J. Zhang, S. Das Gupta and C. Gale, 
      Phys.\ Rev.\ C{\bf 50}, 1617 (1994).

\bibitem{gale90} C. Gale, G.M. Welke, M Prakash, S.J. Lee, and S. Das Gupta,
      Phys.\ Rev.\ C{\bf 41}, 1545 (1990).

\bibitem{lik96} B.A. Li, C.M. Ko and Z.Z. Ren,
      Phys. Rev. Lett. {\bf 78}, 1644 (1997).

\bibitem{csernai92} L.P. Csernai, G. Fai, C. Gale and E. Osnes,
      Phys. Rev. C {\bf 46}, 736 (1992).

\bibitem{fai} V.K. Mishra, G. Fai, L.P. Csernai and E. Osnes,
      Phys. Rev. C{\bf 47}, 1519 (1993).

\bibitem{dem96} J.F. Dempsey {\it et al.}, Phys. Rev. C {\bf 54}, 1710 (1996).

\bibitem{kunde96} G.J. Kunde {\it et al.},
      Phys. Rev. Lett. {\bf 77}, 2897 (1996).

\bibitem{toke96} J. Toke {\it et al.}, Phys. Rev. Lett. {\bf 75}, 2920 (1996).

\bibitem{sob97} L.G. Sobotka, J.F. Dempsey, R.J. Charity, and P. Danielewicz,
      Phys.\ Rev.\ C {\bf 55}, 2109 (1997).

\bibitem{Kor97} G. Kortmeyer, W. Bauer, and G.J. Kunde,
      Phys.\ Rev.\ C {\bf 55}, 2730 (1997).

\bibitem{bsiemens} G.F. Bertsch and P.J. Siemens, 
      Phys. Lett. {\bf B126}, 9 (1983).
 
\bibitem{ayik1} S. Ayik and C. Gregoire, Phys. Lett. {\bf B212}, 269, (1988);
      Nucl. Phys. {\bf A513}, 187 (1990).

\bibitem{randrup1} J. Randrup and B. Remaud, 
      Nucl. Phys. {\bf A514}, 339 (1990).

\bibitem{randrup2} Ph. Chomaz, G.F. Burgio and J. Randrup; 
      Phys. Lett. {\bf B254}, 340 (1991); G.F. Burgio, Ph. Chomaz and 
      J. Randrup; Nucl. Phys. {\bf A529}, 157 91991); 
      Phys. Rev. Lett., {\bf 69}, 1888 (1992).

\bibitem{colonna} M. Colonna, Ph. Chomaz and J. Randrup, 
      Nucl. Phys. {\bf A567}, 637 (1994); M. Colonna and Ph. Chomaz, 
      Phys. Rev. C{\bf 49}, 1908 (1994).

\bibitem{ayik2} S. Ayik and J. Randrup, Phys. Rev. C{\bf 50}, 2947 (1994).

\bibitem{hama96}I. Hamamoto and H. Sagawa, Phys. Rev. C{\bf 53}, R1492 (1996).

\bibitem{ditoro}M. Colonna, M. Di Toro and A.B. Larionov, 
	Catania preprint, 1997.

\bibitem{lam81} D.Q. Lamb, J.M. Lattimer, C.J. Pethick and D.G. Ravenhall, 
      Nucl. Phys. {\bf A360}, 459 (1981).

\bibitem{lat85} J.M. Lattimer, C.J. Pethick, D.G. Ravenhall and D.Q. Lamb,
      Nucl. Phys. {\bf A432}, 646 (1985).

\bibitem{gle92} N.K. Glendenning, Phys. Rev. D{\bf 46}, 1274 (1992).

\bibitem{kuo96} T.T.S. Kuo, S. Ray, J. Shamanna and R.K. Su, 
      Int. Jour. of Modern Phys. E: Nucl. Phys. {\bf 5}, 303 (1996); 
      S. Ray, J. Shamanna and T.T.S. Kuo, 
      Phys. Lett. {\bf B} (1996) in press;
      T.T.S. Kuo, S. Ray, J. Shamanna, R.K. Su, SUNY-preprint, 1996

\bibitem{cha90} S.K. Charagi and S.K. Gupta, 
      Phys. Rev. C{\bf 41}, 1610 (1990).

\bibitem{nndata} G. Alkahzov {\it et al.}, Nucl. Phys. {\bf A280}, 365 (1977).

\bibitem{ber88} G.F. Bertsch, G.E. Brown, V. Koch and Bao-An Li,
      Nucl. Phys. {\bf A490}, 745 (1988).

\bibitem{boh89} A. Bohnet {\it et al.}, Nucl. Phys. {\bf A494}, 349 (1989).

\bibitem{fas89} A. Faessler, Nucl. Phys. {\bf A495}, 103c (1989).

\bibitem{koh91} H.S. K\"ohler, Nucl. Phys. {\bf A529}, 209 (1991).

\bibitem{gqli93} G.Q. Li and R. Machleidt, Phy.\ Rev. C{\bf 48}, 1702 (1993);
      {\it ibid}, C{\bf 49}, 566 (1994).

\bibitem{alm1} T. Alm, G. R\"opke and M. Schmidt, 
      Phys. Rev. C{\bf 50} (1994) 31.

\bibitem{alm2} T. Alm, G. R\"opke, W. Bauer, F. Daffin and M. Schmidt,
      Nucl. Phys. {\bf A587}, 815 (1995).

\bibitem{wes93} G.D. Westfall {\it et al.}, 
      Phys. Rev. Lett. {\bf 71}, 1986 (1993).

\bibitem{kla93} D. Klakow, G. Welke and W. Bauer, 
      Phys. Rev. C{\bf 48}, 1982 (1993).

\bibitem{hun96} M.J. Huang {\it et al.},
      Phys. Rev. Lett. {\bf 77}, 3739 (1996).

\bibitem{yen96} S.J. Yennello {\it et al.}, in Proc. of International
      Workshop on Physics of Unstable Nuclear beams, Serra Negra, Brazil, 
      Aug. 28-31, 1996. (World Scientific, Singapore).

\bibitem{remaud} B. Remaud, C. Gr\"egoire, F. S\"ebille and P. Schuck,
      Nucl.\ Phys.\ {\bf A488} (1988) 423c.

\bibitem{hart88}C. Hartnack, H. St\"ocker and W. Greiner, 
      in Proceedings of the International Workshop on Gross 
      Properties of Nuclei and Nuclear Excitation XVI, 
      Hirschegg, Austria, 1988, ed.\ H. Feldmeier, p.138.

\bibitem{betty} M.B. Tsang, G.F. Bertsch, W.G. Lynch and M. Tohyama, 
      Phys.\ Rev.\ C{\bf 40} (1989) 1685.

\bibitem{libauer1} B.A. Li and W. Bauer, 
      Phys. Lett. {\bf B254}, 335 (1991); Phys. Rev. C{\bf 44}, 450 (1991). 

\bibitem{libauer2} B.A. Li, W. Bauer and G.F. Bertsch, 
      Phys. Rev. C{\bf 44}, 2095 (1991). 

\bibitem{lis95} B.A. Li and S.J. Yennello, Phys. Rev. C{\bf 52}, R1746 (1995).

\bibitem{li96} B.A. Li, Z.Z. Ren, C.M. Ko and S.J. Yennello,
      Phys. Rev. Lett. {\bf 76}, 4492 (1996).

\bibitem{greiner} H. St\"ocker and W. Greiner, 
      Phys. Rep. {\bf 137}, 277 (1986).

\bibitem{bert88} G.F. Bertsch and S. Das Gupta, Phys.\
      Rep.\ {\bf 160}, 189 (1988).

\bibitem{cas90} W. Cassing, V. Metag, U. Mosel and K. Niita, 
      Phys. Rep.\ {\bf 188}, 363 (1990).

\bibitem{bauer92} W. Bauer, C.K. Gelbke, and S. Pratt, Ann.\ Rev.\ 
      Nucl.\ Part.\ Sci.\ {\bf 42} (1992) 77.

\bibitem{aich91} J. Aichelin, Phys. Rep.\ {\bf 202}, 233 (1991).

\bibitem{xu} H.M. Xu, Phys.\ Rev.\ Lett.\ {\bf 67}, 2769 (1992); 
      Phys.\ Rev.\ C{\bf 46}, R392 (1992).

\bibitem{li93} B.A. Li, Phys.\ Rev.\ C{\bf 48}, 2415 (1993).

\bibitem{chr95} J.A. Christley {\it et al.}, 
      Nucl. Phys. {\bf A587}, 390 (1995).

\bibitem{hod63} P.E. Hodgson, Phys. Lett. {\bf 6}, 75 (1963).

\bibitem{len89} H. Lenske, H.H. Wolter and H.G. Bohlen, 
      Phys. Rev. Lett. {\bf 62}, 1457 (1989).

\bibitem{udo} W.U. Schr\"oder, private communication.

\bibitem{ren} Z.Z. Ren, W. Mittig, B.Q. Chen, and Z.Y. Ma, 
      Phys. Rev. C{\bf 52}, R20 (1995).

\bibitem{hils87} D. Hilscher {\it et al.}, Phys.\ Rev.\ C{\bf 36}, 208 (1987).

\bibitem{hils88} D. Hilscher, in Proc. of a Specialists's Meeting on
      Preequilibrium Nuclear Reactions, Semmering, Austria, 10-12th,
      Feb. 1988, Ed.\ B. Strohmaier (OECD, Paris, 1988), p. 245.

\bibitem{hils92} D. Polster {\it et al.}, in Book of abstracts, International 
      Nuclear Physics Conference, Wiesbaden, Germany, 
      July 26-Aug. 1, 1992, p3.3.8. 

\bibitem{hils95} D. Polster {\it et al.}, Phys.\ Rev.\ C{\bf 51}, 1167 (1995).

\bibitem{udo2} D.K. Agnihotri et al., in Proceedings of the $13^{th}$
Winter Workshop of Nuclear Dynamics, Marathon, Florida, Feb. 1997, 
Eds. W. Bauer and A. Mignerey, Plenum Press, to be published.

\bibitem{gon90} M. Gonin {\it et al.}, Phys. Rev. C{\bf 42}, 2125 (1990).

\bibitem{han95} K.A. Hanold {\it et al.}, Phys. Rev. C{\bf 52}, 1462 (1995).

\bibitem{bass94} S.A. Bass, J. Konopka, M. Bleicher, 
      H. St\"ocker and W. Greiner, GSI annual report, p. 66 (1994).

\bibitem{sherry1} S.J. Yennello, B. Young, J. Yee, J.A. Winger, J. S. Winfield,
      G.D. Westfall, A. Vander Molen, B.M. Sherrill, J. Shea, E. Norbeck, 
      D.J. Morrissey, T. Li, E. Gualtieri, D. Craig, W. Benenson and D. Bazin,
      Phys.\ Lett.\ {\bf B321} (1994) 15.

\bibitem{sherry2} S. J. Yennello, J.A. Winger, H. Johnston, T. White, 
      E. Gualtieri, D. Craig, S. Hannuschke, J. Yee, R. Pak, A. Vander Molen, 
      W. Benenson, G.D. Westfall, T. Li, W.J. Liope, D. J. Morrissey, 
      J. S. Winfield and M. Steiner, TAMU Progress in Research, P. I-7, 1994.

\bibitem{sherry3} H. Johnston, J. Winger, T. White, B. Hurst, D. O'Kelly 
      and S. J. Yennello, Phys. Lett. {\bf B371}, 186 (1996).

\bibitem{sherry4} H. Johnston, T. White, Bao-An Li, E. Ramakrishnan, 
	J. Winger, D.J. Rowland, B. Hurst, F. Gimeno-Nogues, D. O'Kelly,
	Y-W. Lui and S. J. Yennello, Phys. Rev. C, (1997) in press.

\bibitem{yariv} Y. Yariv and Z. Frankel, Phys.\ Rev.\ C{\bf 26} (1982) 2138.

\bibitem{gatty} B. Gatty {\it et al.}, Z. Phys.\ {\bf A273} (1975) 65.

\bibitem{borderie} B. Borderie {\it et al.}, 
      in: Proc. of ACS Nuclear Chemistry Award Symposium, Anaheim, USA,
      April 2-4, 1995.

\bibitem{amd} A. Ono and H. Horiuchi, Phys. Rev. C{\bf 53}, 2958 (1996).

\bibitem{bauer88} W. Bauer, Phys. Rev. Lett. {\bf 61}, 2534 (1988).

\bibitem{gary} S. Das Gupta and G.D. Westfall, 
      Physics Today, {\bf 46}(5), 34 (1993).

\bibitem{exp1} D. Krofcheck {\it et al.}, 
      Phys. Rev. Lett. {\bf 63}, 2028 (1989).

\bibitem{exp2} C.A. Ogilvie {\it et al.}, Phys. Rev. C{\bf 40}, 2592; 
      {\it ibid}, C{\bf 42}, R10 (1990);\\ 
      Phys.\ Lett.\ {\bf B231}, 35 (1989).

\bibitem{exp3} J. P\'eter {\it et al.}, Phys.\ Lett.\ {\bf B237}, 187 (1990). 

\bibitem{exp4} J.P. Sullivan {\it et al.}, Phys.\ Lett.\ {\bf B249}, 8 (1990).

\bibitem{exp5} J. P\'eter {\it et al.}, Nucl.\ Phys.\ {\bf A519}, 611 (1990).

\bibitem{exp6} W.M. Zhang {\it et al.}, Phys.\ Rev.\ C{\bf 42}, R491 (1990).

\bibitem{exp7} D. Krofcheck {\it et al.}, Phys.\ Rev. C{\bf 43}, 350 (1991);
      {\it ibid} {\bf 46}, 1416 (1992).

\bibitem{exp8} W.K. Wilson {\it et al.}, Phys.\ Rev.\ C{\bf 45}, 768 (1992).  

\bibitem{exp9} G.D. Westfall {\it et al.}, Phys.\ Rev.\ Lett.\, 
      {\bf 71}, 1986 (1993).

\bibitem{dani85} P. Danielewicz and G. Odyniec, 
      Phys.\ Lett.\ {\bf B157}, 146 (1985).

\bibitem{moli1} J. Molitoris and H. St\"ocker, 
      Phys.\ Lett.\ {\bf B162}, 47 (1985).

\bibitem{moli2} J. Molitoris, D. Hahn and H. St\"ocker, 
      Nucl.\ Phys.\ {\bf A447}, 13c (1986).

\bibitem{bert87} G.F. Bertsch, W.G. Lynch and M.B. Tsang, 
      Phys.\ Lett.\ {\bf B189}, 738 (1987).

\bibitem{dani89} P. Danielewicz {\it et al.}, 
      Phys.\ Rev.\ C{\bf 38}, 120 (1989).

\bibitem{vd} V. de la Mota, F. Sebille, M. Farine, B. Remaud and P. Schuck, 
      Phys.\ Rev.\ C{\bf 46}, 677 (1992).

\bibitem{pan} Q. Pan and P. Danielewicz, 
      Phys.\ Rev.\ Lett.\ {\bf 70}, 2062 (1993).

\bibitem{dani96} P. Danielewicz, private communication.

\bibitem{pak1} R. Pak {\it et al.}, Phys. Rev. Lett. {\bf 78} 1022 (1997).

\bibitem{pak2} R. Pak {\it et al.}, Phys. Rev. Lett. {\bf 78} 1026 (1997).

\bibitem{pak3} R. Pak {\it et al.}, Phys. Rev. C{\bf 53}, 1469 (1996).

\bibitem{kox87} S. Kox {\it et al.}. Phys. Rev. C{\bf 35}, 1678 (1987).

\bibitem{yab92} K. Yabana, Y. Ogawa and Y. Suzuki, 
      Nucl. Phys. {\bf A539}, 295 (1992).

\bibitem{oga92} Y. Ogawa, K. Yabana and Y. Suzuki,
      Nucl. Phys. {\bf A543}, 722 (1992).

\bibitem{shen89} W.Q. Shen {\it et al.}, Nucl. Phys. {\bf A491}, 130 (1989).

\bibitem{kar75} P.J. Karol, Phys. Rev. C{\bf 11}, 1203 (1975).
 
\bibitem{mit87} W. Mittig {\it et al.}, Phys. Rev. Lett. {\bf 59}, 1889 (1987).

\bibitem{suz95} T. Suzuki {\it et al.}, Phys. Rev. Lett. {\bf 75}, 3241 (1995).

\bibitem{cole} A.J. Cole, Z. Phys. {\bf A322}, 315 (1985); 
      Phys. Rev. C{\bf 35}, 117 (1987); {\it ibid}, C{\bf 40}, 1265 (1989).

\bibitem{har85} B.G. Harvey, Nucl. Phys. {\bf A444}, 498 (1985).

\bibitem{jac87}A .D. Jackson and M. Boggild, 
      Nucl. Phys. {\bf A470}, 669 (1987).

\bibitem{shen1} Y.Q. Ma {\it et al.}, Phys. Lett. {\bf B302}, 386 (1993);
      Phys. Rev. C{\bf 48}, 850 (1993).

\bibitem{bertsch89} G.F. Bertsch, B.A. Brown and H. Sagawa, 
      Phys. Rev. C{\bf 39}, 1154 (1989).

\bibitem{stock86} R. Stock, Phys.\ Report, {\bf 135}, 259 (1986).

\bibitem{li95} B.A. Li, Phys.\ Lett. {\bf B346}, 5 (1995). 

\bibitem{nag81} S. Nagamiya, Phys. Rev. C{\bf 24}, 971 (1981).

\bibitem{nore88} B. Noren {\it et al.}, 
      Nucl.\ Phys.\ {\bf A489}, 763 (1988).

\bibitem{ber84} V. Bernard {\it et al.}, Nucl. Phys. {\bf A423}, 511 (1984).

\bibitem{lom88} R.J. Lombard and J.P. Maillet, 
      Europhys. Lett. {\bf 6}, 323 (1988).

\bibitem{libhu} B.A. Li, M.S. Hussein and W. Bauer, 
      Nucl. Phys. {\bf A533}, 749 (1991).

\bibitem{bauLA} W. Bauer {\it et al.}, in Proceedings of the Workshop
      on the Science of Intensive Radioactive Ion Beams, Los Alamos report
      LA-11964-C, p. 57 (1990).

\bibitem{tel87} A. Tellez, R.J. Lombard and J.P. Maillet, 
      J. of Phys. {\bf G13}, 311 (1987).

\bibitem{she94} B.M. Sherrill, Advances in Nuclear Dynamics, 
      eds. J. Harris, A. Mignerey and W. Bauer, 
      (World scientific, Singapore), 1994. P.70.

\end{thebibliography}
\end{document}